\documentclass[paper,toc,nohyper]{JHEP3}
\usepackage{epsfig,cite}
\usepackage{amsbsy} 
\usepackage{epsfig}

\def\ph#1{\phantom{.}}

\def\beq{\begin{equation}}   
\def\eeq{\end{equation}}
\def\bea{\begin{eqnarray}}  
\def\eea{\end{eqnarray}} 
\def\nn{\nonumber}

\def\doubletilde#1{\widetilde{\vphantom{\raise 1.5pt \hbox{#1}}\smash{\kern -2pt\widetilde{#1}}}}

\def\MA#1#2{{\cal M}^{#1}_{A,#2}}
\def\MB#1#2{{\cal M}^{#1}_{B,#2}}
\def\MC#1#2{{\cal M}^{#1}_{C,#2}}

\def\NF{N_F}

\def\Poles{{\cal P}oles}
\def\Finite{{\cal F}inite}
\def\Re{\mbox{Re}}
\def\bom#1{{\mbox{\boldmath $#1$}}}
\def\MSbar{$\overline{{\rm MS}}$}

\def\JET{J}

\def\e{\epsilon}
\def\d{\hbox{d}}

\def\Li{\hbox{Li}}
\def\ln{\hbox{ln}}

\def\nn{\nonumber}

\title{\boldmath 
Infrared structure of $e^+e^- \to 3$~jets at NNLO
}
\author{
A.~Gehrmann--De Ridder\\
Institute for Theoretical Physics, ETH, CH-8093 Z\"urich,
Switzerland\\ 
E-mail: \email{gehra@phys.ethz.ch}}
\author{
T.~Gehrmann\\
Institut f\"ur Theoretische Physik, Universit\"at Z\"urich,
Winterthurerstrasse 190,\\ CH-8057 Z\"urich, Switzerland\\
E-mail: \email{thomas.gehrmann@physik.uzh.ch}}
\author{E.W.N.~Glover\\
Institute of Particle Physics Phenomenology, 
        Department of Physics,\\
        University of Durham, Durham, DH1 3LE, UK\\
	E-mail: \email{e.w.n.glover@durham.ac.uk}}
\author{
G.~Heinrich\\
School of Physics, The University of Edinburgh, Edinburgh EH9 3JZ,
Scotland\\
E-mail: \email{gheinric@ph.ed.ac.uk}}

\abstract{
We describe the calculation of the next-to-next-to-leading order (NNLO)
QCD corrections to three-jet production and related event shape 
observables in electron-positron annihilation. Infrared singularities 
due to double real radiation at tree level and single real radiation 
at one loop are subtracted from the full QCD matrix elements 
using antenna functions, which are then integrated analytically and
added to the two loop contribution.
Using this antenna subtraction method, we obtain numerically finite
contributions from five-parton and four-parton processes, and observe
an explicit analytic cancellation of infrared poles in the 
four-parton and three-parton contributions. All contributions are 
implemented in a flexible parton-level event generator programme, 
allowing the numerical computation of 
any infrared-safe observable related to three-jet 
final states to NNLO accuracy.}

\keywords{QCD, Jets, LEP and ILC Physics, NLO and NNLO Computations}
\preprint{{ZU-TH 18/07}, {IPPP/07/62}, Edinburgh 2007-26}

\begin{document}

\section{Introduction}
\label{sec:intro}
Among jet observables, the three-jet production rate in  electron--positron
annihilation plays a very prominent role.  The initial experimental observation
of three-jet events at PETRA~\cite{petra}, in agreement with the theoretical
prediction~\cite{ellis}, provided  first evidence for the gluon, and thus
strong support for the theory of Quantum Chromodynamics 
(QCD)~\cite{dissertori}. Subsequently
the three-jet rate  and related event shape observables were used
for the precise determination  of the QCD coupling constant $\alpha_s$
(see~\cite{bethke} for a review).  Especially at LEP, three-jet observables
were measured to a very high  precision and the error on the extraction of
$\alpha_s$ from these  data is dominated by the uncertainty inherent in the
theoretical description of the jet observables. This description is 
at present based on a  
  next-to-leading order (NLO)  
calculation~\cite{ERT,ert2,ert3,kn,gg,cs}, combined with next--to-leading 
logarithmic (NLL) resummation~\cite{ctwt,hasko} and  inclusion of power
corrections~\cite{power}.
The calculation of  next-to-next-to-leading order (NNLO), i.e.\  ${\cal
O}(\alpha_s^3)$, corrections to the three-jet rate in $e^+e^-$ annihilation
has therefore been high on the list of priorities for a long time~\cite{kunszt}.

Besides its  phenomenological importance,  the three-jet rate has also served
as a theoretical testing ground for the development of new techniques for
higher order calculations in QCD: both the 
subtraction~\cite{ERT,sub,singleun} 
and the
phase-space slicing~\cite{ert2} methods for the extraction of infrared
singularities from  NLO real radiation  processes were
developed in the context of the first three-jet calculations. The systematic
formulation of  phase-space  slicing~\cite{gg} as  well as the dipole
subtraction~\cite{cs} method were also first demonstrated for three-jet
observables, before being applied to other processes. 

Over the past years, many of the ingredients necessary for 
NNLO calculations of 
jet observables have become available: two-loop corrections
to all
phenomenologically relevant massless $2 \to 2$~\cite{twol} and 
$1 \to 3$~\cite{3jme,muw} reactions 
were computed already several years ago, while one-loop $2\to 3$~\cite{onel-3}
 and 
$1\to 4$~\cite{onel-4} matrix elements are available for even longer. 
Despite all ingredients being available in principle, until recently 
it was not possible to perform NNLO calculations of any kind of exclusive 
observables, since techniques for the extraction of 
multiple real radiation 
singularities at NNLO were not sufficiently developed. 

Up to now, the only general method for handling this problem was 
the sector decomposition technique~\cite{secdec,ggh}
 for the treatment of real radiation singularities. Using this, 
NNLO calculations of exclusive processes 
 were performed for  $e^+e^- \to 2j$~\cite{babis2j}, 
Higgs 
production~\cite{babishiggs}    
and vector boson production~\cite{babisdy} at hadron colliders, 
as well as the QED corrections to muon 
decay~\cite{babismu}.

Furthermore, exploiting the specific kinematic features of the 
observable under consideration, exclusive NNLO results were derived for 
the forward-backward asymmetry in $e^+e^-$ annihilation~\cite{afbnnlo},
for $e^+e^-\to 2j$~\cite{our2j,weinzierl2j} and most recently 
for Higgs production at hadron colliders~\cite{catanihiggs}.

In the present paper, we employ a recently developed 
general technique for the 
treatment of infrared singularities, antenna 
subtraction~\cite{ourant}, to derive the NNLO corrections to three-jet 
production in electron-positron annihilation. The first phenomenological
applications of our results to the thrust distribution were documented 
earlier in~\cite{ourthrust}. 

The paper is structured as follows: in section~\ref{sec:struc}, we 
outline the perturbative calculation of jet observables and summarise the 
antenna subtraction method used here. The implementation of 
this method requires phase space mappings, which 
are described in Section~\ref{sec:map}.
All relevant tree-level,
one-loop and two-loop matrix elements are listed in section~\ref{sec:me}.  
Section~\ref{sec:nlo} briefly summarises how the NLO corrections are 
implemented using antenna subtraction.
Sections~\ref{sec:termA}--\ref{sec:termG} 
contain the subtraction terms constructed for all colour factors relevant 
in this calculation. The numerical implementation of all terms into a 
parton-level event generator is described in Section~\ref{sec:numer}.
As a first example of the implementation, we discuss the NNLO corrections 
to the thrust distribution in section~\ref{sec:thrust}.  
A summary and an outlook on applications is given in Section~\ref{sec:conc}.

\section{Perturbative calculation of
 jet observables in $e^+e^-$-annihilation}
\label{sec:struc}

To obtain the perturbative corrections to a jet observable at a given  order,
all partonic multiplicity channels contributing to that order  have to be
summed. In general, each partonic channel contains both ultraviolet and
infrared (soft and collinear) singularities.  
The ultraviolet poles are removed
by renormalisation, however for suitably inclusive observables,
the soft and collinear infrared poles cancel
among each other when all partonic channels are summed over~\cite{kln}.

While infrared singularities from purely virtual corrections are obtained 
immediately after integration over the loop momenta, their extraction is  more
involved for real emission (or mixed real-virtual) contributions. Here, the
infrared singularities only become explicit after integrating  the real
radiation matrix elements over the phase space appropriate to  the jet
observable under consideration. In general, this integration  involves the
(often iterative) definition of the jet observable, such that  an analytic
integration is not feasible (and also not appropriate). Instead,   one would
like to have a flexible method that can be easily adapted to  different jet
observables or jet definitions. 

Three types of approaches for this task have been developed so far. 
Phase-space slicing 
techniques~\cite{ert2,gg,ggamma} decompose the final state phase space into 
resolved regions, which are integrated 
numerically and unresolved regions, which are integrated analytically. 
The sector decomposition approach~\cite{secdec,ggh}
divides the integration region into sectors containing a single type 
of singularity each. Subsequently, the 
 phase space integration is expanded into 
distributions.
In this approach, 
the coefficients of all 
infrared divergent terms, as well as the finite remainder, can be 
computed numerically. 
Finally, subtraction methods~\cite{ERT,sub,singleun} 
extract infrared singularities of the real radiation contributions 
 using infrared subtraction terms.
These terms are constructed such that they approximate the
full real radiation matrix elements in all singular limits while still
being simple enough to integrate analytically.

To specify the notation, we define 
the tree-level $n$-parton contribution to the $m$-jet cross 
section (for tree
level cross sections $n = m$; we leave $n \neq m$ for later
reference) 
in $d$ dimensions by,
\begin{equation}
{\rm d} \sigma^{B}={\cal N}\,\sum_{{n}}{\rm d}\Phi_{n}(p_{1},\ldots,p_{n};q)
\frac{1}{S_{{n}}}\,|{\cal M}_{n}(p_{1},\ldots,p_{n})|^{2}\; 
\JET_{m}^{(n)}(p_{1},\ldots,p_{n}).
\label{eq:sigm}
\end{equation}
the normalisation factor 
${\cal N}$ includes all QCD-independent factors as well as the 
dependence on the renormalised QCD coupling constant $\alpha_s$,
$\sum_{n}$ denotes the sum over all configurations with $n$ partons,
${\rm d}\Phi_{n}$ is the phase space for an $n$-parton final state with total
four-momentum $q^{\mu}$ in
$d=4-2\e$ space-time dimensions,
\begin{equation}
\d \Phi_n(p_{1},\ldots,p_{n};q) 
= \frac{\d^{d-1} p_1}{2E_1 (2\pi)^{d-1}}\; \ldots \;
\frac{\d^{d-1} p_n}{2E_n (2\pi)^{d-1}}\; (2\pi)^{d} \;
\delta^d (q - p_1 - \ldots - p_n) \,,
\end{equation}
while $S_{n}$ is a
symmetry factor for identical partons in the final state.
The jet function $ \JET_{m}^{(n)}$ defines the procedure for 
building $m$ jets out of $n$ partons.
The main property of $\JET_{m}^{(n)}$ is that the jet observable defined
above is collinear and infrared safe as explained in
\cite{cs,stermanweinberg}. 
In general $\JET_{m}^{(n)}$ contains $\theta$ and $\delta$-functions.  
 $\JET_{m}^{(n)}$ can also represent the definition of the $n$-parton 
contribution to an event shape observable related to $m$-jet final states.

$|{\cal M}_{n}|^2$ denotes a 
squared, colour ordered tree-level $n$-parton matrix element.
Contributions to the squared matrix element which are subleading in the number 
of colours can equally be treated in the same context, noting that 
these subleading terms yield configurations where a 
certain number of essentially non-interacting particles
are emitted between a pair of hard 
radiators. By carrying 
out the colour algebra, it becomes evident that non-ordered gluon emission 
inside a colour-ordered system is equivalent to photon emission off the 
outside legs of the system~\cite{campbell,colord}. For simplicity,
these subleading colour contributions are also denoted as squared matrix 
elements $|{\cal M}_m|^2$, 
although they often correspond purely to interference
terms between different amplitudes.
The precise definition depends on the number and types
of particles involved in the process.   
However, all colour orderings are summed
over in $\sum_{m}$ with the appropriate colour weighting.

From (\ref{eq:sigm}), one obtains the leading order approximation to the 
$m$-jet cross section by integration over the appropriate phase space. 
\begin{equation}
{\rm d}\sigma_{LO}=\int_{{\rm d}\Phi_{m}}{\rm d}\sigma^{B} \;.
\end{equation}
Depending on the jet function used, 
this cross section can still be differential in certain kinematic
quantities.

\subsection{NLO antenna subtraction}
At NLO, we consider the following $m$-jet cross section,
\begin{equation}
{\rm d}\sigma_{NLO}=\int_{{\rm d}\Phi_{m+1}}\left({\rm d}\sigma^{R}_{NLO}
-{\rm d}\sigma^{S}_{NLO}\right) +\left [\int_{{\rm d}\Phi_{m+1}}
{\rm d}\sigma^{S}_{NLO}+\int_{{\rm d}\Phi_{m}}{\rm d}\sigma^{V}_{NLO}\right].
\end{equation}
The cross section ${\rm d}\sigma^{R}_{NLO}$ has the same expression as the 
Born cross section ${\rm d}\sigma^{B}$ (\ref{eq:sigm}) above
except that $n \to m+1$, while 
${\rm d}\sigma^{V}_{NLO}$ is the one-loop virtual correction to the 
$m$-parton Born cross section ${\rm d}\sigma^{B}$.
The cross section ${\rm d}\sigma^{S}_{NLO}$ is a 
(preferably local) counter-term for  
 ${\rm d}\sigma^{R}_{NLO}$. It has the same unintegrated
singular behaviour as ${\rm d}\sigma^{R}_{NLO}$ in all appropriate limits.
Their difference is free of divergences 
and can be integrated over the $(m+1)$-parton phase space numerically.
The subtraction term  ${\rm d}\sigma^{S}_{NLO}$ has 
to be integrated analytically over all singular regions of the 
$(m+1)$-parton phase space. 
The resulting cross section added to the virtual contribution 
yields an infrared finite result.

Several methods for  constructing
NLO subtraction terms systematically  were proposed in the 
literature~\cite{sub,cullen,ant,hadant,cs,singleun}. 
For some of these methods, 
extension to NNLO was discussed~\cite{ourant,Kosower:2002su,nnlosub} 
and partly worked out. 
Up to now, the only method worked out in full detail to NNLO is antenna
subtraction~\cite{cullen,ant}. In our calculation of NNLO corrections 
to three-jet observables, we used this method, which we 
briefly outline in the following. The details of the method, and 
a full definition of the notation, can be found in~\cite{ourant}. 

The basic idea of the antenna subtraction approach is to construct 
the subtraction terms  from antenna functions which encapsulate
all singular limits due to the 
emission of unresolved partons between two colour-connected hard
partons. This construction exploits the universal factorisation of both
phase space and squared matrix elements in all unresolved limits.
The full antenna subtraction term is then constructed by summing 
products of antenna functions with reduced matrix elements over all possible
unresolved configurations.

At NLO, the antenna subtraction term thus reads:
\begin{eqnarray}
\lefteqn{{\rm d}\sigma_{NLO}^{S} =
{\cal N}\,\sum_{m+1}{\rm d}\Phi_{m+1}(p_{1},\ldots,p_{m+1};q)
\frac{1}{S_{{m+1}}} }\nonumber \\ 
&\times &\sum_{j}\;X^0_{ijk}\,
|{\cal M}_{m}(p_{1},\ldots,\tilde{p}_{I},\tilde{p}_{K},\ldots,p_{m+1})|^2\,
\JET_{m}^{(m)}(p_{1},\ldots,\tilde{p}_{I},\tilde{p}_{K},\ldots,p_{m+1})
\;.
\label{eq:sigmasNLO}
\end{eqnarray}

The key ingredient is the phase space mapping which relates the
original momenta $p_{i},p_{j},{p}_{k}$ describing 
the two hard radiator partons $i$ and $k$
and the emitted parton $j$ to a redefined on-shell set 
$\tilde{p}_{I},\tilde{p}_{K}$ which 
 are linear combinations of $p_{i},p_{j},{p}_{k}$. The phase space mapping
yielding this redefinition is described in detail in section~\ref{sec:map} 
below. With this mapping,the phase space factorises,
\begin{equation}
\label{eq:psx3}
\d \Phi_{m+1}(p_{1},\ldots,p_{m+1};q)  = 
\d \Phi_{m}(p_{1},\ldots,\tilde{p}_{I},\tilde{p}_{K},\ldots,p_{m+1};q)\cdot 
\d \Phi_{X_{ijk}} (p_i,p_j,p_k;\tilde{p}_{I}+\tilde{p}_{K})\;.
\end{equation}

The other elements of the subtraction term also depend on {\it either} the
original momenta $p_{i},p_{j},{p}_{k}$ {\it or} 
the redefined on-shell momenta $\tilde{p}_{I},\tilde{p}_{K}$
 but not both.   This enables the subtraction term to completely factorise.
 
To be more specific, both the $m$-parton amplitude and the
 jet function $\JET^{(m)}_m$
depend only on $p_{1},\ldots,\tilde{p}_{I},\tilde{p}_{K},\ldots,p_{m+1}$ 
i.e.\ 
on the redefined on-shell momenta
$\tilde{p}_{I},\tilde{p}_{K}$.
On the other hand, 
the tree-level three-parton antenna function $X^0_{ijk}$ depends only on 
$p_{i},p_{j},{p}_{k}$. 
$X^0_{ijk}$ describes all of the configurations (for this colour ordered
amplitude)
where parton $j$ is unresolved. It can be obtained from appropriately 
normalised tree-level three-parton squared matrix elements. 
The antenna factorisation of
squared matrix element and phase space can be illustrated pictorially, as 
displayed in Figure~\ref{fig:nloant}.
Together particles $i$ and $k$ form a colour connected 
hard antenna that radiates particle $j$. In doing so, 
the momenta of 
the radiators change 
to form particles $I$ and $K$. The type of particle may also change.

\FIGURE[h!]{ 
\epsfig{file=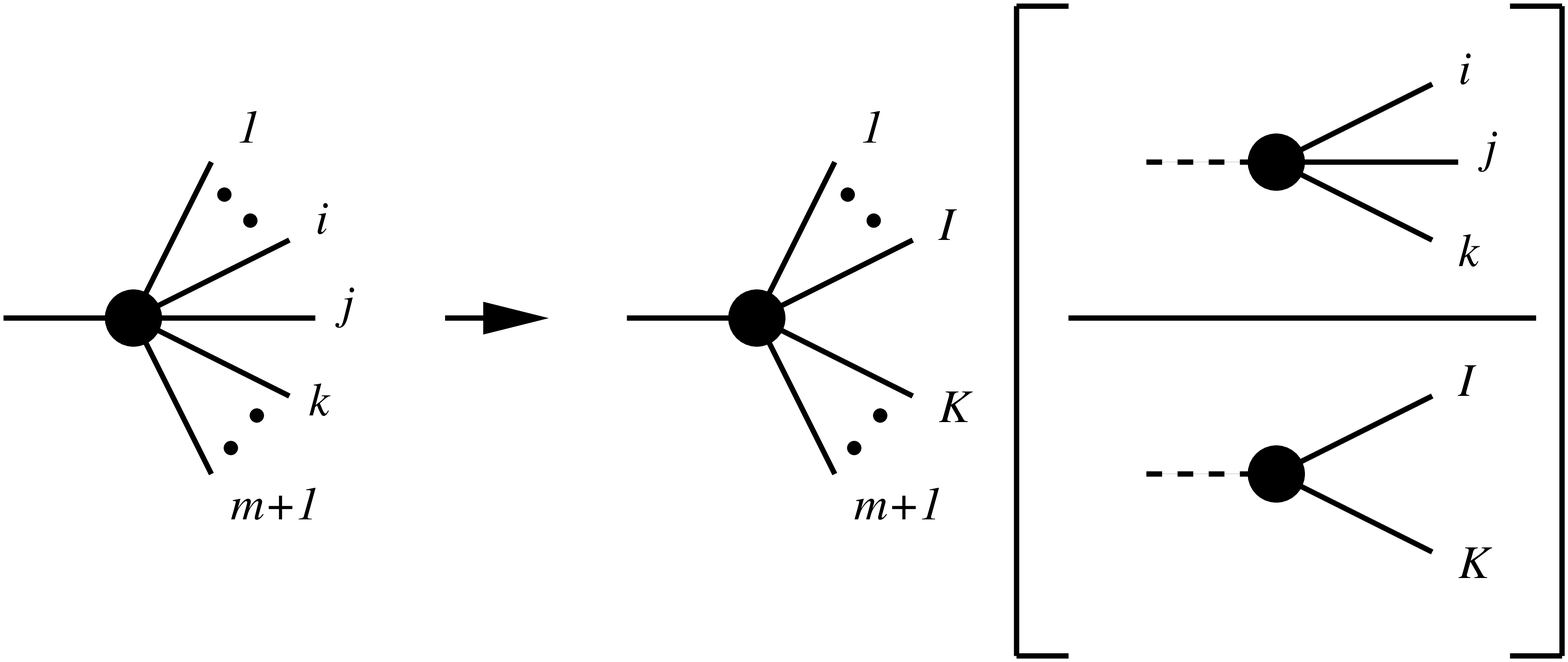,width=10cm}
\label{fig:nloant}
\caption{Illustration of NLO antenna factorisation representing the
factorisation of both the squared matrix elements and the $(m+1)$-particle phase
space. The term in square brackets
represents both the antenna function $X^0_{ijk}$ and the antenna phase space
${\rm d}\Phi_{X_{ijk}}$.}}

One can therefore carry out the integration over the antenna phase space appropriate to 
   ${p}_{i}$, $p_j$ and ${p}_{k}$ analytically, exploiting the 
factorisation of the phase space of eq.~(\ref{eq:psx3}).
The NLO antenna phase space $\d \Phi_{X_{ijk}}$ is proportional to the 
three-particle phase space,
as can be seen by using $m=2$ in the above formula.
For the analytic integration, 
we can use (\ref{eq:psx3}) to rewrite
each of the subtraction terms  in the form, 
\begin{displaymath}
|{\cal M}_{m}|^2\,
\JET_{m}^{(m)}\; 
{\rm d}\Phi_{m}
\int {\rm d} \Phi_{X_{ijk}}\;X^0_{ijk} = |{\cal M}_{m}|^2\,
\JET_{m}^{(m)}\; {\rm d}\Phi_{m}
\;{\cal X}_{ijk}
\end{displaymath}
where $|{\cal M}_{m}|^2$, $\JET_{m}^{(m)}$ and ${\rm d}\Phi_{m}$ depend only on
$p_1,,\ldots,\tilde{p}_{I},\tilde{p}_{K},\ldots,p_{m+1}$
and ${\rm d} \Phi_{X_{ijk}}$ and $X^0_{ijk}$ depend only on $p_i, p_j, p_k$.
This integration is performed analytically in $d$ dimensions,
yielding the integrated three-parton antenna function 
${\cal X}_{ijk}$, 
to make the infrared singularities explicit and
added directly to the one-loop $m$-particle contributions.

\subsection{NNLO antenna subtraction}
\label{sec:NNLOant}
At NNLO, the $m$-jet production is induced by final states containing up to
$(m+2)$ partons, including the one-loop virtual corrections to $(m+1)$-parton final 
states. As at NLO, one has to introduce subtraction terms for the 
$(m+1)$- and $(m+2)$-parton contributions. 
Schematically the NNLO $m$-jet cross section reads,
\begin{eqnarray}
{\rm d}\sigma_{NNLO}&=&\int_{{\rm d}\Phi_{m+2}}\left({\rm d}\sigma^{R}_{NNLO}
-{\rm d}\sigma^{S}_{NNLO}\right) + \int_{{\rm d}\Phi_{m+2}}
{\rm d}\sigma^{S}_{NNLO}\nonumber \\ 
&&+\int_{{\rm d}\Phi_{m+1}}\left({\rm d}\sigma^{V,1}_{NNLO}
-{\rm d}\sigma^{VS,1}_{NNLO}\right)
+\int_{{\rm d}\Phi_{m+1}}{\rm d}\sigma^{VS,1}_{NNLO}  
\nonumber \\&&
+ \int_{{\rm d}\Phi_{m}}{\rm d}\sigma^{V,2}_{NNLO}\;,
\end{eqnarray}
where $\d \sigma^{S}_{NNLO}$ denotes the real radiation subtraction term 
coinciding with the $(m+2)$-parton tree level cross section 
 $\d \sigma^{R}_{NNLO}$ in all singular limits. 
Likewise, $\d \sigma^{VS,1}_{NNLO}$
is the one-loop virtual subtraction term 
coinciding with the one-loop $(m+1)$-parton cross section 
 $\d \sigma^{V,1}_{NNLO}$ in all singular limits. 
Finally, the two-loop correction 
to the $m$-parton cross section is denoted by ${\rm d}\sigma^{V,2}_{NNLO}$.

At NNLO, individual antenna functions are obtained 
from normalised four-parton tree-level and three-parton 
one-loop matrix elements.
The full antenna subtraction term is then constructed by summing 
products of antenna functions with reduced matrix elements over all possible
unresolved configurations.
\FIGURE[t]{
\caption{\label{fig:sub2a} Illustration 
of NNLO antenna factorisation representing the
factorisation of both the squared matrix elements and the $(m+2)$-particle 
phase
space when the unresolved particles are colour connected. }
\parbox{14cm}{\begin{center}\epsfig{file=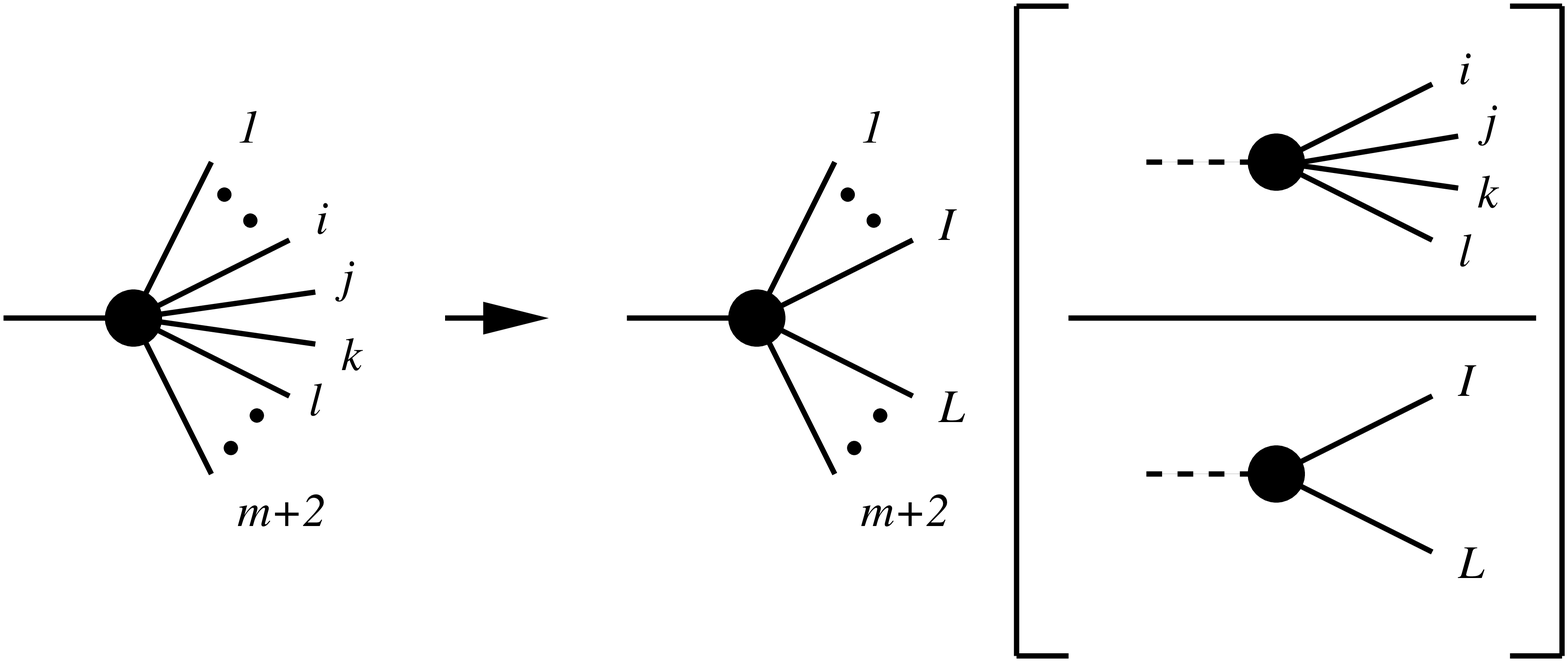,,height=4cm}\end{center}}
}

In ${\rm d}\sigma^{S}_{NNLO}$, we have to distinguish four
different types of unresolved configurations:
\begin{itemize}
\item[(a)] One unresolved parton but the experimental observable selects only
$m$ jets;
\item[(b)] Two colour-connected unresolved partons (colour-connected);
\item[(c)] Two unresolved partons that are not colour connected but share a common
radiator (almost colour-unconnected);
\item[(d)] Two unresolved partons that are well separated from each other 
in the colour chain (colour-unconnected).
\end{itemize}
Among those, configuration (a) is properly 
accounted for by a single tree-level three-parton antenna function 
like used already at NLO. Configuration (b) requires a 
tree-level four-parton antenna function (two unresolved partons emitted 
between a pair of hard partons) 
as shown in Figure~\ref{fig:sub2a}, while (c) and (d) are accounted for by 
products of two tree-level three-parton antenna functions. 
The subtraction terms for these configurations read:
\begin{eqnarray}
{\rm d}\sigma_{NNLO}^{S,a}
&=&  {\cal N}\,\sum_{m+2}{\rm d}\Phi_{m+2}(p_{1},\ldots,p_{m+2};q)
\frac{1}{S_{{m+2}}} \nonumber \\
&\times& \,\Bigg [ \sum_{j}\;X^0_{ijk}\,
|{\cal M}_{m+1}(p_{1},\ldots,\tilde{p}_{I},\tilde{p}_{K},\ldots,p_{m+2})|^2\,
\nonumber \\ &&
\hspace{3cm} \times
\JET_{m}^{(m+1)}(p_{1},\ldots,\tilde{p}_{I},\tilde{p}_{K},\ldots,p_{m+2})\;
\Bigg
]\;,\label{eq:sub2a}\\
{\rm d}\sigma_{NNLO}^{S,b}
&=&  {\cal N}\,\sum_{m+2}{\rm d}\Phi_{m+2}(p_{1},\ldots,p_{m+2};q)
\frac{1}{S_{{m+2}}} \nonumber \\
&\times& \,\Bigg [ \sum_{jk}\;\left( X^0_{ijkl}
- X^0_{ijk} X^0_{IKl} - X^0_{jkl} X^0_{iJL} \right)\nonumber \\
&\times&
|{\cal M}_{m}(p_{1},\ldots,\tilde{p}_{I},\tilde{p}_{L},\ldots,p_{m+2})|^2\,
\JET_{m}^{(m)}(p_{1},\ldots,\tilde{p}_{I},\tilde{p}_{L},\ldots,p_{m+2})\;
\Bigg]\;,
\label{eq:sub2b}\\
{\rm d}\sigma_{NNLO}^{S,c}
&= & - {\cal N}\,\sum_{m+2}{\rm d}\Phi_{m+2}(p_{1},\ldots,p_{m+2};q)
\frac{1}{S_{{m+2}}} \nonumber \\
&\times& \,\Bigg [  \sum_{j,l}\;X^0_{ijk}\;x^0_{mlK}\,
|{\cal M}_{m}(p_{1},\ldots,\tilde{p}_{I},{\doubletilde{p}}_{K},{\doubletilde{p}}_{M},\ldots,p_{m+2})|^2\,\nonumber \\
&&\hspace{3cm}\times
\JET_{m}^{(m)}(p_{1},\ldots,\tilde{p}_{I},{\doubletilde{p}}_{K},{\doubletilde{p}}_{M},\ldots,p_{m+2})\;
\phantom{\Bigg]}
\nonumber \\
&& \,+ \sum_{j,l}\;X^0_{klm}\;x^0_{ijK}\,
|{\cal M}_{m}(p_{1},\ldots,{\doubletilde{p}}_{I},{\doubletilde{p}}_{K},\tilde{p}_{M},\ldots,p_{m+2})|^2\,\nonumber \\
&&\hspace{3cm}\times
\JET_{m}^{(m)}(p_{1},\ldots,{\doubletilde{p}}_{I},{\doubletilde{p}}_{K},\tilde{p}_{M},\ldots,p_{m+2})\;
\Bigg
]\;,
\label{eq:sub2c}\\
{\rm d}\sigma_{NNLO}^{S,d}
&= & - {\cal N}\,\sum_{m+2}{\rm d}\Phi_{m+2}(p_{1},\ldots,p_{m+2};q)
\frac{1}{S_{{m+2}}} \nonumber \\
&\times& \,\Bigg [ \sum_{j,o}\;X^0_{ijk}\;X^0_{nop}\,
|{\cal M}_{m}(p_{1},\ldots,\tilde{p}_{I},\tilde{p}_{K},\ldots,\tilde{p}_{N},\tilde{p}_P,\ldots,p_{m+2})|^2\,\nonumber \\
&\times&
\JET_{m}^{(m)}(p_{1},\ldots,\tilde{p}_{I},\tilde{p}_{K},\ldots,\tilde{p}_{N},\tilde{p}_P,\ldots,p_{m+2})\;\Bigg
]\;.
\label{eq:sub2d}
\end{eqnarray}
Again, the original momenta of the $(m+2)$-parton phase space are denoted by 
$i,j,\ldots$, while the combined momenta obtained from a phase space 
mapping are labelled by $I,J,\ldots$. Only the combined momenta appear in the 
jet function. The phase space mappings appropriate 
to the different cases are described in detail in section~\ref{sec:map} below.
$X^0_{ijkl}$ is a four-parton antenna function, containing all configurations 
where partons $j$ and $k$ are unresolved, while $x^0_{ijk}$ is a three-parton 
sub-antenna function containing only  limits where parton $j$ is unresolved 
with respect to parton $i$, but not limits where parton $j$ is unresolved 
with respect to parton $k$. The factorisation of the phase space is analogous 
to the factorisation at NLO (\ref{eq:psx3}), such that 
integration of these antenna functions 
over the antenna phase space amounts to inclusive three-particle or 
four-particle integrals~\cite{ggh}. 
\FIGURE[t]{
\caption{Illustration of NNLO antenna factorisation representing the
factorisation of both the one-loop
``squared" matrix elements (represented by the white blob)
when the unresolved particles are colour connected. 
\label{fig:subv}}
\epsfig{file=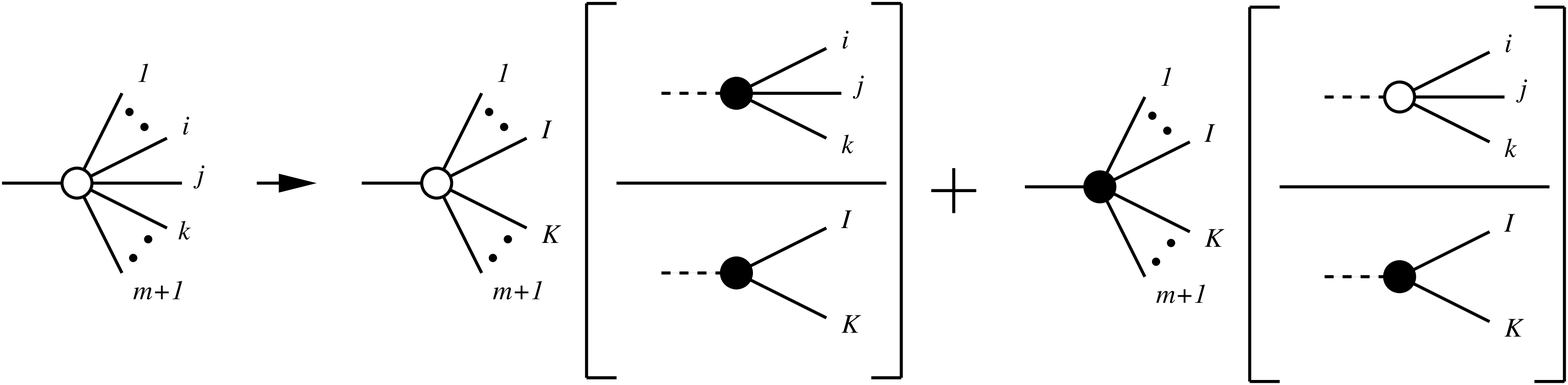,height=3.3cm}}

In single unresolved limits, the one-loop cross section 
$\d \sigma^{V,1}_{NNLO}$ is described by the sum of two terms~\cite{onelstr}: 
a tree-level splitting function times a one-loop cross section 
and a one-loop splitting function times a tree-level cross section. 
Consequently, the 
one-loop single unresolved subtraction term $\d \sigma^{VS,1}_{NNLO}$
is constructed from tree-level and one-loop three-parton antenna functions,
as sketched in Figure~\ref{fig:subv}. Several other terms in  
 $\d \sigma^{VS,1}_{NNLO}$ cancel with the results
from the integration of terms in 
the double real radiation subtraction term  $\d \sigma^{S}_{NNLO}$
over the phase space appropriate to one of the unresolved partons, thus 
ensuring the cancellation of all explicit infrared poles in the difference 
$\d \sigma^{V,1}_{NNLO}-\d \sigma^{VS,1}_{NNLO}$.
Explicitly, the one-loop single unresolved subtraction term is 
given by the sum of the three following contributions:
\begin{eqnarray}
{\rm d}\sigma_{NNLO}^{VS,1,a}
&=&   {\cal N}\,\sum_{m+1}{\rm d}\Phi_{m+1}(p_{1},\ldots,p_{m+1};q)
\frac{1}{S_{{m+1}}} \nonumber \\
&\times& \,\Bigg [ \sum_{ik}\;  - {\cal X}^0_{ijk}(s_{ik}) \,
|{\cal M}_{m+1}(p_{1},\ldots,{p}_{i},{p}_{k},\ldots,p_{m+1})|^2\,
\nonumber \\
&& \hspace{3cm} \times
\JET_{m}^{(m+1)}(p_{1},\ldots,{p}_{i},{p}_{k},\ldots,p_{m+1})\;
\Bigg
]\;, \label{eq:subv2a}\\
{\rm d}\sigma_{NNLO}^{VS,1,b}
&= & {\cal N}\,\sum_{m+1}{\rm d}\Phi_{m+1}(p_{1},\ldots,p_{m+1};q)
\frac{1}{S_{{m+1}}} \nonumber \\
&\times& \,\sum_{j} \Bigg [X^0_{ijk}\,
|{\cal M}^1_{m}(p_{1},\ldots,\tilde{p}_{I},\tilde{p}_{K},\ldots,p_{m+1})|^2\,
\JET_{m}^{(m)}(p_{1},\ldots,\tilde{p}_{I},\tilde{p}_{K},\ldots,p_{m+1})\;
\nonumber \\
&&\phantom{\sum_{j} }+\;X^1_{ijk}\,
|{\cal M}_{m}(p_{1},\ldots,\tilde{p}_{I},\tilde{p}_{K},\ldots,p_{m+1})|^2\,
\JET_{m}^{(m)}(p_{1},\ldots,\tilde{p}_{I},\tilde{p}_{K},\ldots,p_{m+1})\;\Bigg
]\;,\nonumber \\
\label{eq:subv2b}\\
{\rm d}\sigma_{NNLO}^{VS,1,c}
&=&   {\cal N}\,\sum_{m+1}{\rm d}\Phi_{m+1}(p_{1},\ldots,p_{m+1};q)
\frac{1}{S_{{m+1}}} \nonumber \\
&\times& \,\Bigg [ \sum_{ik}\; {\cal X}^0_{ijk}(s_{ik}) \,
\sum_o X^0_{nop} \,
|{\cal M}_{m}(p_{1},\ldots,p_i,p_k,\ldots,\tilde{p}_{N},\tilde{p}_{P},\ldots,p_{m+1})|^2\,
\nonumber \\ 
&& \hspace{3cm} \times 
\JET_{m}^{(m)}(p_{1},\ldots,p_i,p_k,\ldots,\tilde{p}_{N},\tilde{p}_{P},\ldots,p_{m+1})\;
\Bigg
],
\label{eq:subv2c}
\end{eqnarray}
In here, $X^1_{ijk}$ denotes a one-loop three-parton antenna function.

Finally, all remaining terms in 
$\d \sigma^{S}_{NNLO}$ and $\d \sigma^{VS,1}_{NNLO}$ have to be integrated 
over the four-parton and three-parton antenna phase spaces. After   
integration, the infrared poles are rendered explicit and
cancel with the 
infrared pole terms in the two-loop squared matrix element 
$\d \sigma^{V,2}_{NNLO}$. 

The  subtraction terms $\d \sigma^{S}_{NLO}$,
$\d \sigma^{S}_{NNLO}$ 
and $\d \sigma^{VS,1}_{NNLO}$ require three different types of 
antenna functions corresponding to the different pairs of hard partons 
forming the antenna: quark-antiquark, quark-gluon and gluon-gluon antenna 
functions. In the past~\cite{cullen,ant}, NLO antenna functions were 
constructed by imposing definite properties in 
all single unresolved limits (two collinear limits 
and one soft limit for each 
antenna). 
This procedure turns out to be impractical at NNLO, where each antenna 
function must have definite behaviours in a large number of single and 
double unresolved limits. Instead, we derived these antenna functions in 
a systematic manner from physical matrix elements known to possess the 
correct limits. The quark-antiquark antenna functions can be obtained 
directly from 
the $e^+e^- \to 2j$ real radiation corrections at NLO and NNLO~\cite{our2j}. 
For quark-gluon and gluon-gluon antenna functions, 
effective Lagrangians~\cite{hw,hgg}
are used to obtain tree-level processes yielding a quark-gluon or 
gluon-gluon final state. The antenna functions are then derived from 
the real radiation corrections to these processes. 
Quark-gluon antenna functions 
were derived~\cite{chi} from the purely QCD 
(i.e.\ non-supersymmetric) NLO and NNLO corrections to the decay of 
a heavy neutralino into a massless gluino plus partons~\cite{hw}, while 
gluon-gluon antenna functions~\cite{h} result from the QCD corrections 
to Higgs boson decay into partons~\cite{hgg}. 

All tree-level three-parton and four-parton antenna functions 
and  three-parton one-loop antenna functions are listed in~\cite{ourant}.
Their integration over the antenna phase space amounts to 
performing  
inclusive infrared-divergent 
three-parton and four-parton phase space integrals. Techniques for 
evaluating these integrals are described in~\cite{ggh,pstricks}, and all integrated 
antenna functions are documented in~\cite{ourant}.


\section{Phase space mappings}
\label{sec:map}
The subtraction terms for single and double unresolved configurations 
described in the previous section involve the mapping of the 
momenta appearing in the antenna functions into 
combined  momenta, which 
appear in the reduced matrix elements and in the jet function. 

At NLO, one 
needs momentum mappings 
from three partons to two partons, 
$${\cal F}^{(3\to 2)}:\,\{p_i,p_j,p_k\} \to \{\tilde{p}_{I},\tilde{p}_{K}\}.$$
At NNLO, several further mappings are needed. First and foremost, one 
needs 
momentum mappings 
from four partons to two partons,
$${\cal F}^{(4\to 2)}:\,\{p_i,p_j,p_k,p_l\} \to 
\{\tilde{p}_{I},\tilde{p}_{L}\}.$$ In addition, repeated 
mappings from three partons to two partons are also required.

For the subtraction 
and phase space factorisation  to work correctly, 
all mappings 
must satisfy the following requirements (specified here for the 
example of ${\cal F}^{(4\to 2)}$): 
\begin{enumerate}
\item momentum conservation: $\tilde{p}_I+\tilde{p}_L=p_i+p_j+p_k+p_l$.
\item the new momenta should be on-shell:  $\tilde{p}_I^2=0$, 
$\tilde{p}_L^2=0$.  
\item  the new momenta should reduce to the appropriate 
original momenta in the exact singular limits, e.g.\ 
$\tilde{p}_I=p_i+p_j+p_k$, $\tilde{p}_L=p_l$ in the $(i,j,k)$ triple collinear 
limit.
\item the mapping should not introduce spurious singularities.
\end{enumerate}

The momentum mappings we use follow largely those worked out  
in \cite{ant,Kosower:2002su}.
The different types of mappings needed for our calculation are described in detail in the following subsections.

\subsection{Mapping for single unresolved configurations}

In the single unresolved limit where parton $j$
becomes unresolved and $i,k$ are the hard radiators, 
the momenta of the partons $i,j,k$ are mapped to 
$\tilde{p}_I=\widetilde{(ij)}$ and $\tilde{p}_K=\widetilde{(kj)}$ 
in the following way:
\bea
\tilde{p}_I&=&x\,p_i+r\,p_j+z\,p_k\nn\\
\tilde{p}_K&=&(1-x)\,p_i+(1-r)\,p_j+(1-z)\,p_k\;,\nn
\eea
where 
\bea
x&=&\frac{1}{2(s_{ij}+s_{ik})}\Big[(1+\rho)\,s_{ijk} -2\,r\,s_{jk}     \Big]\nn\\
z&=&\frac{1}{2(s_{jk}+s_{ik})}\Big[(1-\rho)\,s_{ijk} -2\,r\,s_{ij}     \Big]\nn\\
&&\nn\\
\rho^2&=&1+\frac{4\,r(1-r)\,s_{ij}s_{jk}}{s_{ijk}s_{ik}}\;.\label{single}
\eea
The parameter $r$ can be chosen conveniently, we use~\cite{ant} 
$$ r=\frac{s_{jk}}{s_{ij}+s_{jk}}\;.$$

\subsection{Mappings from five partons to three partons}
In the construction of the subtraction terms for the five-parton channel,
one encounters, both the four-parton antenna functions
(double unresolved limits), and products 
of two three-parton antenna functions, which compensate for 
the single unresolved 
limits of the double unresolved subtraction terms. The former require 
a $4\to 2$ mapping, while the latter require, in general, a $5\to 3$ mapping 
affecting all momenta. Both types of mappings are described in the following.

\subsubsection{Double unresolved configurations}\label{sec:pmap5to3}


In the double unresolved limit where partons $i_2$ and $i_3$ 
become unresolved and $i_1,i_4$ are the hard radiators, 
the partons $i_1,\ldots,i_4$ are mapped to the partons $j_1,j_2$
with momenta 
\beq
\tilde{p}_{j_1}\equiv \widetilde{(i_1i_2i_3)}\;,\;
\tilde{p}_{j_2}\equiv \widetilde{(i_4i_3i_2)}\label{tilde}\;.
\eeq
We will use the shorthand notation (\ref{tilde}) extensively in the 
following, keeping in mind however  that $\tilde{p}_{j_1}$ and 
$\tilde{p}_{j_2}$
are linear combinations of all four original momenta 
with a mapping ${\cal F}^{(4\to 2)}:\,\{p_{i_1},p_{i_2},p_{i_3},p_{i_4}\} \to 
\{\tilde{p}_{j_1},\tilde{p}_{j_2}\}$
given by:
\bea
\tilde{p}_{j_1}&=&x\,p_{i_1}+r_1\,p_{i_2}+r_2\,p_{i_3}+z\,p_{i_4}\nn\\
\tilde{p}_{j_2}&=&(1-x)\,p_{i_1}+(1-r_1)\,p_{i_2}+(1-r_2)\,p_{i_3}+(1-z)\,p_{i_4}\;.
\eea
Defining $s_{kl}=(p_{i_k}+p_{i_l})^2$, 
the coefficients are given by\,\cite{Kosower:2002su}
\begin{eqnarray}
r_1&=&\frac{s_{23}+s_{24}}{s_{12}+s_{23}+s_{24}}\nonumber\\
r_2&=&\frac{s_{34}}{s_{13}+s_{23}+s_{34}}\nonumber\\
x&=&\frac{1}{2(s_{12}+s_{13}+s_{14})}\Big[(1+\rho)\,s_{1234} \nonumber\\
&&-r_1\,(s_{23}+2\,s_{24})   -r_2\,(s_{23}+2\,s_{34})  \nonumber\\
&&+(r_1-r_2)\frac{s_{12}s_{34}-s_{13}s_{24}}{s_{14}}     \Big]\nonumber\\
z&=&\frac{1}{2(s_{14}+s_{24}+s_{34})}\Big[(1-\rho)\,s_{1234} \nonumber\\
&&-r_1\,(s_{23}+2\,s_{12})   -r_2\,(s_{23}+2\,s_{13})  \nonumber\\
&&-(r_1-r_2)\frac{s_{12}s_{34}-s_{13}s_{24}}{s_{14}}     \Big]\nonumber\\
\rho&=&\Big[1+\frac{(r_1-r_2)^2}{s_{14}^2\,s_{1234}^2}\,\lambda(s_{12}\,
s_{34},s_{14}\,s_{23},s_{13}\,s_{24})\nonumber\\
&&  +\frac{1}{s_{14}\,s_{1234}}\Big\{
2\,\big(r_1\,(1-r_2)+r_2(1-r_1)\big)\big( s_{12}s_{34}+s_{13}s_{24}-s_{23}s_{14} \big)\nonumber\\
&&\qquad\qquad +\,4\,r_1\,(1-r_1)\,s_{12} s_{24}+4\,r_2\,(1-r_2)\,s_{13}s_{34}\Big\}\Big]^{\frac{1}{2}}\;,
\nonumber\\
\lambda(u,v,w)&=&u^2+v^2+w^2-2(uv+uw+vw)\;.\nonumber
\end{eqnarray}

\subsubsection{Iteration of single unresolved configurations}

The four-particle antennae $X^0_{ijkl}$ contain by construction all 
colour-connected double unresolved limits of the $(m+2)$-parton 
matrix element where partons $j$ and $k$ become unresolved. 
However, $X^0_{ijkl}$ can also become singular in  single
unresolved limits, where it does not coincide with limits 
of the matrix element. 
These limits have to be subtracted as indicated in 
eq.~\ref{eq:sub2b}
and as described in section 2.3 
of \cite{ourant}. In the limit where $j$ becomes 
unresolved between $i$ and $k$, the four-particle antenna collapses to 
the three-particle antenna $X^0(i,j,k)$ and a three-particle 
``remainder matrix element" $X^0(I,K,l)$, which also has the form of a 
three-particle antenna. If one of the partons $I,K,l$ 
becomes unresolved in a second single emission, the resulting momenta 
in these limits coincide with those 
from a double unresolved configuration as defined in 
subsection~\ref{sec:pmap5to3}. Therefore we need momentum mappings
corresponding to such a ``two-step" emission in order to 
be able to subtract the spurious singularities of the 
four-particle antennae $X^0_{ijkl}$.

As explained in section \ref{sec:NNLOant} above and in \cite{ourant}, 
we have to distinguish  between colour-connected unresolved partons 
and almost colour-unconnected unresolved partons. 
If the unresolved partons  are colour-connected, 
we use the momentum mappings
${\cal F}_{\rm{B,C}}^{(5\to 4\to 3)}:\,\{p_{i_1},p_{i_2},p_{i_3},p_{i_4},p_{i_5}\} \to 
\{p_{l_1},p_{l_2},p_{l_3},p_{l_4}\} \to 
\{\tilde{p}_{j_1},\tilde{p}_{j_2},\tilde{p}_{j_3}\}$, 
where the different types B,C are 
described in more detail below. 
These two-step mappings 
first map the four partons $i_1,\ldots,i_4$ 
making up the four-particle antenna  
to three ``intermediate" partons $l_1,l_2,l_3$, 
and then map the 
three intermediate partons to two partons $j_1,j_2$. 
The fifth parton $i_5=l_4=j_3$ only acts as a spectator.
An additional type of mapping, denoted by ${\cal F}_{\rm{K}}^{(5\to 4\to 3)}$, 
is needed for subtraction terms of 
two almost colour-unconnected  unresolved partons, as defined 
in eq.~\ref{eq:sub2c}
and in \cite{ourant}, and involves redefinitions
of all five initial partons. All three mappings are depicted 
in Figures~\ref{fig:mapb}--\ref{fig:mape}.


\FIGURE[t]{
\caption{${\cal F}_{\rm{B}}^{(5\to 4\to 3)}(i_1,i_2,i_3,i_4,i_5;j_1,j_2,j_3)$:
both combined partons ($l_1$ and $l_3$)  are radiators in the second step, original parton $i_3$ becomes unresolved.}
\label{fig:mapb}
\epsfig{file=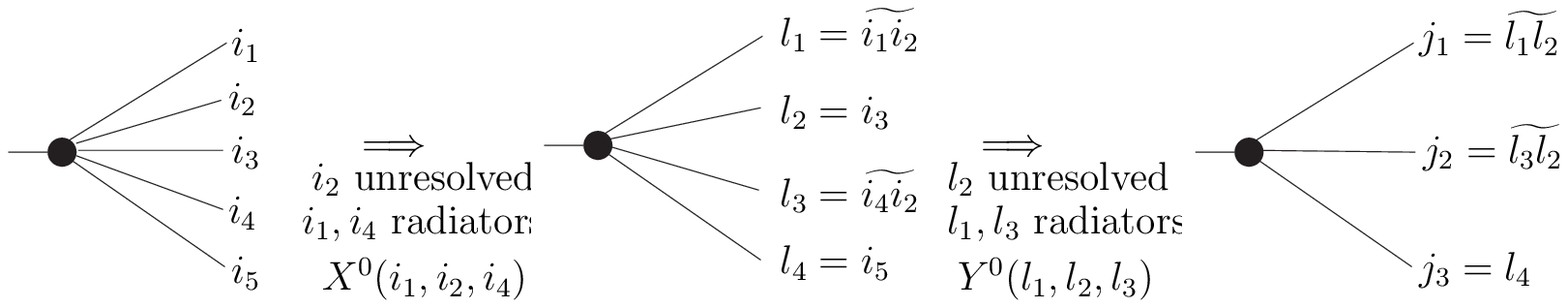,width=14.5cm}
 }

\FIGURE[t]{
\caption{${\cal F}_{\rm{C}}^{(5\to 4\to 3)}(i_1,i_2,i_3,i_4,i_5;j_1,j_2,j_3)$: 
one of the combined partons (here $l_3$) becomes unresolved between $l_1$ and $l_2$ in the second step.}
\label{fig:mapc}
\epsfig{file=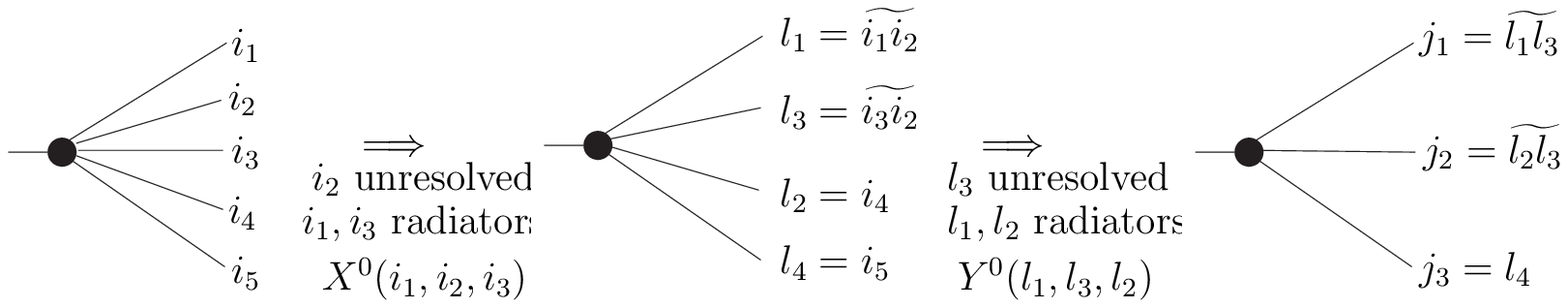,width=14.5cm}
 }

\FIGURE[t]{
\caption{${\cal F}_{\rm{K}}^{(5\to 4\to 3)}(i_1,i_2,i_3,i_4,i_5;j_1,j_2,j_3)$:
$i_3$ is the shared hard radiator, $i_4=l_2$ becomes unresolved between 
$l_3$ and $l_4$ in the second step.}
\label{fig:mape}

\epsfig{file=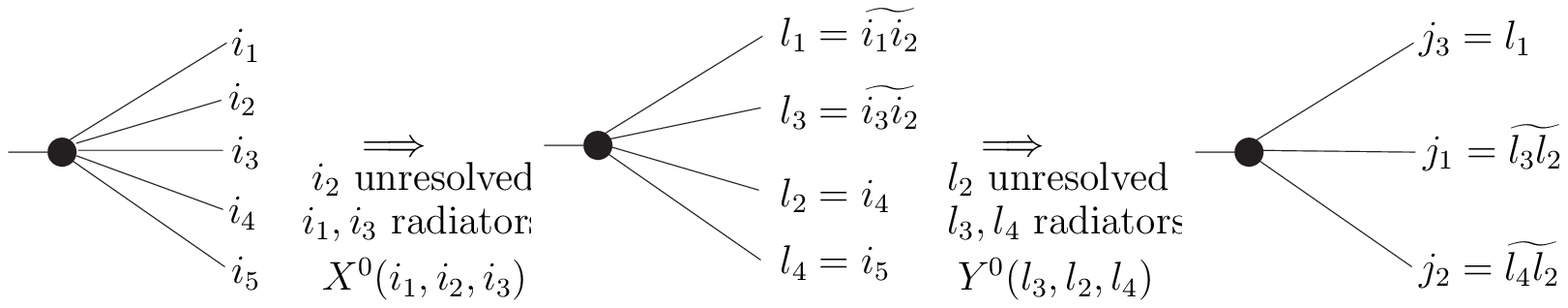,width=14.5cm} }

In the first step, one of the partons $\{i_1,i_2,i_3,i_4\}$ becomes unresolved.
The momentum $i_5=l_4$ is always unaffected by the
mapping.  The precise definition of the resulting momenta 
$l_1,l_2,l_3$ depends on the mapping.
In the first step of the two-step mapping ${\cal F}_{\rm{B}}^{(5\to 4\to 3)}$ we have
\begin{eqnarray}
\label{eq:pmapb}
l_1&=&x_1\,i_1+r_1\,i_2+z_1\,i_4\equiv\widetilde{i_1 i_2}\nonumber \\
l_2&=&i_3\nonumber \\
l_3&=&(1-x_1)\,i_1+(1-r_1)\,i_2+(1-z_1)\,i_4\equiv\widetilde{i_4 i_2}\nonumber \\
l_4&=&i_5\;,
\end{eqnarray}
while for the mappings ${\cal F}_{\rm{C}}^{(5\to 4\to 3)}$ and ${\cal F}_{\rm{K}}^{(5\to 4\to 3)}$, the first step is of the form
\begin{eqnarray}
\label{eq:pmapck}
l_1&=&x_1\,i_1+r_1\,i_2+z_1\,i_3\equiv\widetilde{i_1 i_2}\nonumber \\
l_2&=&i_4\nonumber \\
l_3&=&(1-x_1)\,i_1+(1-r_1)\,i_2+(1-z_1)\,i_3\equiv\widetilde{i_3 i_2}\nonumber \\
l_4&=&i_5\;.
\end{eqnarray}
The momentum $i_2$ is always the unresolved one, denoted 
generically by $i_u$ in the following. The two hard radiators denoted by $i_{a}$ and $i_{b}$
depend on the mapping.
In ${\cal F}_{\rm{B}}^{(5\to 4\to 3)}$, $i_1$ and $i_4$ are the 
hard radiators, while in ${\cal F}_{\rm{C}}^{(5\to 4\to 3)}$ and ${\cal F}_{\rm{K}}^{(5\to 4\to 3)}$,
 $i_1$ and $i_3$ are the 
hard radiators.

Note that mappings of type C apply to all configurations where 
 one of the combined partons  becomes unresolved  in the second step, so they 
also include the cases where the roles of $l_1$ and $l_3$ are interchanged.

The coefficients $r_1,x_1,z_1$ appearing in eqs.~(\ref{eq:pmapb}) and 
(\ref{eq:pmapck})
are given by eq.\,(\ref{single}),
\bea
r_1&=&\frac{s_{u{b}}}{s_{{a}u}+s_{u{b}}}\nn\\
x_1&=&\frac{1}{2(s_{au}+s_{ab})}\Big[(1+\rho)\,s_{aub} -2\,r_1\,s_{ub}     \Big]\nn\\
z_1&=&\frac{1}{2(s_{ub}+s_{ab})}\Big[(1-\rho)\,s_{aub} -2\,r_1\,s_{au}     \Big]\nn\\
\rho&=&\Big[1+\frac{4\,r_1(1-r_1)\,s_{{a}u}s_{u{b}}}{
s_{{a}{b}}s_{{a}u {b}}}\Big]^{\frac{1}{2}}\;\label{DAK}
\eea
where now $s_{ub}=(i_u+i_b)^2$ etc. 

In the second step, one of the intermediate partons $\{l_1,l_2,l_3\}$ becomes unresolved. 
The resulting momenta $j_1,j_2,j_3$ are defined as 
\begin{eqnarray}
j_1&=&x_2\,l_{a}+r_2\,l_u+z_2\,l_{b}\equiv\widetilde{l_{a}l_u}\label{j1}\\
j_2&=&(1-x_2)\,l_{a}+(1-r_2)\,l_u+(1-z_2)\,l_{b}\equiv\widetilde{l_{b}l_u}\label{j2}\\
j_3&=&l_r\;,\nn
\end{eqnarray}
where again $l_u$ denotes the momentum which is unresolved in the second step, 
$l_{a},l_{b}$ are the radiators and $l_r$ does not take part in the 
second recombination step. The coefficients $r_2,x_2,z_2$ are defined 
analogously to eq.~(\ref{DAK}), where now $s_{ub}=(l_u+l_b)^2$ etc. 

\begin{table}[t]
\begin{center}
\begin{tabular}{|c||c|c|c||c|c|c|}
\multicolumn{4}{c}{\phantom{Mapping type~~}First step}&
\multicolumn{3}{c}{Second step}\\
\hline
Mapping type&$i_u$&$i_a,i_b$&$i_s$&$l_u$&$l_{a},l_{b}$&$l_s$\\
&&&&&&\\
\hline 
\hline &&&&&&\\
B&$i_2$&$i_1,i_4$&$i_3$&$l_2$&$l_1,l_3$&$l_4$\\
C&$i_2$&$i_1,i_3$&$i_4$&$l_3$ or $l_1$&($l_1$ or $l_3$), $l_2$&$l_4$\\
K&$i_2$&$i_1,i_3$&$i_4$&$l_2$&$l_3,l_4$&$l_1$\\
&&&&&&\\
\hline
\end{tabular}
\end{center}
\caption{Identification of the unresolved, radiator and spectator 
momenta for both steps of the momentum mappings ${\cal F}_{\rm{B,C,K}}^{(5\to 4\to 3)}$.}
\label{Table:pmaps}
\end{table}

The combination of antenna functions associated with the
repeated unresolved singularity is thus
\begin{equation}
X^0(i_a,i_u,i_b)Y^0(l_a,l_u,l_b)\\
\end{equation}
as indicated 
in figures~\ref{fig:mapb}--\ref{fig:mape} where $X^0$ and $Y^0$ generically stand
for three-particle tree antenna functions.
The identification of the radiator and unresolved momenta 
$i_a,\ldots,l_b$ for the different mappings 
${\cal F}_{\rm{B,C,K}}^{(5\to 4\to 3)}$ can be read off from 
Table~\ref{Table:pmaps} and 
is also illustrated in Figures~\ref{fig:mapb}--\ref{fig:mape}.

\subsection{Decomposition of antenna functions into sub-antennae}
The antenna phase space mappings require two uniquely identified hard 
radiator momenta and an ordered emission of the unresolved partons. 

With the antenna functions of~\cite{ourant}, 
it is not always possible to uniquely identify the hard momenta, especially if
more than one final state parton is a gluon. Moreover, in 
the four-parton antenna functions (containing two unresolved partons) 
at subleading colour, the emission is not colour-ordered. It was already 
outlined in~\cite{ourant} that in both these cases, a further decomposition 
of the antenna functions into different sub-antennae configurations 
is required. 

The tree-level three-parton antenna functions $D_3^0(1_q,3_g,4_g)$ 
(quark-gluon-gluon) and $F_3^0(1_g,2_g,3_g)$ (gluon-gluon-gluon) contain
more than one antenna configuration, since each gluon can become unresolved. 
Their decomposition was discussed in~\cite{ourant}, it reads:
\begin{eqnarray}
D_3^0(1, 3, 4) &=& d_3^0(1, 3, 4) + d_3^0(1, 4, 3) \; ,
\label{eq:bigd}\\
F_3^0(1, 2, 3) &=& f_3^0(1, 3, 2) + f_3^0(3, 2, 1) + f_3^0(2, 1, 3)  \; ,
\label{eq:bigf}
\end{eqnarray}
with
\begin{eqnarray}
d_3^0(1, 3, 4) &=& 
\frac{1}{s_{134}^2} \, \Bigg(
\frac{2 s_{134}^2 s_{14}}{ s_{13}  s_{34}}
+ \frac{s_{14} s_{34}
+   s_{34}^2}{ s_{13}} 
+ \frac{s_{13}s_{14}}{ s_{34}} + \frac{5}{2}  s_{134}
+  \frac{1}{2}s_{34}\Bigg)
+ {\cal O}(\e)\;,
\label{eq:smalld}\\
f_3^0(1, 3, 2) &=& 
\frac{1}{s_{123}^2} \, \Bigg(
2\frac{s_{123}^2 s_{12}}{ s_{13}  s_{23}}
+ \frac{s_{12} s_{13}}{s_{23}}
+ \frac{s_{12} s_{23}}{s_{13}}
+ \frac{8}{3} s_{123} \Bigg) + {\cal O}(\e)\;.
\label{eq:smallf}
\end{eqnarray}
With this decomposition, the sub-antennae $d_3^0(i,j,k)$ and $f_3^0(i,j,k)$
contain only a  soft singularity associated with gluon $j$, and collinear 
singularities $i\parallel j$ and $k\parallel j$, such that $i$ and $k$ 
can be identified as hard radiators. Soft singularities associated with 
$i$ or $k$ and the collinear singularity $i\parallel k$,
which were present in the full antenna functions, are now contained in 
different sub-antennae, 
obtained by permutations of the momenta.
 Therefore, each sub-antenna can have a unique 
phase space mapping $(i,j,k) \to (\widetilde{ij},\widetilde{kj})$. 

The one-loop three-parton antenna functions $D_3^1(1_q,3_g,4_g)$ 
(quark-gluon-gluon) and $F_3^1(1_g,2_g,3_g)$ (gluon-gluon-gluon)
can be decomposed according to the same pattern, exploiting the fact that 
each  can be written as a function proportional to its tree-level 
counterpart plus a function which is not singular in any unresolved 
limit.  

The decomposition of tree-level four-parton antenna functions is more 
involved, especially since both single and double unresolved limits 
have to be accounted for properly. It turns out to be very useful 
to introduce the following combinations of three-parton antenna functions:
\begin{eqnarray}
Q^0_{3} (1,3,2) &=& d_3^0 (1,3,2) - A_3^0(1,3,2)\;,\nonumber \\
R^0_{3} (1,3,2) &=& Q^0_{3}(1,3,2) - Q^0_{3}(1,2,3)\;\nonumber \\
S^0_{3} (1,3,2) &=& Q^0_{3}(1,3,2) + Q^0_{3}(1,2,3)
+ E_3^0(1,3,2) \;.
\end{eqnarray}
None of these contains any soft limit or collinear $1\parallel i$ limit. 
Only $Q^0_{3}$ and $R^0_{3}$ contain a $2\parallel 3$ limit, while 
$S^0_{3}$ is also finite in this limit, owing to the ${\cal N}=1$ 
supersymmetry relation between the tree-level splitting functions,
\begin{equation}
P_{q\bar q \to G} (z) + P_{gg \to G}(z) = P_{qg\to Q} (z) + 
 P_{qg\to Q} (1-z)\;.  
\end{equation}

Among the tree-level four-parton antenna functions, only
$\tilde{A}_4^0(1_q,3_g,4_g,2_{\bar q})$ (quark-gluon-gluon-antiquark at 
subleading colour), $D_4^0(1_q,3_g,4_g,5_g)$ (quark-gluon-gluon-gluon), 
$E_4^0(1_q,3_{q'},4_{\bar q'},5_g)$ (quark-quark-antiquark-gluon at 
leading colour),  $F_4^0(1_g,2_g,3_g,4_g)$ (gluon-gluon-gluon-gluon) as 
well as  
$G_4^0 (1_g,3_q,4_{\bar q},2_g)$ and
 $\tilde{G}_4^0 (1_g,3_q,4_{\bar q},2_g)$
(gluon-quark-antiquark-gluon at leading and subleading colour) 
must be decomposed into sub-antennae. In the context of the 
three-jet calculation discussed here, $F_4^0$, $G_4^0$ and 
$\tilde{G}_4^0$ do not contribute and will not be discussed further.

The decomposition of $\tilde{A}_4^0(1_q,3_g,4_g,2_{\bar q})$  is needed 
because both gluons can become collinear either with quark $1_q$
or with antiquark $2_{\bar q}$. The two possible phase space mappings 
are of the ${\cal F}^{(5\to 3)}$ type described in section~\ref{sec:pmap5to3}
\begin{displaymath}
 \mbox{(a): }(1,3,4,2) \to (\widetilde{134},\widetilde{243}) \qquad
 \mbox{(b): }(1,4,3,2) \to (\widetilde{143},\widetilde{234})
\end{displaymath}
each allow only 
gluons adjacent to quark or antiquark to become collinear. 
To disentangle the different sub-antennae, it is sufficient to partial-fraction 
the antenna function in the different collinear denominators, 
as done in~\cite{ourant}. With these, we decompose 
\begin{equation}
\tilde{A}_4^0(1,3,4,2) 
= \tilde{A}_{4,a}^0(1,3,4,2) + 
\tilde{A}_{4,b}^0(1,3,4,2), 
\end{equation}
with
\begin{eqnarray}
\tilde{A}_{4,a}^0(1,3,4,2) &=& 
\tilde{a}_{4}^0(1,3,4,2) + 
\tilde{a}_{4}^0(2,4,3,1)\;,\nonumber \\
\tilde{A}_{4,b}^0(1,3,4,2) &=& 
\tilde{A}_{4,a}^0(1,4,3,2)\;,
\end{eqnarray}
where $\tilde{a}_{4}^0(i,j,k,l)$ contains only singularities for $i\parallel
j$ or $k\parallel l$
 and was defined in~\cite{ourant}. With this 
decomposition,
$\tilde{A}_{4,a}^0(1,3,4,2)$ contains only $1\parallel 3$ and 
$2\parallel 4$ singularities, and can be used with the 
$(1,3,4,2) \to (\widetilde{134},\widetilde{243})$ mapping. Since this 
decomposition is straightforward, we refrain from spelling it out explicitly 
in the subtraction terms presented in subsequent sections. 

The leading-colour quark-quark-antiquark-gluon antenna function
 $E_4^0(1_q,3_{q'},4_{\bar q'},5_g)$ contains limits where either 
the quark-antiquark pair $(3_{q'},4_{\bar q'})$ or the gluon $5_g$ 
can become soft. Since these limits yield different hard radiator partons, 
they can not be accounted for in a single phase space mapping, but require 
two separate ${\cal F}^{(5\to 3)}$  mappings:
$$ \mbox{(a): } (1,3,4,5) \to (\widetilde{134},\widetilde{543})\,,\qquad
 \mbox{(b): } (1,5,4,3) \to (\widetilde{154},\widetilde{345})\,.$$
By analysing the different triple and single collinear limits of 
 $E_4^0$, one finds the following decomposition:
\begin{eqnarray}
 E_{4,a}^0(1,3,4,5) & = & B_4^0(1,3,4,5) + E_3^0(5,4,3)\, Q^0_3
(1,\widetilde{(34)},\widetilde{(54)})\;,\\
 E_{4,b}^0(1,3,4,5) & = &   E_{4}^0(1,3,4,5) - E_{4,a}^0(1,3,4,5)\;.
\end{eqnarray}
After this decomposition, $E_{4,a}^0$ can be used with mapping (a) and 
$E_{4,b}^0$ with mapping (b). The above decomposition also ensures a 
well-defined behaviour in all double and single unresolved limits
(see~\cite{ourant} for a definition of the splitting factors):
\begin{eqnarray}
E_{4,a}^0(1,3,4,5)&\stackrel{ 3_{q'} \to 0, 4_{\bar{q}'} \to 0}
{\longrightarrow}& S_{15}(3,4)\;, \nonumber \\
E_{4,b}^0(1,3,4,5)&\stackrel{ 3_{q'}\parallel 4_{\bar q'}, 5_g \to 0}
{\longrightarrow}&   S_{1;543}(z) \;
\frac{1}{s_{34}}\; P_{q\bar q\to G}(z)\;,\nonumber \\
E_{4,a}^0(1,3,4,5)&\stackrel{ 1_q\parallel 3_{q'}  \parallel 4_{\bar q'}}
{\longrightarrow} &
P^{{\rm non-ident.}}_{134 \to Q}(w,x,y) 
\;,\nonumber\\
E_{4,a}^0(1,3,4,5)+E_{4,b}^0(1,3,4,5)& 
\stackrel{ 3_{q'}  \parallel 4_{\bar q'}\parallel 5_g }
{\longrightarrow}& P_{543 \to G}(w,x,y) \;,\nonumber \\
E_{4,b}^0(1,3,4,5)&\stackrel{ 1_q\parallel 5_g , 3_{q'}  \parallel 4_{\bar q'}}
{\longrightarrow}& 
\frac{1}{s_{34} s_{15}}\;
P_{qg\to Q}(z)\; P_{q \bar{q}\to G}(y)\;,
\end{eqnarray}
\begin{eqnarray}
E_{4,b}^0(1,3,4,5)&\stackrel{5_g\to 0}
{\longrightarrow}& S_{154}\; E_3^0(1,3,4)\;, \nonumber \\
E_{4,a}^0(1,3,4,5)&\stackrel{3_{q'} \parallel 4_{\bar q'}}
{\longrightarrow}&
 \frac{1}{s_{34}}\;P_{q\bar q\to G}(z)\; 
d_3^0(1,(34),5) + {\rm ang.} \;,\nonumber\\
E_{4,b}^0(1,3,4,5)&\stackrel{3_{q'} \parallel 4_{\bar q'}}
{\longrightarrow}&
 \frac{1}{s_{34}}\;P_{q\bar q\to G}(z)\; 
d_3^0(1,5,(34)) + {\rm ang.} \;, \nonumber \\
E_{4,b}^0(1,3,4,5)&\stackrel{ 1_q \parallel 5_g}
{\longrightarrow}&  
\frac{1}{s_{15}}\;P_{qg\to Q}(z)\; E_3^0((15),3,4)\;,  \nonumber \\
E_{4,b}^0(1,3,4,5)&\stackrel{ 4_{\bar q'}\parallel 5_g}
{\longrightarrow}&  
\frac{1}{s_{45}}\;P_{qg\to Q}(z)\; E_3^0(1,3,(45))\;,
\end{eqnarray}
while all other limits are zero. It can be seen that only the triple collinear 
$3\parallel 4 \parallel 5$ and the single collinear $3\parallel 4$
limits receive contributions from both phase space mappings. This 
is unavoidable, since both these limits match onto the double soft 
$(3_{q'},4_{\bar q'})\to 0$ and the soft $5_g$ limits, which belong to 
different mappings.  

The decomposition of the quark-gluon-gluon-gluon 
antenna function  $D_4^0(1_q,3_g,4_g,5_g)$, is more involved, since any pair 
of two gluons can become soft.  We consider four different ${\cal F}^{(5\to 3)}$  mappings:
\begin{eqnarray*}
 \mbox{(a): } (1,3,4,5) \to (\widetilde{134},\widetilde{543})\,,&\qquad&
 \mbox{(b): } (1,5,4,3) \to (\widetilde{154},\widetilde{345})\,, \\
 \mbox{(c): } (1,3,5,4) \to (\widetilde{135},\widetilde{453})\,,&\qquad&
 \mbox{(d): } (1,5,3,4) \to (\widetilde{153},\widetilde{435})\, .
\end{eqnarray*}
The numerous different double and single unresolved limits of this 
antenna function can be disentangled very elegantly by repeatedly 
exploiting the ${\cal N} =1$ supersymmetry relation~\cite{campbell} among the 
different triple collinear splitting functions~\cite{campbell,ggamma,doubleun}.
Using this relation, one can show that the following 
left-over combination is 
finite in all single unresolved and double unresolved limits:
\begin{eqnarray}
D_{4,l}^0(1,3,4,5) &=& D_{4}^0(1,3,4,5) - \Bigg[
A_4^0(1,3,4,5)  + A_4^0(1,5,4,3)\nonumber \\
&&  -\frac{1}{2} \left(
\tilde{E}_4^0(1,3,5,4) + \tilde{E}_4^0(1,5,3,4) \right)
+ \tilde{A}_4^0(1,3,5,4) \nonumber \\
&&
- E_4^0(1,5,4,3) + B_4^0(1,5,4,3) + C_4^0(1,4,5,3)\nonumber \\
&&
- E_4^0(1,3,4,5) + B_4^0(1,3,4,5) + C_4^0(1,4,3,5)\nonumber \\
&&
+ A_3^0(1,3,4)\, S_{3}^0 (\widetilde{(13)},\widetilde{(43)},5)
+ A_3^0(1,5,4)\, S_{3}^0 (\widetilde{(15)},\widetilde{(45)},3)\nonumber \\
&&
+ A_3^0(3,4,5)\, S_{3}^0 (1,\widetilde{(54)},\widetilde{(34)}) \Bigg] \;.
\end{eqnarray}
Starting from the terms in this expression, the following 
sub-antennae can be constructed:
\begin{eqnarray}
D_{4,a}^0(1,3,4,5) &=& \frac{1}{2} D_{4,l}^0(1,3,4,5)+ A_4^0(1,3,4,5) 
- \frac{1}{2} \tilde{E}_4^0(1,3,5,4) 
\nonumber \\
&&+ A_3^0(1,3,4)\, S_{3}^0 (\widetilde{(13)},\widetilde{(43)},5)
\nonumber \\ &&
+ \frac{1}{2} A_3^0(3,4,5)\,\left(
S_{3}^0 (1,\widetilde{(54)},\widetilde{(34)})
-  R_{3}^0 (1,\widetilde{(54)},\widetilde{(34)})\right)
\nonumber \\ &&-  E_3^0(5,4,3) Q_{3}^0 (1,\widetilde{(34)},\widetilde{(54)})
- A_3^0(1,3,4)\, E_{3}^0 (\widetilde{(13)},\widetilde{(43)},5)
\nonumber \\ &&- A_3^0(1,3,4)\, Q_{3}^0 (\widetilde{(13)},5,\widetilde{(43)})
\;, \nonumber \\
D_{4,b}^0(1,3,4,5) &=& D_{4,a}^0(1,5,4,3)\;, \nonumber \\
D_{4,c}^0(1,3,4,5) &=& \tilde{A}_{4,a}^0(1,3,5,4)
- E_4^0(1,5,4,3) + B_4^0(1,5,4,3) + C_4^0(1,4,5,3) \nonumber \\
&&
+ E_3^0(3,4,5) Q_{3}^0 (1,\widetilde{(54)},\widetilde{(34)})
+ A_3^0(1,3,4)\, E_{3}^0 (\widetilde{(13)},\widetilde{(43)},5)\nonumber \\
&&
+ a_3^0(1,3,4)\, Q_{3}^0 (\widetilde{(13)},5,\widetilde{(43)})
+ a_3^0(4,5,1)\, Q_{3}^0 (\widetilde{(15)},3,\widetilde{(45)})\;,
\nonumber \\
D_{4,d}^0(1,3,4,5) &=& D_{4,c}^0(1,5,4,3)\;.
\end{eqnarray}
With this decomposition, each $D_{4,i}^0$ 
contains only singularities appropriate to
phase space mapping (i). The sum of the  $D_{4,i}^0$  adds to 
 $D_{4}^0$:
\begin{equation}
D_{4,a}^0 + D_{4,b}^0 + D_{4,c}^0 + D_{4,d}^0 = D_4^0\;,
\end{equation}
such that only $D_{4}^0$ must be integrated analytically over the antenna 
phase space. 

The above decomposition disentangles the different double and single 
unresolved limits:
\begin{eqnarray}
D_{4,a}^0(1,3,4,5)&\stackrel{ 3_g \to 0, 4_g \to 0}
{\longrightarrow}&\; S_{1345}\;,\nonumber\\
D_{4,b}^0(1,3,4,5)&\stackrel{ 4_g \to 0, 5_g \to 0}
{\longrightarrow}&\; S_{1543}\;,\nonumber\\
D_{4,c}^0(1,3,4,5)+ D_{4,d}^0(1,3,4,5)&\stackrel{ 3_g \to 0, 5_g \to 0}
{\longrightarrow}&\; S_{134}\;S_{154}\;,\nonumber \\
D_{4,c}^0(1,3,4,5)+D_{4,d}^0(1,3,4,5)&\stackrel{ 1_q\parallel 5_g, 3_g \to 0}
{\longrightarrow}& S_{4;315}(z) \;
\frac{1}{s_{15}}\; P_{qg\to Q}(1-z)
\;,\nonumber \\
D_{4,a}^0(1,3,4,5)+ 
D_{4,c}^0(1,3,4,5) + D_{4,d}^0(1,3,4,5)
&\stackrel{ 4_g\parallel 5_g, 3_g \to 0}
{\longrightarrow}& S_{1;345}(z) \;
\frac{1}{s_{45}}\; P_{gg\to G}(z)
\;,\nonumber \\
D_{4,a}^0(1,3,4,5)&\stackrel{ 1_q\parallel 3_g, 4_g \to 0}
{\longrightarrow}& S_{5;431}(z) \;
\frac{1}{s_{13}}\; P_{qg\to Q}(z)
\;,\nonumber \\
D_{4,b}^0(1,3,4,5)&\stackrel{ 1_q\parallel 5_g, 4_g \to 0}
{\longrightarrow}& S_{3;451}(z) \;
\frac{1}{s_{15}}\; P_{qg\to Q}(z)
\;,\nonumber \\
D_{4,c}^0(1,3,4,5)+D_{4,d}^0(1,3,4,5)&\stackrel{ 1_q\parallel 3_g, 5_g \to 0}
{\longrightarrow}& S_{4;513}(z) \;
\frac{1}{s_{13}}\; P_{qg\to Q}(1-z)
\;,\nonumber \\
D_{4,b}^0(1,3,4,5)+ 
D_{4,c}^0(1,3,4,5) + D_{4,d}^0(1,3,4,5)&\stackrel{ 3_g\parallel 4_g, 5_g \to 0}
{\longrightarrow}& S_{1;543}(z) \;
\frac{1}{s_{34}}\; P_{gg\to G}(z)\;, \nonumber \\ 
D_{4,a}^0(1,3,4,5)&\stackrel{ 1_q\parallel 3_g  \parallel 4_g }
{\longrightarrow}& P_{134 \to Q}(w,x,y)\;,\nonumber\\
D_{4,b}^0(1,3,4,5)&\stackrel{ 1_q\parallel 5_g \parallel 4_g}
{\longrightarrow}& P_{154 \to Q}(w,x,y)\;,\nonumber\\
D_{4,c}^0(1,3,4,5)+ D_{4,d}^0(1,3,4,5)
&\stackrel{ 1_q \parallel 3_g \parallel 5_g}
{\longrightarrow}& \tilde{P}_{135 \to Q}(w,x,y)\;,\nonumber\\
D_4^0(1,3,4,5)&\stackrel{ 3_g\parallel 4_g \parallel 5_g}
{\longrightarrow}& P_{345 \to G}(w,x,y)\;,\nonumber \\
D_{4,a}^0(1,3,4,5)+D_{4,c}^0(1,3,4,5)
&\stackrel{ 1_q \parallel 3_g,  4_g \parallel 5_g}
{\longrightarrow}& 
\frac{1}{s_{13} s_{45}}\;
P_{qg\to Q}(z)\; P_{g g\to G}(y)\;,\nonumber \\
D_{4,b}^0(1,3,4,5)+D_{4,d}^0(1,3,4,5)
&\stackrel{1_q \parallel 5_g,  3_g \parallel 4_g }
{\longrightarrow}& 
\frac{1}{s_{15} s_{34}}\;
P_{qg\to Q}(z)\; P_{ g g\to G }(y)\;,\nonumber \\
\end{eqnarray}
\begin{eqnarray}
D_{4,a}^0(1,3,4,5)&\stackrel{{3}_g\to 0}
{\longrightarrow}&S_{134}\; d_3^0(1,4,5)\;,\nonumber\\
D_{4,c}^0(1,3,4,5)+D_{4,d}^0(1,3,4,5)&\stackrel{{3}_g\to 0}
{\longrightarrow}&S_{134}\; d_3^0(1,5,4)\;,\nonumber\\
D_{4,a}^0(1,3,4,5)&\stackrel{{4}_g\to 0}
{\longrightarrow}&S_{345}\; d_3^0(1,3,5)\;,\nonumber\\
D_{4,b}^0(1,3,4,5)&\stackrel{{4}_g\to 0}
{\longrightarrow}&S_{345}\; d_3^0(1,5,3)\;,\nonumber\\
D_{4,b}^0(1,3,4,5)&\stackrel{{5}_g\to 0}
{\longrightarrow}&S_{154}\; d_3^0(1,4,3)\;,\nonumber\\
D_{4,c}^0(1,3,4,5)+D_{4,d}^0(1,3,4,5)&\stackrel{{5}_g\to 0}
{\longrightarrow}&S_{154}\; d_3^0(1,3,4)\;,\nonumber\\
D_{4,a}^0(1,3,4,5)&\stackrel{1_q \parallel 3_g}
{\longrightarrow}&\frac{1}{s_{13}}\;P_{qg\to Q}(z)\; 
d_3^0((13),4,5)\;,\nonumber\\
D_{4,c}^0(1,3,4,5)&\stackrel{1_q \parallel 3_g}
{\longrightarrow}&\frac{1}{s_{13}}\;P_{qg\to Q}(z)\; 
d_3^0((13),5,4)\;,\nonumber\\
D_{4,b}^0(1,3,4,5)&\stackrel{1_q \parallel 5_g}
{\longrightarrow}&\frac{1}{s_{15}}\;P_{qg\to Q}(z)\; 
d_3^0((15),4,3)\;,\nonumber\\
D_{4,d}^0(1,3,4,5)&\stackrel{1_q \parallel 5_g}
{\longrightarrow}&\frac{1}{s_{15}}\;P_{qg\to Q}(z)\; 
d_3^0((15),3,4)\;,\nonumber\\
D_{4,a}^0(1,3,4,5)&\stackrel{3_g \parallel 4_g}
{\longrightarrow}&
\frac{1}{s_{34}}\;P_{gg\to G}(z)\; 
d_3^0(1,(34),5) + {\rm ang.}\;,\nonumber\\
D_{4,b}^0(1,3,4,5)+D_{4,d}^0(1,3,4,5)&\stackrel{3_g \parallel 4_g}
{\longrightarrow}&
\frac{1}{s_{34}}\;P_{gg\to G}(z)\; 
d_3^0(1,5,(34)) + {\rm ang.}\;,\nonumber\\
D_{4,b}^0(1,3,4,5)&\stackrel{4_g \parallel 5_g}
{\longrightarrow}&
\frac{1}{s_{45}}\;P_{gg\to G}(z)\; 
d_3^0(1,(45),3) + {\rm ang.}\;,\nonumber \\
D_{4,a}^0(1,3,4,5)+D_{4,c}^0(1,3,4,5)&\stackrel{4_g \parallel 5_g}
{\longrightarrow}&
\frac{1}{s_{45}}\;P_{gg\to G}(z)\; 
d_3^0(1,3,(45)) + {\rm ang.}\;.
\end{eqnarray}
All other limits are vanishing. It can be seen that certain limits are shared 
among several antenna functions, which can be largely understood due to two 
reasons: 
\begin{enumerate}
\item in a gluon-gluon collinear splitting, either gluon can become 
soft, and the gluon-gluon splitting function is always shared between two 
sub-antennae, as in (\ref{eq:bigd}), (\ref{eq:bigf}) to disentangle the two 
soft limits.
\item the unresolved emission of gluons $3_g$ and $5_g$
is shared between the mappings (c) and (d) according to the decomposition 
of the non-ordered antenna function $\tilde{A}_4^0$, which distributes 
the soft limit of either gluon between both mappings. 
\end{enumerate}

\subsection{Angular terms}
The angular terms in the single unresolved limits 
are associated with a gluon splitting into two gluons or 
into a quark-antiquark pair. They 
 average to zero after integration over the antenna 
phase space. To ensure numerical stability and reliability, this average 
has to take place within each phase space mapping. We have checked this to be 
the case for the above decompositions of $E_4^0$ and $D_4^0$. The angular 
average in single collinear limits can be made using the standard 
momentum parametrisation~\cite{ap,cs} for the $i\parallel j$  limit:
\begin{eqnarray*}
p_i^\mu = z p^\mu + k_\perp^\mu - \frac{k_\perp^2}{z}\frac{n^\mu}{2p\cdot n}
\;, &\qquad&
 p_j^\mu = (1-z) p^\mu - k_\perp^\mu 
- \frac{k_\perp^2}{1-z}\frac{n^\mu}{2p\cdot n}\;, \\
\mbox{with } 2p_i\cdot p_j = -\frac{k_\perp^2}{z(1-z)}\;, &\qquad& p^2 
= n^2=0\;.
\end{eqnarray*}
In this $p^\mu$ denotes the collinear momentum direction, and $n^\mu$ is an 
auxiliary vector. The collinear limit is approached by $k_\perp^2\to 0$. 

In the simple collinear $i\parallel j$ 
limit of the four-parton antenna functions
$X_4^0(i,j,k,l)$, 
one chooses $n=p_k$ to be one of the non-collinear momenta, such that the 
antenna function can be expressed in terms of $p$, $n$, $k_\perp$ and $p_l$. 
Expanding in $k_\perp^\mu$ yields only non-vanishing  scalar products of the 
form $p_l\cdot k_\perp$. Expressing the integral over the antenna phase 
space in the $(p,n)$ centre-of-mass frame, the angular average can be 
carried out as
\begin{equation}
\frac{1}{2\pi} \int_0^{2\pi} \d \phi\, (p_l\cdot k_\perp) =0 \;, \qquad 
 \frac{1}{2\pi} \int_0^{2\pi} \d \phi\, (p_l\cdot k_\perp)^2 = - 
k_\perp^2\, \frac{p\cdot p_l\, n\cdot p_l}{p\cdot n}\;.
\end{equation}
Higher powers of $k_\perp^\mu$ are not sufficiently singular to contribute to 
the collinear limit. Using the above average, we could analytically verify the 
cancellation of angular terms within each single phase space mapping, 
which is independent on the choice of the reference vector $n_\mu$. 

\noindent {\it The remainder of this section
 has been modified compared to the original 
version of the paper:}
In the $N^2$ and $N^0$ colour factor, 
the angular averaging is not sufficient to cancel the $1/\e$ poles in the
four-parton one loop subtraction terms~\cite{weinzierl}.
In either of these colour factors, 
the difference $\d\sigma^{V,1}_{NNLO}-\d\sigma^{VS,1}_{NNLO}$ contains 
left-over poles of the form 
\begin{equation} 
\frac{1}{\e^2} X_3^0(1,i,2)\, Y_3^0(\widetilde{1i},j,\widetilde{2i}) 
J_3^{(3)}(\widetilde{1i},j,\widetilde{2i}) \left\{
s_{\widetilde{1i}j}^{-\e} + s_{\widetilde{2i}j}^{-\e} -
s_{1j}^{-\e} - s_{2j}^{-\e} - s_{1i2}^{-\e} + s_{12}^{-\e}\right\}\;,
\label{eq:ir4p1lfull}
\end{equation}
where $X_3^0$ and $Y_3^0$ are tree-level three-parton antenna functions. 
Contrary to statements made in~\cite{weinzierl}, these
terms do not appear in the  colour factor $N_F\, N$ in our implementation. 

Furthermore, for these two colour factors the five-parton 
subtraction terms themselves do introduce spurious 
limits from large angle soft radiation. 
The single soft limit of $i$ or $k$ in (6.3) is non-vanishing.
 Instead, it yields (soft $i$):
\begin{eqnarray}
&+&\frac{1}{2} d_3^0(2,k,j) A_3^0(1,\widetilde{jk},\widetilde{2k})\left[
S_{1i\widetilde{(jk)}} + S_{1i\widetilde{(2k)}} - 
S_{\widetilde{(2k)}i\widetilde{(jk)}} - S_{1ij} - S_{1i2} + S_{2ij}\right] 
+(1\leftrightarrow 2) 
\nonumber \\
&-&\frac{1}{2} A_3^0(1,k,2) A_3^0(\widetilde{(1k)},j,\widetilde{(2k)})
\left[S_{\widetilde{(1k)}ij} +S_{\widetilde{(2k)}ij}-
S_{\widetilde{(1k)}i\widetilde{(2k)}} - S_{1ij} - S_{2ij} + S_{1i2}\right]
\end{eqnarray}
with
\begin{equation}
S_{abc} = 2\frac{s_{ac}}{s_{ab}s_{bc}}
\end{equation}
To account for this large angle soft radiation, a new subtraction term 
$\d \sigma_{NNLO}^A$ is introduced. This term is added to 
the five-parton subtraction term $\d \sigma_{NNLO}^S$, 
and its integrated form 
is subtracted from the four-parton subtraction term $\d \sigma_{NNLO}^{VS,1}$,
cancelling the left -over $1/\e$ terms and adding new finite contributions to
the  four-parton and the five-parton subtraction term.

The new subtraction term $\d \sigma_{NNLO}^A$ contributes only in the 
$N^2$ and $N^0$ colour factors. Its contribution to $N^2$ reads:
\begin{eqnarray}
\lefteqn{{\rm d}\sigma_{NNLO,N^2}^A=
N_{{5}}\,{N^2} \,  {\rm d}\Phi_{5}(p_{1},\ldots,p_{5};q)
 \, \frac{1}{3!}\,
\sum_{(i,j,k)\in P_C(3,4,5) } \, \Bigg\{ }\nonumber \\
\ph{(v)}&& 
+ \frac{1}{2}\Big(S_{\widetilde{((1i)k)}i\widetilde{((ji)k)}}
          -S_{\widetilde{(1i)}i\widetilde{(ji)}}
          -  S_{2i\widetilde{((ji)k)}} + S_{2i\widetilde{(ji)}}
 - S_{2i\widetilde{((1i)k)}} + S_{2i\widetilde{(1i)}}
\Big)\, \nonumber \\ 
&&       \hspace{5mm}    \times   d_3^0(\widetilde{(1i)}_{q},k_g,\widetilde{(ji)}_g)\,
{A}^0_{3}(\widetilde{((1i)k)}_q,\widetilde{((ji)k)}_g,2_{\bar q})\,
{J}_{3}^{(3)}(\widetilde{p_{(1i)k}},\widetilde{p_{(ji)k}},p_2) \nonumber \\
\ph{(w)}&& 
+ \frac{1}{2}\Big(S_{\widetilde{((1k)i)}k\widetilde{((jk)i)}}
          -S_{\widetilde{(1k)}k\widetilde{(jk)}}
          -  S_{2k\widetilde{((jk)i)}} + S_{2k\widetilde{(jk)}}
 - S_{2k\widetilde{((1k)i)}} + S_{2k\widetilde{(1k)}}
\Big)\, \nonumber \\ 
&&       \hspace{5mm}    \times   d_3^0(\widetilde{(1k)}_{\bar q},i_g,\widetilde{(jk)}_g)\,
{A}^0_{3}(\widetilde{((1k)i)}_q,\widetilde{((jk)i)}_g,2_{\bar q})\,
{J}_{3}^{(3)}(\widetilde{p_{(1k)i}},\widetilde{p_{(jk)i}},p_2) \nonumber \\
\ph{(x)}&& 
+ \frac{1}{2}\Big(S_{\widetilde{((2i)k)}i\widetilde{((ji)k)}}
          -S_{\widetilde{(2i)}i\widetilde{(ji)}}
          -  S_{1i\widetilde{((ji)k)}} + S_{1i\widetilde{(ji)}}
 - S_{1i\widetilde{((2i)k)}} + S_{1i\widetilde{(2i)}}
\Big)\, \nonumber \\ 
&&       \hspace{5mm}    \times   d_3^0(\widetilde{(2i)}_{\bar q},k_g,\widetilde{(ji)}_g)\,
{A}^0_{3}(1_q,\widetilde{((ji)k)}_g,\widetilde{((2i)k)}_{\bar q})\,
{J}_{3}^{(3)}(p_1,\widetilde{p_{(ji)k}},\widetilde{p_{(2i)k}}) \nonumber \\
\ph{(y)}&& 
+ \frac{1}{2}\Big(S_{\widetilde{((2k)i)}k\widetilde{((jk)i)}}
          -S_{\widetilde{(2k)}k\widetilde{(jk)}}
          -  S_{1k\widetilde{((jk)i)}} + S_{1k\widetilde{(jk)}}
 - S_{1k\widetilde{((2k)i)}} + S_{1k\widetilde{(2k)}}
\Big)\, \nonumber \\ 
&&       \hspace{5mm}    \times   d_3^0(\widetilde{(2k)}_{\bar q},i_g,\widetilde{(jk)}_g)\,
{A}^0_{3}(1_q,\widetilde{((jk)i)}_g,\widetilde{((2k)i)}_{\bar q})\,
{J}_{3}^{(3)}(p_1,\widetilde{p_{(jk)i}},\widetilde{p_{(2k)i}}) 
\; \nonumber \\
\ph{(aj)}&& 
-\frac{1}{2}\Big(
S_{\widetilde{((1i)k)}i\widetilde{((2i)k)}} 
-S_{\widetilde{((1i)k)}ij} 
-S_{\widetilde{((2i)k)}ij}
+S_{\widetilde{(1i)}ij}  +S_{\widetilde{(2i)}ij} 
- S_{\widetilde{(1i)}i\widetilde{(2i)}}\Big)\nonumber\\
&&      \hspace{5mm}    \times   
A_3^0(\widetilde{(1i)}_{q},k_g,\widetilde{(2i)}_{\bar q})\,
{A}^0_{3}(\widetilde{((1i)k)}_{q},j_g,\widetilde{((2i)k)}_{\bar q})\,
{J}_{3}^{(3)}(\widetilde{p_{(1i)k}},p_j,\widetilde{p_{(2i)k}})\nonumber \\
\ph{(ai)}&&
-\frac{1}{2}\Big(
S_{\widetilde{((1k)i)}k\widetilde{((2k)i)}} 
-S_{\widetilde{((1k)i)}kj} 
-S_{\widetilde{((2k)i)}kj}
+S_{\widetilde{(1k)}kj}  +S_{\widetilde{(2k)}kj} 
- S_{\widetilde{(1k)}k\widetilde{(2k)}}\Big)\nonumber\\
 &&     \hspace{5mm}    \times   
A_3^0(\widetilde{(1k)}_{q},i_g,\widetilde{(2k)}_{\bar q})\,
{A}^0_{3}(\widetilde{((1k)i)}_{q},j_g,\widetilde{((2k)i)}_{\bar q})\,
{J}_{3}^{(3)}(\widetilde{p_{(1k)i}},p_j,\widetilde{p_{(2k)i}})\Bigg\}
\end{eqnarray}

The new contribution to the $N^0$ five-parton subtraction term is:
\begin{eqnarray}
\lefteqn{{\rm d}\sigma_{NNLO,N^0}^A= 
N_{{5}}\,{N^0} \,  {\rm d}\Phi_{5}(p_{1},\ldots,p_{5};q)}
 \nonumber \\ 
\ph{(s)}&&\frac{1}{2} \sum_{(i,j) \in (3,4)} 
 \left(
S_{\widetilde{((1i)5)}ij} +
S_{\widetilde{((2i)5)}ij} 
-S_{\widetilde{((1i)5)}i\widetilde{((2i)5)}}
- S_{\widetilde{(1i)}ij} - S_{\widetilde{(2i)}ij}
+  S_{\widetilde{(1i)}i\widetilde{(2i)}}
\right) \nonumber \\ \, 
&& \hspace{5mm}
\times A_3^0(\widetilde{(1i)}_q,5_g,\widetilde{(2i)}_{\bar q}) \,
A_3^0(\widetilde{((1i)5)}_q,j_g,\widetilde{((2i)5)}_{\bar q})
\, J_3^{(3)} (\widetilde{p_{(1i)5}},\widetilde{p_{(2i)5}},p_j)
\end{eqnarray}
These large-angle soft subtraction terms contain 
soft antenna functions of the form $S_{ajc}$
which is simply the eikonal factor for a soft gluon $j$ emitted between hard
partons $a$ and $c$. Those soft factors are associated with an 
antenna phase space mapping $(i,j,k)\to(I,K)$. The hard momenta $a$, $c$
do not need to be equal to the hard momenta $i$, $k$ in the
antenna phase space - they can be arbitrary on-shell momenta. 

The integral of each of these soft antenna functions over 
the antenna phase space can be written as
\begin{eqnarray}
{\cal S}_{ac;ik} &=& \int \d \Phi_{X_{ijk}} S_{ajc} \nonumber \\
&=& \left( s_{IK} \right)^{-\e}\;  \frac{\Gamma^2(1-\e)e^{\e \gamma}}
{\Gamma(1-3 \e)}\; 
\left(-\frac{2}{\e}\right)
 \left[-\frac{1}{\e} +\ln\left(x_{ac,IK}\right)  
+ \e\,\Li_{2}\left(-\frac{1-x_{ac,IK}}{x_{ac,IK}}
\right) \right]\;,\nonumber \\
\end{eqnarray}
where we have defined
\begin{equation}
x_{ac,IK} = \frac{s_{ac}s_{IK}}{(s_{aI}+s_{aK})(s_{cI}+s_{cK})}\;.
\end{equation}
So that the integration of the new $N^{2}$ subtraction terms reads
\begin{eqnarray}
\int \d \Phi_{X_{ijk}}{\rm d}\sigma_{NNLO,N^2}^A=
 {N_{{4}}}\,{N^2}\, \left(\frac{\alpha_s}{2\pi}\right)\,
 {\rm d}\Phi_{4}(p_{1},\ldots,p_{4};q) \times \nonumber \\
\frac{1}{4} \sum_{(i,j)\in (3,4)} \Bigg\{  
\left( {\cal S}_{\widetilde{(1i)}\widetilde{(ji)};1j} 
- {\cal S}_{1j;1j}
- {\cal S}_{2\widetilde{(ji)};1j} + {\cal S}_{2j;1j} -
{\cal S}_{2\widetilde{(1i)};1j}  
 + {\cal S}_{12;1j} \right) \nonumber \\
d_3^0(1_q,i_{g},j_{g})\,
 A_3^0(\widetilde{(1i)}_q,\widetilde{(ji)}_g,2_{\bar q})
\JET_{3}^{(3)}(\widetilde{p_{1i}},\widetilde{p_{ji}},p_2) 
+(1\leftrightarrow 2) \nonumber \\
-\left(
 {\cal S}_{\widetilde{(1i)}\widetilde{(2i)};12} 
- {\cal S}_{12;12}
- {\cal S}_{\widetilde{(2i)}j;12} + {\cal S}_{2j;12} -
{\cal S}_{\widetilde{(1i)}j;12}  
 + {\cal S}_{1j;12} \right)\nonumber \\
A_3^0(1_q,i_{g},2_{\bar q})\,
 A_3^0(\widetilde{(1i)}_q,j_g,\widetilde{(2i)}_{\bar q})
\JET_{3}^{(3)}(\widetilde{p_{1i}},p_j,\widetilde{p_{2i}}) \Bigg\}
\end{eqnarray}
while for the $N^0$ term the integration of the 5-parton contribution over the
antenna phase space yields
\begin{eqnarray}
\lefteqn{\int \d \Phi_{X_{ijk}}{\rm d}\sigma_{NNLO,N^0}^A= 
 {N_{{4}}}\,{N^0}\, \left(\frac{\alpha_s}{2\pi}\right)\,
 {\rm d}\Phi_{4}(p_{1},p_2,p_3,p_{4};q)} \nonumber \\ 
&&
\frac{1}{2}  \sum_{(i,j) \in (3,4)} 
\left( {\cal S}_{\widetilde{(1i)}j;12} + 
 {\cal S}_{\widetilde{(2i)}j;12} -  {\cal S}_{\widetilde{(1i)}\widetilde{(2i)};12}
-  {\cal S}_{1j;12} -  {\cal S}_{2j;12} +  {\cal S}_{12;12}
\right)
\end{eqnarray}

\section{Parton-level contributions to $e^+e^- \to 3$ jets up to NNLO}
\label{sec:me}
Three-jet production at tree-level is induced by the decay of a virtual
photon (or other neutral gauge boson) into a quark-antiquark-gluon final
state. At higher orders, this process receives corrections from extra
real or virtual particles. The individual partonic channels
that contribute through to NNLO
are shown in Table~\ref{table:partons}.  
\begin{table}[t]
\begin{center}
\begin{tabular}{lll}
\hline\\
LO & $\gamma^*\to q\,\bar qg$ & tree level \\[2mm]
NLO & $\gamma^*\to q\,\bar qg$ & one loop \\
 & $\gamma^*\to q\,\bar q\, gg$ & tree level \\
 & $\gamma^*\to q\,\bar q\, q\bar q$ & tree level \\[2mm]
NNLO & $\gamma^*\to q\,\bar qg$ & two loop \\
 & $\gamma^*\to q\,\bar q\, gg$ & one loop \\
& $\gamma^*\to q\,\bar q\, q\,\bar q$ & one loop \\
& $\gamma^*\to q\,\bar q\, q\,\bar q\, g$ & tree level \\
& $\gamma^*\to q\,\bar q\, g\,g\,g$ & tree level\\
& $\gamma^*\to ggg$ & (one loop)$^2$ 
\\[2mm]
\hline
\end{tabular}
\end{center}
\label{table:partons}
\caption{The partonic channels contributing to $e^+e^- \to 3$~jets.}
\end{table} 

According to the structure of the coupling to the external vector boson, one 
distinguishes non-singlet and singlet contributions. The
non-singlet contributions arise from the interference of amplitudes where 
the external gauge boson couples to the same quark lines, while the 
pure singlet contribution is due to the interference of amplitudes 
where 
the external gauge boson couples to different quark lines.
Up to NLO,
only non-singlet contributions appear. It is only at NNLO, that the first 
non-vanishing singlet terms are allowed. These appear in
the tree-level $\gamma^*\to q\,\bar q\, q\,\bar q\, g$ process, the 
one-loop  $\gamma^*\to q\,\bar q\, gg$ and 
$\gamma^*\to q\,\bar q\, q\,\bar q$ processes and the two-loop 
$\gamma^*\to q\,\bar qg$ process. All these processes yield both 
non-singlet and singlet contributions.
The $\gamma^*\to ggg$ process, which is 
mediated by a closed quark loop, is entirely a singlet contribution. 
In four-jet observables at 
 ${\cal O}(\alpha_s^3)$, the singlet contributions were found to be 
extremely small~\cite{dixonsigner}.
Also, the singlet contribution from three-gluon final states 
to three-jet observables was found to be  negligible~\cite{nigeljochum}. 

Matrix elements and subtraction terms at NLO and NNLO can be naturally 
decomposed according to their colour structure. The cross section at NLO 
receives contributions from three different colour factors:
\begin{equation}
\d \sigma_{NLO} =
 \d \sigma_{NLO,N}
 + \d \sigma_{NLO,1/N} +   \d \sigma_{NLO,N_F} \; .
\end{equation}

The NNLO contribution to the cross section 
receives contributions from seven different colour 
factors:
\begin{eqnarray}
\d \sigma_{NNLO} &=&
 \d \sigma_{NNLO,N^2}
 + \d \sigma_{NNLO,N^0} 
+ \d \sigma_{NNLO,1/N^2}\nonumber
\\ && 
+   \d \sigma_{NNLO,N_F\,N} 
+ \d \sigma_{NNLO,N_F/N}
+ \d \sigma_{NNLO,N_F^2} 
+ \d \sigma_{NNLO,N_{F,\gamma}}\;.
\end{eqnarray}
The first six terms in this equation are non-singlet contributions, 
the last term is the numerically unimportant 
singlet contribution.

In the following, we list the 
matrix elements for the contributing partonic channels shown in 
Table~\ref{table:partons} and discuss their structure. 

\subsection{Tree-level matrix elements for up to five partons}

The tree-level amplitude $M^0_{q\bar q(n-2)g}$ 
for a virtual photon to produce a quark-antiquark pair
and $(n-2)$-gluons, 
$$\gamma^*(q) \to q(p_1) \bar q(p_2)  g(p_3)\ldots
g(p_n) $$
can be expressed as sum over the permutations of the colour ordered amplitude
$\MA{0}{n}$ of the possible 
orderings for the gluon colour indices
\begin{equation}
\label{eq:qqnm2g}
M^0_{q\bar q(n-2)g} 
= i e (\sqrt{2}g)^{n-2} \sum_{(i,\ldots,k) \in P(3,\ldots, n)} \left(
T^{a_i}\cdots T^{a_n}\right)_{i_1i_2} \MA{0}{n} (p_1,p_3,\ldots,p_n,p_2)\;.
\end{equation}

The 
squared matrix elements for $n=3,\ldots,5$, summed over gluon polarisations, 
but excluding symmetry factors for identical particles,
are given by,
\begin{eqnarray}
\left|M^0_{q\bar qg}\right|^2
&=& N_3 \, A_3^0(1_q,3_g,2_{\bar q}) \, ,\\
\left|M^0_{q\bar qgg}\right|^2
&=& N_4 \,  
\left[
\sum_{(i,j) \in P(3,4)}
N A_{4}^0 (1_q,i_g,j_g,2_{\bar q})  
-\frac{1}{N}
 \tilde A_{4}^0 (1_q,3_g,4_g,2_{\bar q})  \right]\, 
,\\
\left|M^0_{q\bar qggg}\right|^2
&=& N_5 \,  
\Bigg[
\sum_{(i,j,k) \in P(3,4,5)}\left(
N^2 A_{5}^0 (1_q,i_g,j_g,k_g,2_{\bar q}) 
-
\tilde A_{5}^0 (1_q,i_g,j_g,k_g,2_{\bar q}) \right)
\nonumber \\
&& \hspace{3cm}+\left(\frac{N^2+1}{N^2}\right)
\bar A_{5}^0 (1_q,3_g,4_g,5_g,2_{\bar q}) \Bigg]\, ,
\end{eqnarray}
where,
\begin{equation}
N_n = 4 \pi  \alpha\, 
\sum_q e_q^2 \left(g^2\right)^{(n-2)}
\left(N^2-1\right)\,\left|{\cal M}^0_{q\bar q}\right|^2,
\end{equation}
and
\begin{equation}
\left|{\cal M}^0_{q\bar q}\right|^2 = 4 (1-\epsilon) q^2.
\end{equation}
The squared colour-ordered matrix elements $A^0_3$, $A^0_4$ and $\tilde A^0_4$
are given in~\cite{ourant}. 
For the five parton case~\cite{fivep},
\begin{eqnarray}
 A_5^{0}(1_q,i_g,j_g,k_g,2_{\bar q}) \left|{\cal M}^0_{q\bar q}\right|^2&=&
\bigg|\MA{0}{5}(p_1,p_i,p_j,p_k,p_2)\bigg|^2\\
\tilde A_5^{0}(1_q,i_g,j_g,k_g,2_{\bar q}) \left|{\cal M}^0_{q\bar q}\right|^2
&=&
\nonumber \\
&&
\hspace{-4.4cm}\bigg| \MA{0}{5}(p_1,p_i,p_j,p_k,p_2)
+\MA{0}{5}(p_1,p_i,p_k,p_j,p_2)
+\MA{0}{5}(p_1,p_k,p_i,p_j,p_2)
\bigg|^2,
\\ 
\bar A_{5}^0(1_q,i_g,j_g,k_g,2_{\bar q}) \left|{\cal M}^0_{q\bar q}\right|^2 
&=& 
\left|\sum_{(i,j,k) \in P(3,\ldots, 5)}
\MA{0}{5}(p_1,p_i,p_j,p_k,p_2)\right|^2\, .
\end{eqnarray}
In the subleading colour contribution $\tilde{A}_5^0$, gluon $k$ is 
effectively photon-like, while in the sub-subleading colour contribution
(also called Abelian contribution),
$\bar{A}_5^0$, all three gluons are effectively photon-like.   
Photon-like gluons do not couple to 
three- and four-gluon vertices, and there are no simple collinear 
limits as any two photon-like gluons become
collinear.  As a consequence, the only colour-connected pair 
in $\bar{A}_5^0$ are the quark and
antiquark.  


The tree-level amplitude for 
$$\gamma^*(q) \to q(p_1) \bar q(p_2) q' (p_3) \bar q'(p_4)$$
is given by
\begin{eqnarray}
M_{q\bar q q'\bar q'}^0
&=& i e_1 g^2 
\delta_{q_1q_2}\delta_{q_3q_4}
\left(\delta_{i_1i_4}\delta_{i_3i_2}-\frac{1}{N}
\delta_{i_1i_2}\delta_{i_3i_4}\right) 
\MB{0}{4} (p_1,p_2,p_3,p_4) \nonumber \\
&& \hspace{1.6cm}
+ (1\leftrightarrow 3,2\leftrightarrow 4)\;,
\end{eqnarray}
where $\delta_{q_1q_2}\delta_{q_3q_4}$ indicates the quark
flavours.
The amplitude $\MB{0}{4}(p_1,p_2,p_3,p_4)$ thus denotes the contribution 
from the $q_1 \bar q_2$--pair coupling to the vector boson.  
The identical quark amplitude is obtained
\begin{equation}
M_{q\bar q q\bar q}^0 =
M_{q\bar q q'\bar q'}^0
 - (2\leftrightarrow 4).
\end{equation}

The resulting four-quark 
squared matrix elements, summed over final state quark 
flavours and including symmetry factors are given by 
\begin{eqnarray}
\left|M_{4q}^0\right|^2 
&=&\sum_{q,q'}
\left|M_{q\bar q q'\bar q'}\right|^2 
+\sum_{q}
\left|M_{q\bar q q\bar q}\right|^2 \nonumber \\
&=& N_4\, \Bigg [  
N_F B_{4}^0 (1_q,3_q,4_{\bar q},2_{\bar q}) 
- \frac{1}{N} \, \left(
C_{4}^0 (1_q,3_q,4_{\bar q},2_{\bar q})
+C_{4}^0 (2_{\bar q},4_{\bar q},3_q,1_q) \right) \nonumber \\
&&+ N_{F,\gamma} \,\hat B_{4}^0 (1_q,3_q,4_{\bar q},2_{\bar q})
\Bigg ]\, ,
\end{eqnarray}
where 
\begin{eqnarray}
B_{4}^0 (1_q,3_q,4_{\bar q},2_{\bar q})\left|{\cal M}^0_{q\bar q}\right|^2
&=& \left| 
 \MB{0}{4}(p_1,p_2,p_3,p_4)\right|^2,\nonumber \\
C_{4}^0 (1_q,3_q,4_{\bar q},2_{\bar q})\left|{\cal M}^0_{q\bar q}\right|^2
&=& - \Re \bigg(
 \MB{0}{4}(p_1,p_2,p_3,p_4)\MB{0,\dagger}{4}(p_1,p_4,p_3,p_2)
\bigg)\, , \\
\hat B_{4}^0 (1_q,3_q,4_{\bar q},2_{\bar q})\left|{\cal M}^0_{q\bar q}\right|^2
&=& \mbox{Re} \bigg(
 \MB{0}{4}(p_1,p_2,p_3,p_4)\MB{0,\dagger}{4}(p_3,p_4,p_1,p_2)
\bigg).
\end{eqnarray}
Explicit expressions for 
$B_4^0$ and $C_4^0$ are given in~\cite{ourant}.
The last term, $\hat B_{4}^0$, is proportional to the 
charge weighted sum of the quark flavours,
$N_{F,\gamma}$, which for electromagnetic interactions is given by,
\begin{equation}
N_{F,\gamma} = \frac{(\sum_q e_q)^2}{\sum_q e_q^2} .
\end{equation}
It is relevant only for observables where the final state quark 
charge can be determined.

There are four colour structures in the 
tree-level amplitude for $$\gamma^*(q) \to q(p_1) \bar q(p_2) 
q' (p_3) \bar q'(p_4) g(p_5)$$
which reads 
\begin{eqnarray}
M_{q\bar q q'\bar q' g}
&=& i e_1  g^3 \sqrt{2}  
\delta_{q_1q_2} \delta_{q_3q_4} 
\nonumber \\
&\times&
\bigg[
T^{a_5}_{i_1i_4} \delta_{i_3i_2}  
\MB{0,a}{5} (p_1,p_2,p_3,p_4,p_5)
- \frac{1}{N} T^{a_5}_{i_1i_2} \delta_{i_3i_4} 
\MB{0,c}{5} (p_1,p_2,p_3,p_4,p_5)
  \nonumber \\
&& 
+ T^{a_5}_{i_3i_2} \delta_{i_1i_4}  
\MB{0,b}{5} (p_1,p_2,p_3,p_4,p_5)
- \frac{1}{N} T^{a_5}_{i_3i_4} \delta_{i_1i_2} 
\MB{0,d}{5} (p_1,p_2,p_3,p_4,p_5)\bigg]
  \nonumber \\
&& \hspace{1.6cm}
+ (1\leftrightarrow 3,2\leftrightarrow 4) \;.
\end{eqnarray}
The amplitude $\MB{0,x}{5}(p_1,p_2,p_3,p_4,p_5)$ for $x=a,\ldots,d$
denotes the contribution 
from the $q_1 \bar q_2$--pair coupling to the vector boson.  
Due to the colour decomposition,
the following relation holds between the leading and subleading colour 
amplitudes:
\begin{eqnarray}
\MB{0,e}{5}(p_1,p_2,p_3,p_4,p_5)&=&
\MB{0,a}{5} (p_1,p_2,p_3,p_4,p_5) + 
\MB{0,b}{5} (p_1,p_2,p_3,p_4,p_5) \nonumber \\
&=& 
\MB{0,c}{5} (p_1,p_2,p_3,p_4,p_5) +
\MB{0,d}{5} (p_1,p_2,p_3,p_4,p_5) \;.
\end{eqnarray}
As before, the identical quark matrix element is obtained by permuting the
antiquark momenta,
\begin{equation}
M_{q\bar q q\bar qg}^0 =
M_{q\bar q q'\bar q'g}^0
 - (2\leftrightarrow 4).
\end{equation}

The squared matrix element, summed over flavours and including 
symmetry factors is given by,
\begin{eqnarray}
\lefteqn{\left|M_{4qg}^0\right|^2 
=
\sum_{q,q'}
\left|M_{q\bar q q'\bar q' g}\right|^2 
+\sum_q \left|M_{q\bar q q\bar q g}\right|^2} \nonumber \\
&=&
N_5\, \Bigg [
N\NF \left(
B_5^{0,a}(1_q,5_g,4_{\bar q'};3_{q'},2_{\bar q})
+B_5^{0,b}(1_q,4_{\bar q'};3_{q'},5_g,2_{\bar q})
\right)\nonumber\\
&& + \frac{\NF}{N}
\left(
 B_5^{0,c}(1_q,5_g,2_{\bar q};3_{q'},4_{\bar q'})
+B_5^{0,d}(1_q,2_{\bar q};3_{q'},5_g,4_{\bar q'})
-2 B_5^{0,e}(1_q,2_{\bar q};3_{q'},4_{\bar q'};5_g)
\right)\nonumber \\
&&-C_5^{0}(1_q,3_{q},4_{\bar q},5_g,2_{\bar q})
+\left(\frac{N^2+1}{N^2}\right) \left(
\tilde C_5^{0}(1_q,3_{q},4_{\bar q},5_g,2_{\bar q}) 
+ \tilde C_5^{0}(2_{\bar q},4_{\bar q},3_{q},5_g,1_q)\right)
\nonumber \\
&&- N N_{F,\gamma}\left(
 \hat B_5^{0,a}(1_q,5_g,4_{\bar q'};3_{q'},2_{\bar q})
+\hat B_5^{0,b}(1_q,4_{\bar q'};3_{q'},5_g,2_{\bar q})
-\hat B_5^{0,e}(1_q,4_{\bar q'};3_{q'},2_{\bar q},5_g)
\right)\nonumber \\
&&+\frac{N_{F,\gamma}}{N}
\left(
 \hat B_5^{0,c}(1_q,5_g,2_{\bar q};3_{q'},4_{\bar q'})
+\hat B_5^{0,d}(1_q,2_{\bar q};3_{q'},5_g,4_{\bar q'})
+\hat B_5^{0,e}(1_q,2_{\bar q};3_{q'},4_{\bar q'};5_g)
\right)\Bigg ],
\label{eq:4q1g}
\nonumber \\
\end{eqnarray}
where for $x=a,\ldots,e$
\begin{eqnarray}
B_5^{0,x}(\ldots) \left|{\cal M}^0_{q\bar q}\right|^2 &=& |\MB{0,x}{5}(p_1,p_2,p_3,p_4,p_5)|^2,\\
\hat B_5^{0,x}(\ldots) \left|{\cal M}^0_{q\bar q}\right|^2 &=& \mbox{Re} \left(\MB{0,x}{5}(p_1,p_2,p_3,p_4,p_5)
\MB{0,x,\dagger}{5}(p_3,p_4,p_1,p_2,p_5)\right),
\end{eqnarray}
and
\begin{eqnarray}
C_5^{0}(1_q,3_{q},4_{\bar q},5_g,2_{\bar q}) \left|M^0_{q\bar q}\right|^2 &=& 
-2\mbox{Re} \bigg(
 \MB{0,a}{5}(p_1,p_2,p_3,p_4,p_5)\MB{0,c,\dagger}{5}(p_1,p_4,p_3,p_2,p_5)
\nonumber \\&&
\phantom{2\mbox{Re}}+\MB{0,b}{5}(p_1,p_2,p_3,p_4,p_5)\MB{0,d,\dagger}{5}(p_1,p_4,p_3,p_2,p_5)
\nonumber \\&&
\phantom{2\mbox{Re}}+\MB{0,a}{5}(p_1,p_2,p_3,p_4,p_5)\MB{0,d,\dagger}{5}(p_3,p_2,p_1,p_4,p_5)
\nonumber \\&&
\phantom{2\mbox{Re}}+\MB{0,b}{5}(p_1,p_2,p_3,p_4,p_5)\MB{0,c,\dagger}{5}(p_3,p_2,p_1,p_4,p_5)
\bigg),\nonumber \\
&&\\
\tilde{C}_5^{0}(1_q,3_{q},4_{\bar q},5_g,2_{\bar q})
\left|M^0_{q\bar q}\right|^2
&=& - \mbox{Re} \bigg(
 \MB{0,e}{5}(p_1,p_2,p_3,p_4,p_5)\MB{0,e,\dagger}{5}(p_1,p_4,p_3,p_2,p_5)
\bigg).\nonumber \\
\end{eqnarray}

\subsection{One-loop matrix elements for up to four partons}

The renormalised one-loop  amplitude $M^1_{q\bar qg}$ 
for a virtual photon to produce a quark-antiquark pair
 together with a single gluon, 
$$\gamma^*(q) \to q(p_1) \bar q(p_2)  g(p_3)$$
contains a single colour structure such that
\begin{equation}
\label{eq:ggponeloop}
M^1_{q\bar qg} 
= i e \sqrt{2} g \left(\frac{g^2}{16\pi^2}\right)
T^{a_3}_{i_1i_2} \MA{1}{3} (p_1,p_3,p_2)\;.
\end{equation}
Unless stated otherwise, the renormalisation scale is set to 
$\mu^2 = q^2$. 

The interference of the one-loop amplitude with the three-parton
tree-level amplitude (\ref{eq:qqnm2g}) is given by
\begin{eqnarray}
2\Re\left(M^{0,\dagger}_{q\bar qg}M^1_{q\bar qg} \right)
&=& N_3 \left(\frac{\alpha_s}{2\pi}\right)
A_3^{(1\times 0)}(1_q,3_g,2_{\bar q})\,,
\end{eqnarray}
where
\begin{eqnarray}
\label{eq:M31}
A_3^{(1\times 0)}(1_q,3_g,2_{\bar q})
&=&
\, \bigg( N \left[
A_3^1(1_q,3_g,2_{\bar q})
+{\cal A}_2^1(s_{123}) A_3^0(1_q,3_g,2_{\bar q})\right]\nonumber \\
&&
-\frac{1}{N}  \left[
\tilde
A_3^1(1_q,3_g,2_{\bar q})
+{\cal A}_2^1(s_{123}) A_3^0(1_q,3_g,2_{\bar q})\right]
+N_F \hat A_3^1(1_q,3_g,2_{\bar q})\bigg)\, ,\nonumber \\
\end{eqnarray}
where $A^1_3$, $\tilde A^1_3$ and $\hat A^1_3$ 
are given up to ${\cal O}(\epsilon^0)$ in~\cite{ourant}.

Moreover, the one-loop process $\gamma^*\to ggg$ also yields three-jet final 
states. Since this process has no tree-level counterpart, 
it does only contribute at NNLO. Its amplitude 
can be denoted as~\cite{nigeljochum} 
\begin{equation}
\label{eq:gggoneloop}
M^1_{ggg} 
= i \sum_q e_q \sqrt{2} g \left(\frac{g^2}{16\pi^2}\right)
d^{a_1a_2a_3} \MC{1}{3} (p_1,p_2,p_3)\;.
\end{equation}

The one-loop corrections to $\gamma^* \to 4$~partons
 have been available for
some time~\cite{onel-4}.   
The one-loop amplitude for 
$$\gamma^*(q) \to q(p_1) \bar q(p_2)  g(p_3) 
g(p_4)$$ contains two colour structures,
\begin{eqnarray}
M_{q\bar q gg}^1 
&=& i e 2 g^2 \left(\frac{g^2}{16\pi^2}\right)\nonumber \\
&\times&
\Bigg[
\sum_{(i,j) \in P(3,4)} 
\left(T^{a_i}T^{a_j}\right)_{i_1i_2} 
\bigg(N\MA{1,a}{4} (p_1,p_i,p_j,p_2)
-\frac{1}{N}\MA{1,b}{4} (p_1,p_i,p_j,p_2)\nonumber \\
&&\hspace{4cm}
+N_F\MA{1,c}{4} (p_1,p_i,p_j,p_2)
+ \frac{\sum_q e_q}{e}  
 \MA{1,e}{4}(p_1,p_i,p_j,p_2) \bigg)\nonumber \\
&&\hspace{4cm}
+\frac{1}{2}\delta^{a_ia_j}\delta_{i_1i_2}
\MA{1,d}{4} (p_1,p_3,p_4,p_2)
\Bigg]\;,
\end{eqnarray}
where
\begin{eqnarray}
\MA{1,d}{4} (p_1,p_3,p_4,p_2)&=&
\MA{1,d}{4} (p_1,p_4,p_3,p_2).
\end{eqnarray}

The ``squared" matrix element is the interference between the tree-level and
one-loop amplitudes,
\begin{eqnarray}
\lefteqn{2 \left | M^{0,\dagger}_{q\bar q gg} M^1_{q\bar q gg}\right|
=N_4 \left(\frac{\alpha_s}{2\pi}\right)\,} \nonumber
\\
&\times&  
\Bigg[
\sum_{(i,j) \in P(3,4)} 
\bigg(
N^2 
A_{4}^{1,a} (1_q,i_g,j_g,2_{\bar q})
-
A_{4}^{1,b} (1_q,i_g,j_g,2_{\bar q})
+NN_F A_{4}^{1,c} (1_q,i_g,j_g,2_{\bar q})\nonumber \\
&&\hspace{2.2cm}
+NN_{F,\gamma}\, A_{4}^{1,e} (1_q,i_g,j_g,2_{\bar q})
\bigg)\nonumber \\
&&-
\Bigg(
\tilde A_{4}^{1,a} (1_q,3_g,4_g,2_{\bar q})
-\tilde A_{4}^{1,d} (1_q,3_g,4_g,2_{\bar q})
-\frac{1}{N^2}
\tilde A_{4}^{1,b} (1_q,3_g,4_g,2_{\bar q}) \nonumber \\ && \hspace{2.2cm}
+\frac{N_F}{N}
\tilde A_{4}^{1,c} (1_q,3_g,4_g,2_{\bar q})
+\frac{N_{F,\gamma}}{N}
\tilde A_{4}^{1,e} (1_q,3_g,4_g,2_{\bar q})
\Bigg) \Bigg]\,,
\end{eqnarray}
where for  $x=a,\ldots,d$,
\begin{eqnarray}
A_{4}^{1,x} (1_q,i_g,j_g,2_{\bar q}) \left|{\cal M}^0_{q\bar q}\right|^2&=& 
\Re \left(\MA{1,x}{4} (p_1,p_i,p_j,p_2) 
\MA{0,\dagger}{4} (p_1,p_i,p_j,p_2)\right),\\
\tilde A_{4}^{1,x} (1_q,3_g,4_g,2_{\bar q}) \left|{\cal M}^0_{q\bar q}\right|^2&=& 
\Re \left(\tilde\MA{1,x}{4} (p_1,p_3,p_4,p_2) 
\tilde\MA{0,\dagger}{4} (p_1,p_3,p_4,p_2)\right),
\end{eqnarray}
and
\begin{eqnarray}
\tilde\MA{1,x}{4} (p_1,p_3,p_4,p_2)&=&
\MA{1,x}{4} (p_1,p_3,p_4,p_2)+\MA{1,x}{4} (p_1,p_4,p_3,p_2).
\end{eqnarray}

The renormalised
singularity structure of the various contributions can be easily written in
terms of the tree-level squared matrix elements multiplied
by combinations of infrared singularity operators~\cite{catani}, for 
which we use the notation defined in~\cite{ourant}.
Explicitly, we find
\begin{eqnarray}
\Poles(A_{4}^{1,a}(1_q,i_g,j_g,2_{\bar q}))&=&
2\left(
 {\bf I}^{(1)}_{qg} (\e,s_{1i})
+{\bf I}^{(1)}_{gg} (\e,s_{ij})
+{\bf I}^{(1)}_{g\bar q} (\e,s_{j2})\right)A_{4}^0 (1_q,i_g,j_g,2_{\bar q})
,\nonumber  \\ \\
\Poles( A_{4}^{1,b}(1_q,i_g,j_g,2_{\bar q}))&=&
2{\bf I}^{(1)}_{q\bar q} (\e,s_{12})
A_{4}^0 (1_q,i_g,j_g,2_{\bar q}),\\
\Poles(A_{4}^{1,c}(1_q,i_g,j_g,2_{\bar q}))&=&2\left(
 {\bf I}^{(1)}_{qg,F} (\e,s_{1i})
+{\bf I}^{(1)}_{gg,F} (\e,s_{ij})
+{\bf I}^{(1)}_{g\bar q,F} (\e,s_{j2})\right)A_{4}^0 (1_q,i_g,j_g,2_{\bar q})
, \nonumber \\ \\
\Poles(\tilde A_{4}^{1,a}(1_q,3_g,4_g,2_{\bar q}))&=&\nonumber \\
&&\hspace{-4cm}
\left(
2{\bf I}^{(1)}_{gg} (\e,s_{34})
+{\bf I}^{(1)}_{qg} (\e,s_{14})
+{\bf I}^{(1)}_{g\bar q} (\e,s_{23})
+{\bf I}^{(1)}_{qg} (\e,s_{13})
+{\bf I}^{(1)}_{g\bar q} (\e,s_{24})\right)\tilde A_{4}^0 (1_q,3_g,4_g,2_{\bar q}),\nonumber \\ \\
\Poles( \tilde A_{4}^{1,b}(1_q,3_g,4_g,2_{\bar q}))&=&
2 {\bf I}^{(1)}_{q\bar q} (\e,s_{12}) 
\tilde A_{4}^0 (1_q,3_g,4_g,2_{\bar q}),
\label{eq:At41b}
\\
\Poles(\tilde A_{4}^{1,c}(1_q,3_g,4_g,2_{\bar q}))&=&\nonumber \\
&&\hspace{-4cm}\left(
2{\bf I}^{(1)}_{gg,F} (\e,s_{34})
+{\bf I}^{(1)}_{qg,F} (\e,s_{14})
+{\bf I}^{(1)}_{g\bar q,F} (\e,s_{23})
+{\bf I}^{(1)}_{qg,F} (\e,s_{13})
+{\bf I}^{(1)}_{g\bar q,F} (\e,s_{24})\right)
\nonumber \\ && \times \tilde A_{4}^0 (1_q,3_g,4_g,2_{\bar q}), \\
\Poles(\tilde A_{4}^{1,d}(1_q,3_g,4_g,2_{\bar q}))&=&\nonumber \\
&&\hspace{-4cm}
\left(
 2{\bf I}^{(1)}_{q\bar q} (\e,s_{12})
+2{\bf I}^{(1)}_{gg} (\e,s_{34})
-{\bf I}^{(1)}_{qg} (\e,s_{14})
-{\bf I}^{(1)}_{g\bar q} (\e,s_{23})
-{\bf I}^{(1)}_{qg} (\e,s_{13})
-{\bf I}^{(1)}_{g\bar q} (\e,s_{24})\right)\nonumber \\
&&\times
\tilde A_{4}^0 (1_q,3_g,4_g,2_{\bar q}).
\end{eqnarray}

As at tree-level, the one-loop amplitude for 
$$\gamma^*(q) \to q(p_1) \bar q(p_2)  q'(p_3) 
\bar q'(p_4)$$ contains two colour structures,
\begin{eqnarray}
\lefteqn{
M_{q\bar q q'\bar q'}^1
= i e_1 g^2 
\left(\frac{g^2}{16\pi^2}\right)
\delta_{q_1q_2}\delta_{q_3q_4}}\nonumber \\
&\times&
\Bigg[
\delta_{i_1i_4}\delta_{i_3i_2}
\left(
N \MB{1,a}{4}(p_1,p_2,p_3,p_4)  
- \frac{1}{N}\MB{1,b}{4}(p_1,p_2,p_3,p_4) 
+N_F \MB{1,c}{4}(p_1,p_2,p_3,p_4) 
\right)\nonumber \\
&&
-\frac{1}{N}
\delta_{i_1i_2}\delta_{i_3i_4}
\Bigg(N \MB{1,d}{4}(p_1,p_2,p_3,p_4)  
- \frac{1}{N}\MB{1,e}{4}(p_1,p_2,p_3,p_4) \nonumber \\ &&
\hspace{2.2cm} +N_F \MB{1,f}{4}(p_1,p_2,p_3,p_4)  
\Bigg) \Bigg]
+ (1\leftrightarrow 3,2\leftrightarrow 4)  \;,
\end{eqnarray}
where
\begin{equation}
\MB{1,a}{4}(p_1,p_2,p_3,p_4) + \MB{1,e}{4}(p_1,p_2,p_3,p_4) 
=
\MB{1,b}{4}(p_1,p_2,p_3,p_4) + \MB{1,d}{4}(p_1,p_2,p_3,p_4). 
\end{equation}
As before, the identical quark matrix element is obtained by permuting the antiquark momenta,
\begin{equation}
M_{q\bar q q\bar q}^1 =
M_{q\bar q q'\bar q'}^1
 - (2\leftrightarrow 4).
\end{equation}

Summing over flavours and including 
symmetry factors, we find that the ``squared" matrix element, is given by
\begin{eqnarray}
\lefteqn{2 \left | M_{4q}^{0,\dagger} M_{4q}^{1}\right|=
\sum_{q,q'}
2 \left | M_{q\bar q q'\bar q'}^{0,\dagger} M_{q\bar q q'\bar q'}^{1}\right|
+
\sum_q 2 \left | M_{q\bar q q\bar q}^{0,\dagger} M_{q\bar q q\bar q}^{1}\right|
}  \nonumber \\
&=&N_4 \left(\frac{\alpha_s}{2\pi}\right)\,
\Bigg[
N \NF B_4^{1,a}(1_q,3_q,4_{\bar q},2_{\bar q})
- \frac{\NF}{N} B_4^{1,b}(1_q,3_q,4_{\bar q},2_{\bar q})
+ N_F^2 B_4^{1,c}(1_q,3_q,4_{\bar q},2_{\bar q})
\nonumber \\
&&\hspace{0cm}-C_4^{1,d}(1_q,3_q,4_{\bar q},2_{\bar q})
+\frac{1}{N^2}C_4^{1,e}(1_q,3_q,4_{\bar q},2_{\bar q})
-\frac{N_F}{N}C_4^{1,f}(1_q,3_q,4_{\bar q},2_{\bar q})
\nonumber \\
&&\hspace{0cm}-C_4^{1,d}(2_{\bar q},4_{\bar q},3_q,1_q)
+\frac{1}{N^2}C_4^{1,e}(2_{\bar q},4_{\bar q},3_q,1_q)
-\frac{N_F}{N}C_4^{1,f}(2_{\bar q},4_{\bar q},3_q,1_q)
\nonumber \\
&&\hspace{0cm}
+N N_{F,\gamma} \hat B_4^{1,a}(1_q,3_q,4_{\bar q},2_{\bar q})
- \frac{N_{F,\gamma}}{N} \hat B_4^{1,b}(1_q,3_q,4_{\bar q},2_{\bar q})
+ N_F N_{F,\gamma}  \hat B_4^{1,c}(1_q,3_q,4_{\bar q},2_{\bar q})
\Bigg ],  \nonumber \\
\label{eq:4q1lall}
\end{eqnarray}
where for $x=a,b,c$
\begin{eqnarray}
B_4^{1,x}(1_q,3_q,4_{\bar q},2_{\bar q})\left|{\cal M}^0_{q\bar q}\right|^2 &=&
\Re \left( \MB{1,x}{4}(p_1,p_2,p_3,p_4)\MB{0,\dagger}{4}(p_1,p_2,p_3,p_4)\right), \\
\hat B_4^{1,x}(1_q,3_q,4_{\bar q},2_{\bar q})\left|{\cal M}^0_{q\bar q}\right|^2 &=&
\Re \left( \MB{1,x}{4}(p_1,p_2,p_3,p_4)\MB{0,\dagger}{4}(p_3,p_4,p_1,p_2)\right),  
\end{eqnarray}
and for $x=d,e,f$
\begin{eqnarray}
C_4^{1,x}(1_q,3_q,4_{\bar q},2_{\bar q})\left|{\cal M}^0_{q\bar q}\right|^2 &=&
- \Re \left( \MB{1,x}{4}(p_1,p_2,p_3,p_4)\MB{0,\dagger}{4}(p_1,p_4,p_3,p_2)\right).  
\end{eqnarray}

Using the infrared singularity operators of~\cite{catani},
 we can extract the singular contributions of the 
renormalised one-loop contribution as,
\begin{eqnarray}
\Poles(B_{4}^{1,a}(1_q,3_q,4_{\bar q},2_{\bar q}))&=&
2\left( 
 {\bf I}^{(1)}_{q\bar q} (\e,s_{14})
+{\bf I}^{(1)}_{q\bar q} (\e,s_{23})\right)
B_{4}^0 (1_q,3_q,4_{\bar q},2_{\bar q}),\\
\Poles(B_{4}^{1,b}(1_q,3_q,4_{\bar q},2_{\bar q}))&=&\nonumber \\
&&\hspace{-4cm}
2\left( 
 2{\bf I}^{(1)}_{q\bar q} (\e,s_{14})
-2{\bf I}^{(1)}_{q\bar q} (\e,s_{13})
+2{\bf I}^{(1)}_{q\bar q} (\e,s_{23})
-2{\bf I}^{(1)}_{q\bar q} (\e,s_{24})
+{\bf I}^{(1)}_{q\bar q} (\e,s_{12})
+{\bf I}^{(1)}_{q\bar q} (\e,s_{34})
\right)\nonumber \\
&&\hspace{2cm}\times
B_{4}^0 (1_q,3_q,4_{\bar q},2_{\bar q}),\\
\Poles(C_{4}^{1,d}(1_q,3_q,4_{\bar q},2_{\bar q}))&=&
 2\left( 
 {\bf I}^{(1)}_{q\bar q} (\e,s_{13})
+{\bf I}^{(1)}_{q\bar q} (\e,s_{24})\right)
C_{4}^0 (1_q,3_q,4_{\bar q},2_{\bar q}),\\
\Poles(C_{4}^{1,e}(1_q,3_q,4_{\bar q},2_{\bar q}))&=&\nonumber \\
&&\hspace{-4cm}
2\left( 
{\bf I}^{(1)}_{q\bar q} (\e,s_{12})
+{\bf I}^{(1)}_{q\bar q} (\e,s_{14})
+{\bf I}^{(1)}_{q\bar q} (\e,s_{23})
+{\bf I}^{(1)}_{q\bar q} (\e,s_{34})
-{\bf I}^{(1)}_{q\bar q} (\e,s_{13})
-{\bf I}^{(1)}_{q\bar q} (\e,s_{24})
\right)\nonumber \\
&&\hspace{2cm}\times
C_{4}^0 (1_q,3_q,4_{\bar q},2_{\bar q}),
\label{eq:C41e} \\
\Poles(\hat B_{4}^{1,a}(1_q,3_q,4_{\bar q},2_{\bar q}))&=&
2\left( 
 {\bf I}^{(1)}_{q\bar q} (\e,s_{14})
+{\bf I}^{(1)}_{q\bar q} (\e,s_{23})\right)
\hat B_{4}^0 (1_q,3_q,4_{\bar q},2_{\bar q}),\\
\Poles(\hat B_{4}^{1,b}(1_q,3_q,4_{\bar q},2_{\bar q}))&=&\nonumber \\
&&\hspace{-4cm}
2\left( 
 2{\bf I}^{(1)}_{q\bar q} (\e,s_{14})
-2{\bf I}^{(1)}_{q\bar q} (\e,s_{13})
+2{\bf I}^{(1)}_{q\bar q} (\e,s_{23})
-2{\bf I}^{(1)}_{q\bar q} (\e,s_{24})
+{\bf I}^{(1)}_{q\bar q} (\e,s_{12})
+{\bf I}^{(1)}_{q\bar q} (\e,s_{34})
\right)\nonumber \\
&&\hspace{2cm}\times
\hat B_{4}^0 (1_q,3_q,4_{\bar q},2_{\bar q})\,.
\end{eqnarray}

\subsection{Two-loop matrix elements for three partons}
\label{sec:3jme}

The renormalised two-loop  amplitude $M^2_{q\bar qg}$ 
for a virtual photon to produce a quark-antiquark pair
 together with a single gluon, 
$$\gamma^*(q) \to q(p_1) \bar q(p_2)  g(p_3)$$
contains a single colour structure such that
\begin{equation}
M^2_{q\bar qg} 
= i e \sqrt{2}g \left(\frac{g^2}{16\pi^2}\right)^2
T^{a_3}_{i_1i_2} \MA{2}{3} (p_1,p_3,p_2)\;.
\end{equation}

At NNLO, there are two contributions.   One from the interference of the
two-loop and tree-level amplitudes (\ref{eq:qqnm2g}), the other from the square of the one-loop
amplitudes given in (\ref{eq:ggponeloop}).
These terms were computed in~\cite{3jme} by reducing the large number of
two-loop Feynman integrals to a small number of master integrals, using 
integration-by-parts~\cite{chet} and Lorentz-invariance~\cite{gr} 
identities, solved using the Laporta algorithm~\cite{laporta}. The relevant 
master integrals (two-loop four-point functions with one off-shell leg)
were then derived~\cite{3jmi} 
from their differential equations~\cite{kotikov,remiddi,gr}.

The resulting virtual three-parton contributions are given by,
\begin{eqnarray}
2\Re\left(M^{0,\dagger}_{q\bar qg}M^2_{q\bar qg} \right)
&=& N_3 \left(\frac{\alpha_s}{2\pi}\right)^2
\, A_3^{(2\times 0)}(1_q,3_g,2_{\bar q}) \, ,\\
\Re\left(M^{1,\dagger}_{q\bar qg}M^1_{q\bar qg} \right)
&=& N_3 \left(\frac{\alpha_s}{2\pi}\right)^2
\, A_3^{(1\times 1)}(1_q,3_g,2_{\bar q}) \, .
\end{eqnarray}
Following~\cite{catani}, we organise the 
infrared pole structure of the NNLO contributions renormalised in the 
\MSbar\ scheme in terms of the tree and renormalised one-loop amplitudes such
that,
\begin{eqnarray}
\label{eq:twolooppoles}
\lefteqn{\Poles\left(A_3^{(2\times 0)}(1_q,3_g,2_{\bar q})+A_3^{(1\times 1)}(1_q,3_g,2_{\bar q})\right)
}\nonumber \\
 &=&
2\bigg[-\left({\bom I}_{q\bar qg}^{(1)}(\epsilon)\right)^2 
- \frac{\beta_0}{\epsilon}{\bom I}_{q\bar qg}^{(1)}(\epsilon)\nonumber \\
&&
+e^{-\epsilon\gamma}\frac{\Gamma(1-2\epsilon)}{\Gamma(1-\epsilon)}
\left(\frac{\beta_0}{\epsilon} + K\right) {\bom I}_{q\bar qg}^{(1)}(2\epsilon)
+{\bom H}_{q\bar qg}^{(2)}\bigg] A_3^0(1_q,3_g,2_{\bar q}) \nonumber \\
&&
+2\, {\bom I}_{q\bar qg}^{(1)}(\epsilon)   \, 
A_3^{(1 \times 0)}(1_q,3_g,2_{\bar q})\;.
\end{eqnarray}
Here, 
\begin{eqnarray}
{\bom I}_{q\bar qg}^{(1)}(\epsilon) &=& N \left(
{\bom I}^{(1)}_{qg}(\epsilon,s_{13})
+{\bom I}^{(1)}_{qg}(\epsilon,s_{23})
\right)
-\frac{1}{N} {\bom I}^{(1)}_{q\bar q}(\epsilon,s_{12}) \nonumber \\
&& +N_F  
\left(
{\bom I}^{(1)}_{qg,F}(\epsilon,s_{13})
+{\bom I}^{(1)}_{qg,F}(\epsilon,s_{23})
\right),
\end{eqnarray}
with the individual ${\bom I}^{(1)}_{ij}$ defined in~\cite{ourant} and
\begin{eqnarray}
{\bom H}^{(2)}_{q\bar qg} &= & 
\frac{e^{\epsilon \gamma}}{4\,\epsilon\,\Gamma(1-\epsilon)} \Bigg[
\left(4\zeta_3+\frac{589}{432}- \frac{11\pi^2}{72}\right)N^2
+\left(-\frac{1}{2}\zeta_3-\frac{41}{54}-\frac{\pi^2}{48} \right)\nonumber \\
&&
+\left(-3\zeta_3 -\frac{3}{16} + \frac{\pi^2}{4}\right) \frac{1}{N^2}
+\left(-\frac{19}{18}+\frac{\pi^2}{36} \right) N N_F 
+\left(-\frac{1}{54}-\frac{\pi^2}{24}\right) \frac{N_F}{N}+ \frac{5}{27} N_F^2.
\Bigg]\;.\nonumber \\
\label{eq:Htwo}
\end{eqnarray}

We denote the finite contributions as,
\begin{eqnarray}
\label{eq:twoloopfinite}
\Finite{(A_3^{(2\times 0)}(1_q,3_g,2_{\bar q}))} &=& 
N^2 A_{3,N^2}^{(2\times 0),finite}
+A_{3,1}^{(2\times 0),finite}
+ \frac{1}{N^2}A_{3,1/N^2}^{(2\times 0),finite}\nonumber \\
&&+NN_F A_{3,NN_F}^{(2\times 0),finite}
+ \frac{N_F}{N}A_{3,N_F/N}^{(2\times 0),finite}\nonumber \\
&&
+N_F^2 A_{3,N_F^2}^{(2\times 0),finite}
+N_{F,\gamma} \left(\frac{4}{N}-N\right)A_{3,N_{F,\gamma}}^{(2\times 0),finite},\\
\Finite{(A_3^{(1\times 1)}(1_q,3_g,2_{\bar q}))} &=& 
N^2 A_{3,N^2}^{(1\times 1),finite}
+A_{3,1}^{(1\times 1),finite}
+ \frac{1}{N^2}A_{3,1/N^2}^{(1\times 1),finite}\nonumber \\
&&+NN_F A_{3,NN_F}^{(1\times 1),finite}
+ \frac{N_F}{N}A_{3,N_F/N}^{(1\times 1),finite}\nonumber \\
&&
+N_F^2 A_{3,N_F^2}^{(1\times 1),finite}.
\end{eqnarray}
Explicit formulae for the 
finite remainders have been given in~\cite{3jme}. These are expressed in 
terms of one-dimensional and two-dimensional harmonic polylogarithms
(HPLs and 2dHPLs)~\cite{hpl,3jmi}, which are generalisations of the 
well-known Nielsen polylogarithms~\cite{nielsen}.
A numerical implementation, which is 
required for all practical applications, is available for HPLs and 
2dHPLs~\cite{hplnum}.

Finally, a finite NNLO contribution arises from the squared one-loop amplitude 
(\ref{eq:gggoneloop})
for $\gamma^*\to ggg$:
\begin{equation}
C_3^{(1\times 1)}(1_g,2_g,3_g) = N_{F,\gamma} \left( \frac{4}{N} - N \right)
C_{3,N_{F,\gamma}}^{(1\times 1),finite}\;.
\end{equation}

\section{Construction of the NLO subtraction term}
\label{sec:nlo}

Three-jet production at the leading order is given by:
\begin{equation}
{\rm d}\sigma_{LO}^{R}= 
 N_{3} \,{\rm d}\Phi_{3}(p_{1},p_2,p_{3};q) \,
A^{0}_{3}(1_{q},3_{g},2_{\bar{q}})\, \JET_{3}^{(3)}(p_{1},p_2,p_{3})\;. 
\end{equation}
This leading order cross section defines the normalisation for all 
higher order corrections discussed in the following.

At NLO, the tree-level four-parton processes $\gamma^* \to q\bar q gg$, 
$\gamma^* \to q\bar q q' \bar q'$ (non-identical quarks) and 
$\gamma^* \to q\bar q q \bar q$ (identical quarks) 
yield three-jet final states. Only the two former processes require 
subtraction, since the third process is infrared finite. 

The four-parton real radiation contribution to the NLO cross section is
\begin{eqnarray}
\lefteqn{{\rm d}\sigma_{NLO}^R= 
N_{{4}}\,  {\rm d}\Phi_{4}(p_{1},\ldots,p_{4};q) }
\nonumber \\
&& \times \Bigg\{ \frac{N}{2} 
\, \sum_{(i,j) \in P(3,4)}
A_{4}^0 (1_q,i_g,j_g,2_{\bar q})  
-\frac{1}{2N}
 \tilde A_{4}^0 (1_q,3_g,4_g,2_{\bar q}) +
N_F B_{4}^0 (1_q,3_q,4_{\bar q},2_{\bar q})\nonumber \\
&& \hspace{1.2cm} 
- \frac{1}{N} \, \left(
C_{4}^0 (1_q,3_q,4_{\bar q},2_{\bar q})
+C_{4}^0 (2_{\bar q},4_{\bar q},3_q,1_q) \right) \Bigg\}
\JET_{3}^{(4)}(p_{1},\ldots,p_{4})
\;.
\end{eqnarray}
The antenna subtraction term is then constructed as:
\begin{eqnarray}
{\rm d}\sigma_{NLO}^S&=& 
N_{{4}}\,  {\rm d}\Phi_{4}(p_{1},\ldots,p_{4};q) 
\nonumber \\
&& \times \Bigg\{  
 \sum_{(i,j) \in P(3,4)} \bigg[ \frac{N}{2}\, d_3^0(1_q,i_g,j_g)\, 
A_3^0(\widetilde{(1i)}_q,\widetilde{(ji)}_g,2_{\bar q})\, 
{J}_{3}^{(3)}(\widetilde{p_{1i}},\widetilde{p_{ji}},p_2) 
\nonumber \\
&& \hspace{1.2cm}
+ \frac{N}{2}\, d_3^0(2_{\bar q},i_g,j_g) \,
A_3^0(1_{q},\widetilde{(ji)}_g,\widetilde{(2i)}_{\bar q}) \,
{J}_{3}^{(3)}(p_1,\widetilde{p_{ji}},\widetilde{p_{2i}})  \nonumber \\
&& \hspace{1.2cm} - \frac{1}{2N}\,  A_3^0(1_q,i_g,2_{\bar q}) \,
A_3^0(\widetilde{(1i)}_q,j_g,\widetilde{(2i)}_{\bar q}) \,
{J}_{3}^{(3)}(\widetilde{p_{1i}},p_j,\widetilde{p_{2i}})\bigg] \nonumber \\
&& \hspace{0.6cm} + N_F \,\frac{1}{2} \bigg[ E_3^0(1_q,3_{q'},4_{\bar q'})\, 
A_3^0(\widetilde{(13)}_q,\widetilde{(43)}_g,2_{\bar q})\, 
{J}_{3}^{(3)}(\widetilde{p_{13}},\widetilde{p_{43}},p_2) 
\nonumber \\
&& \hspace{1.8cm}
+  E_3^0(2_{\bar q},3_{q'},4_{\bar q'}) \,
A_3^0(1_{q},\widetilde{(34)}_g,\widetilde{(24)}_{\bar q}) \,
{J}_{3}^{(3)}(p_1,\widetilde{p_{34}},\widetilde{p_{24}})  \bigg]\Bigg\}\;.
\end{eqnarray}
Integration of this subtraction term over the antenna phase spaces 
yields:
\begin{eqnarray}
{\rm d}\sigma_{NLO}^S&=& 
N_{{3}}\, \left(\frac{\alpha_s}{2\pi}\right)\,
 {\rm d}\Phi_{3}(p_{1},\ldots,p_{3};q) 
\nonumber \\
&& \times \left\{ \frac{N}{2}\,\left[  
{\cal D}_3^0(s_{13})+{\cal D}_3^0(s_{23})\right] - \frac{1}{N} 
{\cal A}_3^0(s_{12}) +\frac{N_F}{2}\,\left[  
{\cal E}_3^0(s_{13})+{\cal E}_3^0(s_{23})\right]\right\}\nonumber \\
&& \times
A_3^0(1_{q},3_g,2_{\bar q}) \,
{J}_{3}^{(3)}(p_1,p_2,p_3)
\end{eqnarray}
Together with the virtual one-loop contribution to $\gamma^*\to q\bar q g$,
\begin{eqnarray}
{\rm d}\sigma_{NLO}^V &=& 
N_{{3}}\, \left(\frac{\alpha_s}{2\pi}\right)\,
 {\rm d}\Phi_{3}(p_{1},\ldots,p_{3};q)\, {J}_{3}^{(3)}(p_1,p_2,p_3) 
\nonumber \\
&& \times
\, \bigg( N \left[
A_3^1(1_q,3_g,2_{\bar q})
+{\cal A}_2^1(s_{123}) A_3^0(1_q,3_g,2_{\bar q})\right]\nonumber \\
&&
-\frac{1}{N}  \left[
\tilde
A_3^1(1_q,3_g,2_{\bar q})
+{\cal A}_2^1(s_{123}) A_3^0(1_q,3_g,2_{\bar q})\right]
+N_F \hat A_3^1(1_q,3_g,2_{\bar q})\bigg)\, ,
\end{eqnarray}
one obtains 
\begin{equation}
\Poles\left({\rm d}\sigma_{NLO}^{S}\right) +
\Poles\left({\rm d}\sigma_{NLO}^{V}\right) = 0 \,,
\end{equation}
thus yielding an infrared-finite result.

\section{Construction of the $N^2$ colour factor}
\label{sec:termA}
The $N^2$ colour factor receives contributions from five-parton 
tree-level $\gamma^* \to q \bar q g g g$, four-parton one-loop 
$\gamma^* \to q \bar q g g$ and tree-level two-loop $\gamma^* \to q \bar q g$.
The multiple gluon emissions are colour-ordered, and the squared 
matrix elements do not contain interference amplitudes between 
different orderings. In the loop contributions to this colour factor,
non-planar momentum arrangements are absent. 
  
\subsection{Five-parton contribution}
At leading colour, the five parton contribution to 
 three-jet final states arises from the colour-ordered emission of three 
gluons in $\gamma^* \to q \bar q g g g$. The matrix element
for one ordering is:
\begin{eqnarray}
\left|M^0_{q\bar q3g}\right|^2
&=&  N_5 \, N^2 A_{5}^0 (1_q,i_g,j_g,k_g,2_{\bar q})\;.
\label{eq:2q3glc}
\end{eqnarray}
The real radiation contribution to the cross section
 is obtained by averaging over all possible six 
orderings:
 \begin{eqnarray}
{\rm d}\sigma_{NNLO,N^2}^R&=& 
N_{{5}}\,{N^2} \,  {\rm d}\Phi_{5}(p_{1},\ldots,p_{5};q)
 \, \frac{1}{3!}\,
\sum_{(i,j,k)\in P(3,4,5) }
A_{5}^0 (1_q,i_g,j_g,k_g,2_{\bar q}) \JET_{3}^{(5)}(p_{1},\ldots,p_{5})
\nonumber \\
&=& N_{{5}}\,{N^2} \,  {\rm d}\Phi_{5}(p_{1},\ldots,p_{5};q)
 \, \frac{1}{3!}\,
\nonumber \\ && \sum_{(i,j,k)\in P_C(3,4,5)} 
\left[
A_{5}^0 (1_q,i_g,j_g,k_g,2_{\bar q}) + A_{5}^0 (1_q,k_g,j_g,i_g,2_{\bar q}) 
\right]\JET_{3}^{(5)}(p_{1},\ldots,p_{5})\;,
\nonumber \\
\label{eq:a50} 
\end{eqnarray}
where the second expression is obtained by restricting the summation to 
the three 
cyclic permutations of the gluon momenta, while making the corresponding 
three non-cyclic permutations explicit. The cyclic form 
(\ref{eq:a50}) is more appropriate for the construction of 
the real radiation subtraction term, since this form matches 
onto the full quark-gluon antenna functions of~\cite{ourant}, which 
have a cyclic ambiguity in their momentum arrangements. 

The real radiation subtraction term for this colour factor reads
\begin{eqnarray}
\lefteqn{{\rm d}\sigma_{NNLO,N^2}^S= 
N_{{5}}\,{N^2} \,  {\rm d}\Phi_{5}(p_{1},\ldots,p_{5};q)
 \, \frac{1}{3!}\,
\sum_{(i,j,k)\in P_C(3,4,5) } \, \Bigg\{ }\nonumber \\
\ph{(a)}&&\phantom{+} d^0_3(1_q,i_g,j_g)\,
A^0_{4}(\widetilde{(1i)}_q,\widetilde{(ji)}_g,k_g,2_{\bar q}) \,
{J}_{3}^{(4)}(\widetilde{p_{1i}},\widetilde{p_{ji}},p_k,p_2) 
\nonumber \\
\ph{(b)}&&+ f^0_3(i_g,j_g,k_g)\,
{A}^0_{4}(1_q,\widetilde{(ij)}_g,\widetilde{(kj)}_g,2_{\bar q})\,
{J}_{3}^{(4)}(p_1,\widetilde{p_{ij}},\widetilde{p_{kj}},p_2) 
\nonumber \\
\ph{(c)}&&+ d^0_3(2_{\bar q},k_g,j_g)\,
{A}^0_{4}(1_q,i_g,\widetilde{(jk)}_g,\widetilde{(2k)}_{\bar q})\,
{J}_{3}^{(4)}(p_1,p_i,\widetilde{p_{jk}},\widetilde{p_{2k}}) 
\nonumber \\
\ph{(d)}&&+ d^0_3(1_q,k_g,j_g)\,
A^0_{4}(\widetilde{(1k)}_q,\widetilde{(jk)}_g,i_g,2_{\bar q}) \,
{J}_{3}^{(4)}(\widetilde{p_{1k}},\widetilde{p_{jk}},p_i,p_2) 
\nonumber \\
\ph{(e)}&&+ f^0_3(k_g,j_g,i_g)\,
{A}^0_{4}(1_q,\widetilde{(kj)}_g,\widetilde{(ij)}_g,2_{\bar q})\,
{J}_{3}^{(4)}(p_1,\widetilde{p_{kj}},\widetilde{p_{ij}},p_2) 
\nonumber \\
\ph{(f)}&&+ d^0_3(2_{\bar q},i_g,j_g)\,
{A}^0_{4}(1_q,k_g,\widetilde{(ji)}_g,\widetilde{(2i)}_{\bar q})\,
{J}_{3}^{(4)}(p_1,p_k,\widetilde{p_{ji}},\widetilde{p_{2i}}) 
\nonumber \\
\ph{(g1,h1)}&&+
\Bigg( D_{4,a}^0(1_q,i_g,j_g,k_g) - d_3^0(1_q,i_g,j_g)\,
 d_3^0(\widetilde{(1i)}_q,\widetilde{(ji)}_g,k_g)
\nonumber \\
\ph{(i1)}&& \hspace{0.3cm}
- f_3^0(i_g,j_g,k_g)\,
 d_3^0(1_q,\widetilde{(ij)}_g,\widetilde{(kj)}_g)
\Bigg) A^0_{3}(\widetilde{(1ij)}_q,\widetilde{(kji)}_g,2_{\bar q})\,
{J}_{3}^{(3)}(\widetilde{p_{1ij}},\widetilde{p_{kji}},p_2)
  \nonumber \\
\ph{(g2,i2)}&&+
\Bigg( D_{4,b}^0(1_q,i_g,j_g,k_g) 
- f_3^0(i_g,j_g,k_g)\,
 d_3^0(1_q,\widetilde{(kj)}_g,\widetilde{(ij)}_g)
\nonumber \\
\ph{(j2)}&& \hspace{0.3cm}
- d_3^0(1_q,k_g,j_g)\,
 d_3^0(\widetilde{(1k)}_q,\widetilde{(jk)}_g,i_g)
\Bigg) A^0_{3}(\widetilde{(1kj)}_q,\widetilde{(ijk)}_g,2_{\bar q})\,
{J}_{3}^{(3)}(\widetilde{p_{1kj}},\widetilde{p_{ijk}},p_2)
  \nonumber \\
\ph{(g3)}&&+
\Bigg( D_{4,c}^0(1_q,i_g,j_g,k_g) 
\nonumber \\
\ph{(h3)}&& \hspace{0.3cm}- d_3^0(1_q,i_g,j_g)\,
 d_3^0(\widetilde{(1i)}_q,k_g,\widetilde{(ji)}_g)
\Bigg) A^0_{3}(\widetilde{(1ik)}_q,\widetilde{(jki)}_g,2_{\bar q})\,
{J}_{3}^{(3)}(\widetilde{p_{1ik}},\widetilde{p_{jki}},p_2)
  \nonumber \\
\ph{(g4)}&&+
\Bigg( D_{4,d}^0(1_q,i_g,j_g,k_g) 
\nonumber \\
\ph{(j4)}&& \hspace{0.3cm}
- d_3^0(1_q,k_g,j_g)\,
 d_3^0(\widetilde{(1k)}_q,i_g,\widetilde{(jk)}_g)
\Bigg) A^0_{3}(\widetilde{(1ki)}_q,\widetilde{(jik)}_g,2_{\bar q})\,
{J}_{3}^{(3)}(\widetilde{p_{1ki}},\widetilde{p_{jik}},p_2)
\nonumber \\
\ph{(k1,l1)}&&+
\Bigg( D_{4,a}^0(2_q,i_g,j_g,k_g) - d_3^0(2_q,i_g,j_g)\,
 d_3^0(\widetilde{(2i)}_q,\widetilde{(ji)}_g,k_g)
\nonumber \\
\ph{(m1)}&& \hspace{0.3cm}
- f_3^0(i_g,j_g,k_g)\,
 d_3^0(2_q,\widetilde{(ij)}_g,\widetilde{(kj)}_g)
\Bigg) A^0_{3}(\widetilde{(2ij)}_q,\widetilde{(kji)}_g,1_{\bar q})\,
{J}_{3}^{(3)}(\widetilde{p_{2ij}},\widetilde{p_{kji}},p_1)
  \nonumber \\
\ph{(k2,m2)}&&+
\Bigg( D_{4,b}^0(2_q,i_g,j_g,k_g) 
- f_3^0(i_g,j_g,k_g)\,
 d_3^0(2_q,\widetilde{(kj)}_g,\widetilde{(ij)}_g)
\nonumber \\
\ph{(n2)}&& \hspace{0.3cm}
- d_3^0(2_q,k_g,j_g)\,
 d_3^0(\widetilde{(2k)}_q,\widetilde{(jk)}_g,i_g)
\Bigg) A^0_{3}(\widetilde{(2kj)}_q,\widetilde{(ijk)}_g,1_{\bar q})\,
{J}_{3}^{(3)}(\widetilde{p_{2kj}},\widetilde{p_{ijk}},p_1)
  \nonumber \\
\ph{(k3)}&&+
\Bigg( D_{4,c}^0(2_q,i_g,j_g,k_g) 
\nonumber \\
\ph{(l3)}&& \hspace{0.3cm}- d_3^0(2_q,i_g,j_g)\,
 d_3^0(\widetilde{(2i)}_q,k_g,\widetilde{(ji)}_g)
\Bigg) A^0_{3}(\widetilde{(2ik)}_q,\widetilde{(jki)}_g,1_{\bar q})\,
{J}_{3}^{(3)}(\widetilde{p_{2ik}},\widetilde{p_{jki}},p_1)
  \nonumber \\
\ph{(k4)}&&+
\Bigg( D_{4,d}^0(2_q,i_g,j_g,k_g) 
\nonumber \\
\nonumber \\
\ph{(n4)}&& \hspace{0.3cm}
- d_3^0(2_q,k_g,j_g)\,
 d_3^0(\widetilde{(2k)}_q,i_g,\widetilde{(jk)}_g)
\Bigg) A^0_{3}(\widetilde{(2ki)}_q,\widetilde{(jik)}_g,1_{\bar q})\,
{J}_{3}^{(3)}(\widetilde{p_{2ki}},\widetilde{p_{jik}},p_1)
  \nonumber \\
\ph{(o,p)}&&-
\Bigg( \tilde{A}_4^0(1_q,i_g,k_g,2_{\bar q}) 
- A_3^0(1_q,i_g,2_{\bar q})
 A_3^0(\widetilde{(1i)}_q,k_g,\widetilde{(2i)_{\bar q}})
\nonumber \\
\ph{(q)}&& \hspace{0.3cm} - A_3^0(1_q,k_g,2_{\bar q})
 A_3^0(\widetilde{(1k)}_q,i_g,\widetilde{(2k)_{\bar q}})
\Bigg)\,{A}^0_{3}(\widetilde{(1ik)}_q,j_g,\widetilde{(2ki)}_{\bar q})\,
{J}_{3}^{(3)}(\widetilde{p_{1ik}},p_j,\widetilde{p_{2ki}})
  \nonumber \\
\ph{(r)}&& 
- \frac{1}{2}d_3^0(1_q,i_g,j_g) 
             d_3^0(2_{\bar q},k_g,\widetilde{(ji)}_g)\,
{A}^0_{3}(\widetilde{(1i)}_q,\widetilde{((ji)k)}_g,\widetilde{(2k)}_{\bar q})\,
{J}_{3}^{(3)}(\widetilde{p_{1i}},\widetilde{p_{(ji)k}},\widetilde{p_{2k}}) 
  \nonumber \\
\ph{(s)}&& - \frac{1}{2}d_3^0(2_{\bar q},k_g,j_g) 
                d_3^0(1_{q},i_g,\widetilde{(jk)}_{g})\,
{A}^0_{3}(\widetilde{(1i)}_q,\widetilde{((jk)i)}_g,\widetilde{(2k)}_{\bar q})\,
{J}_{3}^{(3)}(\widetilde{p_{1i}},\widetilde{p_{(jk)i}},\widetilde{p_{2k}}) 
  \nonumber \\
\ph{(t)}&& - \frac{1}{2}d_3^0(1_q,k_g,j_g) 
                d_3^0(2_{\bar q},i_g,\widetilde{(jk)}_g)\,
{A}^0_{3}(\widetilde{(1k)}_q,\widetilde{((jk)i)}_g,\widetilde{(2i)}_{\bar q})\,
{J}_{3}^{(3)}(\widetilde{p_{1k}},\widetilde{p_{(jk)i}},\widetilde{p_{2i}}) 
  \nonumber \\
\ph{(u)}&& - \frac{1}{2}d_3^0(2_{\bar q},i_g,j_g) 
                d_3^0(1_{q},k_g,\widetilde{(ji)}_{g})\,
{A}^0_{3}(\widetilde{(1k)}_q,\widetilde{((ji)k)}_g,\widetilde{(2i)}_{\bar q})\,
{J}_{3}^{(3)}(\widetilde{p_{1k}},\widetilde{p_{(ji)k}},\widetilde{p_{2i}}) 
  \nonumber \\
\ph{(v)}&& 
+ \frac{1}{2}d_3^0(1_q,i_g,j_g) 
             d_3^0(\widetilde{(1i)}_{q},k_g,\widetilde{(ji)}_g)\,
{A}^0_{3}(\widetilde{((1i)k)}_q,\widetilde{((ji)k)}_g,2_{\bar q})\,
{J}_{3}^{(3)}(\widetilde{p_{(1i)k}},\widetilde{p_{(ji)k}},{p_{2}}) 
  \nonumber \\
\ph{(w)}&& 
+ \frac{1}{2}d_3^0(1_q,k_g,j_g) 
             d_3^0(\widetilde{(1k)}_{q},i_g,\widetilde{(jk)}_g)\,
{A}^0_{3}(\widetilde{((1k)i)}_q,\widetilde{((jk)i)}_g,2_{\bar q})\,
{J}_{3}^{(3)}(\widetilde{p_{(1k)i}},\widetilde{p_{(jk)i}},{p_{2}}) 
  \nonumber \\
\ph{(x)}&& 
+ \frac{1}{2}d_3^0(2_{\bar q},i_g,j_g) 
             d_3^0(\widetilde{(2i)}_{\bar q},k_g,\widetilde{(ji)}_g)\,
{A}^0_{3}(1_q,\widetilde{((ji)k)}_g,\widetilde{((2i)k)}_{\bar q})\,
{J}_{3}^{(3)}(p_1,\widetilde{p_{(ji)k}},\widetilde{p_{(2i)k}}) 
  \nonumber \\
\ph{(y)}&& 
+ \frac{1}{2}d_3^0(2_{\bar q},k_g,j_g) 
             d_3^0(\widetilde{(2k)}_{\bar q},i_g,\widetilde{(jk)}_g)\,
{A}^0_{3}(1_q,\widetilde{((jk)i)}_g,\widetilde{((2k)i)}_{\bar q})\,
{J}_{3}^{(3)}(p_1,\widetilde{p_{(jk)i}},\widetilde{p_{(2k)i}}) 
  \nonumber \\
\ph{(aa)}&& 
- \frac{1}{2}A_3^0(1_q,i_g,2_{\bar q}) 
             d_3^0(\widetilde{(1i)}_{q},k_g,{j}_g)\,
{A}^0_{3}(\widetilde{((1i)k)}_q,\widetilde{(jk)}_g,\widetilde{(2i)}_{\bar q})\,
{J}_{3}^{(3)}(\widetilde{p_{(1i)k}},\widetilde{p_{jk}},\widetilde{p_{2i}}) 
  \nonumber \\
\ph{(ab)}&& 
+ \frac{1}{2}d_3^0(1_q,k_g,j_g) 
             A_3^0(\widetilde{(1k)}_{q},i_g,{2}_{\bar q})\,
{A}^0_{3}(\widetilde{((1k)i)}_q,\widetilde{(jk)}_g,\widetilde{(2i)}_{\bar q})\,
{J}_{3}^{(3)}(\widetilde{p_{(1k)i}},\widetilde{p_{jk}},\widetilde{p_{2i}}) 
  \nonumber \\
\ph{(ac)}&& 
- \frac{1}{2}A_3^0(1_q,k_g,2_{\bar q}) 
             d_3^0(\widetilde{(1k)}_{q},i_g,{j}_g)\,
{A}^0_{3}(\widetilde{((1k)i)}_q,\widetilde{(ji)}_g,\widetilde{(2k)}_{\bar q})\,
{J}_{3}^{(3)}(\widetilde{p_{(1k)i}},\widetilde{p_{ji}},\widetilde{p_{2k}}) 
  \nonumber \\
\ph{(ad)}&& 
+ \frac{1}{2}d_3^0(1_q,i_g,j_g) 
             A_3^0(\widetilde{(1i)}_{q},k_g,{2}_{\bar q})\,
{A}^0_{3}(\widetilde{((1i)k)}_q,\widetilde{(ji)}_g,\widetilde{(2k)}_{\bar q})\,
{J}_{3}^{(3)}(\widetilde{p_{(1i)k}},\widetilde{p_{ji}},\widetilde{p_{2k}}) 
  \nonumber \\
\ph{(ae)}&& 
- \frac{1}{2}A_3^0(1_{q},i_g,2_{\bar q}) 
             d_3^0(\widetilde{(2i)}_{\bar q},k_g,{j}_g)\,
{A}^0_{3}(\widetilde{(1i)}_q,\widetilde{(jk)}_g,\widetilde{((2i)k)}_{\bar q})\,
{J}_{3}^{(3)}(\widetilde{p_{1i}},\widetilde{p_{jk}},\widetilde{p_{(2i)k}}) 
  \nonumber \\
\ph{(af)}&& 
+ \frac{1}{2}d_3^0(2_{\bar q},k_g,j_g) 
             A_3^0({1}_{q},i_g,\widetilde{(2k)}_{\bar q})\,
{A}^0_{3}(\widetilde{(1i)}_q,\widetilde{(jk)}_g,\widetilde{((2k)i)}_{\bar q})\,
{J}_{3}^{(3)}(\widetilde{p_{1i}},\widetilde{p_{jk}},\widetilde{p_{(2k)i}}) 
  \nonumber \\
\ph{(ag)}&& 
- \frac{1}{2}A_3^0(1_{q},k_g,2_{\bar q}) 
             d_3^0(\widetilde{(2k)}_{\bar q},i_g,{j}_g)\,
{A}^0_{3}(\widetilde{(1k)}_q,\widetilde{(ji)}_g,\widetilde{((2k)i)}_{\bar q})\,
{J}_{3}^{(3)}(\widetilde{p_{1k}},\widetilde{p_{ji}},\widetilde{p_{(2k)i}}) 
  \nonumber \\
\ph{(ah)}&& 
+ \frac{1}{2}d_3^0(2_{\bar q},i_g,j_g) 
             A_3^0({1}_{q},k_g,\widetilde{(2i)}_{\bar q})\,
{A}^0_{3}(\widetilde{(1k)}_q,\widetilde{(ji)}_g,\widetilde{((2i)k)}_{\bar q})\,
{J}_{3}^{(3)}(\widetilde{p_{1k}},\widetilde{p_{ji}},\widetilde{p_{(2i)k}}) 
  \nonumber \\
\ph{(ai)}&& 
- \frac{1}{2}A_3^0(1_q,k_g,2_{\bar q}) 
             A_3^0(\widetilde{(1k)}_{q},i_g,\widetilde{(2k)}_{\bar q})\,
{A}^0_{3}(\widetilde{((1k)i)}_q,j_g,\widetilde{((2k)i)}_{\bar q})\,
{J}_{3}^{(3)}(\widetilde{p_{(1k)i}},{p_{j}},\widetilde{p_{(2k)i}}) 
  \nonumber \\
\ph{(aj)}&& 
- \frac{1}{2}A_3^0(1_q,i_g,2_{\bar q}) 
             A_3^0(\widetilde{(1i)}_{q},k_g,\widetilde{(2i)}_{\bar q})\,
{A}^0_{3}(\widetilde{((1i)k)}_q,j_g,\widetilde{((2i)k)}_{\bar q})\,
{J}_{3}^{(3)}(\widetilde{p_{(1i)k}},{p_{j}},\widetilde{p_{(2i)k}}) 
\Bigg\}. \nonumber \\ 
\label{eq:nnlolc}
\end{eqnarray}

\subsection{Four-parton contribution}
The leading colour four-parton contribution comes from the one-loop 
correction to $\gamma^* \to q\bar q gg$, where the gluonic emissions are
colour-ordered. It reads
\begin{eqnarray}
{\rm d}\sigma_{NNLO,N^2}^{V,1}&=&
N_{{4}}\, {N^2}\, \left(\frac{\alpha_s}{2\pi}\right)\,
{\rm d}\Phi_{4}(p_{1},\ldots,p_{4};q)\, \nonumber \\ &&
\times \frac{1}{2} \sum_{(i,j)\in (3,4)}  
A_{4}^{1,a} (1_q,i_g,j_g,2_{\bar q}) 
\,
\JET_{3}^{(4)}(p_{1},p_2,p_3,p_{4}),
\label{eq:sigvA}
\end{eqnarray}
The one-loop single unresolved subtraction term for this colour factor is 
\begin{eqnarray}
\lefteqn{{\rm d}\sigma_{NNLO,N^2}^{VS,1}
= 
 {N_{{4}}}\,{N^2}\, \left(\frac{\alpha_s}{2\pi}\right)\,
 {\rm d}\Phi_{4}(p_{1},\ldots,p_{4};q)\,}
\nonumber \\
 \ph{(a)}&& \Bigg\{
- \frac{1}{2} \sum_{(i,j)\in (3,4)}\left[  
\frac{1}{2}{\cal D}_3^0(s_{1i})+\frac{1}{3}{\cal F}_3^0(s_{ij}) 
+\frac{1}{2}{\cal D}_3^0(s_{2j}) \right]\,    
A_{4}^{0} (1_q,i_g,j_g,2_{\bar q})\,
\JET_{3}^{(4)}(p_{1},\ldots,p_{4}) \nonumber \\ 
 \ph{(b)}&&
+\Bigg\{
\frac{1}{2}\, \sum_{(i,j)\in (3,4)} \bigg[
d_3^0(1_q,i_{g},j_{g})\left[
A_3^1(\widetilde{(1i)}_q,\widetilde{(ji)}_g,2_{\bar q})
+{\cal A}_2^1(s_{1234})
A_3^0(\widetilde{(1i)}_q,\widetilde{(ji)}_g,2_{\bar q})\right] \nonumber \\
 \ph{(c)}
&&\hspace{4mm}+d_3^1(1_q,i_{g},j_{g})\,
A_3^0(\widetilde{(1i)}_q,\widetilde{(ji)}_g,2_{\bar q}) \nonumber \\
 \ph{(d)}&&\hspace{4mm}+ \frac{1}{2} \left( 
{\cal D}_3^0(s_{1ij}) + 
 {\cal D}_3^0(s_{2\widetilde{(ji)}})
\right) \,
d_3^0(1_q,i_{g},j_{g})\,
 A_3^0(\widetilde{(1i)}_q,\widetilde{(ji)}_g,2_{\bar q}) 
 \nonumber \\
 \ph{(e,f)}&&\hspace{4mm}
+ \left( \frac{1}{2} {\cal D}_3^0(s_{1i}) + \frac{1}{3} {\cal F}_3^0(s_{ij})
+ \frac{1}{2} {\cal D}_3^0(s_{1j})  
-   {\cal D}_3^0(s_{1ij})
\right) \,
d_3^0(1_q,i_{g},j_{g})\,
 A_3^0(\widetilde{(1i)}_q,\widetilde{(ji)}_g,2_{\bar q})
\nonumber\\
 \ph{(g)}&&\hspace{4mm}
+ b_{0} \log\frac{q^2}{s_{1ij}} d_3^0(1_q,i_{g},j_{g})\,
A_3^0(\widetilde{(1i)}_q,\widetilde{(ji)}_g,2_{\bar q})
\nonumber\\
 \ph{(h,i)}&&\hspace{4mm}
- \frac{1}{2} \left(\frac{1}{2}
{\cal D}_3^0(s_{2\widetilde{(ji)}}) - \frac{1}{2} {\cal D}_3^0(s_{1ij})
+ {\cal A}_3^0(s_{\widetilde{(1i)}2}) 
-\frac{1}{2} {\cal D}_3^0(s_{2j}) + \frac{1}{2} {\cal D}_3^0(s_{1j})
- {\cal A}_3^0(s_{12}) 
 \right)\nonumber \\
&& \hspace{1cm}
d_3^0(1_q,i_{g},j_{g})\,
 A_3^0(\widetilde{(1i)}_q,\widetilde{(ji)}_g,2_{\bar q})
\bigg]
\JET_{3}^{(3)}(\widetilde{p_{1i}},\widetilde{p_{ji}},p_2) 
+(1\leftrightarrow 2) \Bigg\}
\nonumber \\
 \ph{(j)}&&
-\frac{1}{2}\,\sum_{(i,j)\in (3,4)}\bigg[
\tilde{A}_3^1(1_q,i_g,2_{\bar q})
A_3^0(\widetilde{(1i)}_q,j_g,\widetilde{(2i)}_{\bar q})
 \nonumber \\
 \ph{(k,l)}&&\hspace{4mm}
+  \left( {\cal A}_3^0(s_{12}) - 
{\cal A}_3^0(s_{12i}) 
\right) \,
A_3^0(1_q,i_{g},2_{\bar q})\,
 A_3^0(\widetilde{(1i)}_q,j_g,\widetilde{(2i)}_{\bar q})
 \nonumber \\
 \ph{(m,n)}&&\hspace{4mm}
+\frac{1}{2} \left( 
{\cal A}_3^0 (s_{12i}) 
- \frac{1}{2}
 {\cal D}_3^0(s_{\widetilde{(1i)}j}) 
- \frac{1}{2} {\cal D}_3^0(s_{\widetilde{(2i)}j})
- {\cal A}_3^0(s_{12})  + \frac{1}{2}
 {\cal D}_3^0(s_{1j}) + \frac{1}{2} {\cal D}_3^0(s_{2j})
\right) \nonumber \\ && \hspace{1cm}
A_3^0(1_q,i_{g},2_{\bar q})\,
 A_3^0(\widetilde{(1i)}_q,j_g,\widetilde{(2i)}_{\bar q})
 \bigg]
\JET_{3}^{(3)}(\widetilde{p_{1i}},\widetilde{p_{2i}},p_j)
\end{eqnarray}
\ph{The cancellations are as follows: (5abcdef,4a), (5hij,4e), 
(5lmn,4e), (5pq,4k), (5r-y,4i), (5aa,5ac,5ae,5ag,4i),
(5ab,5ad,5af,5ag,5ah,5ai,5aj,4n). 
The terms (5g,5k,5o) and
(4b,4c,4d,4f,4i,4j,4l,4m) integrate to the three-parton.   Note that 
cancellations at the three-parton level take place among (4d,4f,4h) and
(4h,4l,4m), such that some of the 
structures appearing in individual terms are absent in the three-parton 
contribution. }

\subsection{Three-parton contribution}
The three-parton contribution consists of the two-loop three-parton matrix
element together with the integrated forms of 
the five-parton and four-parton
subtraction terms,
\begin{eqnarray}
\lefteqn{{\rm d}\sigma_{NNLO,N^2}^{S}+{\rm d}\sigma_{NNLO,N^2}^{VS,1} = N^2
 \,}\nonumber \\
&&  \times \Bigg\{  \bigg[
\frac{1}{2}\, \left({\cal D}_4^0  (s_{13}) 
+{\cal D}_4^0  (s_{23})\right) - \frac{1}{8}
\left({\cal D}_3^0  (s_{13}) - {\cal D}_3^0  (s_{23})\right)^2 
-\frac{1}{2} \left(\tilde{{\cal A}}_4^0(s_{12})
- {\cal A}_3^0(s_{12})\, {\cal A}_3^0(s_{12})
\right) 
\nonumber \\ &&\hspace{4mm}
+\frac{1}{2}\, {\cal D}_3^1  (s_{13}) + 
 \frac{1}{2}\, {\cal D}_3^1  (s_{23}) - \tilde{{\cal A}}_3^1(s_{12})
\bigg]\,A_3^0({1}_q,{3}_g, 2_{\bar q}) 
\nonumber \\ &&\hspace{4mm}
+ \frac{1}{2}
 \left( {\cal D}_3^0(s_{13}) +
{\cal D}_3^0(s_{23}) \right) \, 
{A}_3^1(1_q,3_g
,2_{\bar q}) 
\nonumber \\ &&\hspace{4mm}
+ \frac{b_{0}}{2\e} \,  \bigg[  
{\cal D}_3^0(s_{13})
\left(({s_{13}})^{-\e} -  ({s_{123}})^{-\e}\right) 
+{\cal D}_3^0(s_{23})
\left(({s_{23}})^{-\e} -  ({s_{123}})^{-\e}\right) 
\bigg]\,A_3^0({1}_q,{3}_g,
2_{\bar q}) 
\Bigg\} \d \sigma_3\;,
\nonumber\\
\end{eqnarray}
where we 
defined the three-parton normalisation factor
\begin{equation}
{\rm d}\sigma_3 = 
N_{{3}}\, \left(\frac{\alpha_s}{2\pi}\right)^2\,
{\rm d}\Phi_{3}(p_{1},p_2,p_{3};q) \, \JET_{3}^{(3)}(p_1,p_2,p_3)\;.
\end{equation}

Combining the infrared poles of this expression with the two loop
matrix element, we obtain the cancellation of all 
infrared poles in this colour factor,
\begin{equation}
\Poles\left({\rm d}\sigma_{NNLO,N^2}^{S}\right) +
\Poles\left({\rm d}\sigma_{NNLO,N^2}^{VS,1}\right) +
\Poles\left({\rm d}\sigma_{NNLO,N^2}^{V,2}\right) = 0 \,.
\end{equation}

\section{Construction of the $N^0$ colour factor}
\label{sec:termB}

The contribution for the $N^0$ colour factor to three-jet final states 
is more involved than all other colour factors. It receives 
contributions from all partonic subprocesses: $\gamma^* \to q\bar q ggg$ 
and  $\gamma^* \to q\bar q q\bar q g$ at tree-level, 
 $\gamma^* \to q\bar q gg$ and  $\gamma^* \to q\bar q q\bar q$ at one loop 
and $\gamma^* \to q\bar q g$ at two loops. All contributions contain a 
mixture of colour ordered and non-ordered emissions.

\subsection{Five-parton contribution}

The subleading colour $N^0$  
contribution of five-parton final states to three jet final states is
\begin{eqnarray}
\lefteqn{{\rm d}\sigma_{NNLO,N^0}^R= 
{N_{{5}}} \,  {\rm d}\Phi_{5}(p_{1},\ldots,p_{5};q) }
\nonumber
\\ 
&\times&\Bigg[ \frac{1}{3!}\,
\Bigg( \bar A_{5}^0 (1_q,3_g,4_g,5_g,2_{\bar q})
 - \sum_{(i,j,k)\in P(3,4,5)} \tilde A_{5}^0 (1_q,i_g,j_g,k_g,2_{\bar q})
\Bigg) \nonumber \\&&
+ \tilde C_{5}^0 (1_q,3_q,4_{\bar q},5_g,2_{\bar q})
+\tilde C_{5}^0 (2_{\bar q},4_{\bar q},3_q,5_g,1_q)
-   C_{5}^0 (1_q,3_q,4_{\bar q},5_g,2_{\bar q})
\Bigg]
\JET_{3}^{(5)}(p_{1},\ldots,p_{5})\nonumber \\
&=& 
{N_{{5}}} \,  {\rm d}\Phi_{5}(p_{1},\ldots,p_{5};q) 
\nonumber
\\ 
&\times&\Bigg[ \frac{1}{3!}\,
\bar A_{5}^0 (1_q,3_g,4_g,5_g,2_{\bar q})
 - \frac{1}{2}
\sum_{(i,j)\in P(3,4)} \tilde A_{5}^0 (1_q,i_g,j_g,5_g,2_{\bar q})
 \nonumber \\&&
+ \tilde C_{5}^0 (1_q,3_q,4_{\bar q},5_g,2_{\bar q})
+\tilde C_{5}^0 (2_{\bar q},4_{\bar q},3_q,5_g,1_q)
-   C_{5}^0 (1_q,3_q,4_{\bar q},5_g,2_{\bar q})
\Bigg]
\JET_{3}^{(5)}(p_{1},\ldots,p_{5})
\,,\nonumber \\
\label{eq:sigrB}\end{eqnarray}
where the symmetry factor in front of 
$\bar{A}_5^0$ is due to the inherent 
indistinguishability of gluons. 
In $\tilde{A}_5^0$, 
gluon $(5_g)$ is effectively photon-like.  It does not participate in any 
three-gluon 
or four-gluon vertices, and there are no simple collinear limits as 
$(i)_g||(5)_g$ and $(j)_g||(5)_g$.

The real radiation subtraction term for this colour factor is:
\begin{eqnarray}
\lefteqn{{\rm d}\sigma_{NNLO,N^0}^S= 
N_{{5}}\,{N^0} \,  {\rm d}\Phi_{5}(p_{1},\ldots,p_{5};q)}
\, \nonumber \\ 
\ph{(a)}&\times&
\bigg\{-\frac{1}{2} \sum_{(i,j)\in (3,4)} \bigg[  A_3^0(1_q,5_g,2_{\bar q})
\, A_4^0(\widetilde{(15)}_q,i_g,j_g,\widetilde{(25)}_{\bar q}) 
J_3^{(4)} (\widetilde{p_{15}},p_i,p_j,\widetilde{p_{25}})\nonumber \\
\ph{(b)}&& \hspace{4mm}
+ d_3^0(1,i,j)\,
 \tilde{A}_4^0(\widetilde{(1i)}_q,\widetilde{(ji)}_{g},5_g,2_{\bar q}) 
J_3^{(4)} (\widetilde{p_{1i}},\widetilde{p_{ji}},p_5,p_2)\nonumber \\
\ph{(c)}&& \hspace{4mm}
+ d_3^0(2,j,i)\,
 \tilde{A}_4^0(1_q,5_g,\widetilde{(ij)}_g,\widetilde{(2j)}_{\bar q}) 
J_3^{(4)} (p_1,p_5,\widetilde{p_{ij}},\widetilde{p_{2j}})\bigg]\nonumber \\
\ph{(d)}&& + \frac{1}{3!}  \sum_{(i,j,k)\in P_C(3,4,5)} 
A_3^0(1_q,i_g,2_{\bar q})
\, \tilde{A}_4^0(\widetilde{(1i)}_q,j_g,k_g,\widetilde{(2i)}_{\bar q}) 
J_3^{(4)} (\widetilde{p_{1i}},p_j,p_k,\widetilde{p_{2i}})\nonumber \\
\ph{(e)}&&- A_3^0(1_q,5_g,3_{q})
\, \left[C_4^0(\widetilde{(15)}_q,\widetilde{(35)}_{q},4_{\bar q},2_{\bar q}) 
+ C_4^0(2_{\bar q},4_{\bar q},\widetilde{(35)}_{q},\widetilde{(15)}_q)
\right]
J_3^{(4)} (\widetilde{p_{15}},\widetilde{p_{35}},p_4,p_2)\nonumber \\
\ph{(f)}&&- A_3^0(2_{\bar q},5_g,4_{\bar q})\left[
\, C_4^0(1_q,3_q,\widetilde{(45)}_{\bar q},\widetilde{(25)}_{\bar q})+
C_4^0(\widetilde{(25)}_{\bar q},\widetilde{(45)}_{\bar q},3_q,1_q)
\right] 
J_3^{(4)} (p_1,p_3,\widetilde{p_{25}},\widetilde{p_{45}})\nonumber \\
\ph{(g,h)}&&- \frac{1}{2} \,\sum_{(i,j) \in (3,4)}
\Bigg( A_4^0(1_q,i_g,j_g,2_{\bar q})
- d_3^0(1_q,i_g,j_g)\,
 A_3^0(\widetilde{(1i)}_q,\widetilde{(ji)}_g,2_{\bar q})
\nonumber \\
\ph{(i)}&& \hspace{4mm}- d_3^0(2_{\bar q},j_g,i_g)\,
 A_3^0(1_q,\widetilde{(ij)}_g,\widetilde{(2j)}_{\bar q})
\Bigg) \,
A^0_{3}(\widetilde{(1ij)}_q,5_g,\widetilde{(2ji)}_{\bar q})\,
{J}_{3}^{(3)}(\widetilde{p_{1ij}},p_5,\widetilde{p_{2ji}})
  \nonumber \\
\ph{(j,k)}&&- \frac{1}{3}  \,\sum_{(i,j) \in (3,4)}
\Bigg( \tilde{A}_4^0(1_q,i_g,5_g,2_{\bar q}) 
- A_3^0(1_q,i_g,2_{\bar q})\,
 A_3^0(\widetilde{(1i)}_q,5_g,\widetilde{(2i)}_{\bar q})
\nonumber \\
\ph{(l)}&& \hspace{4mm}- A_3^0(1_{q},5_g,2_{\bar q})\,
 A_3^0(\widetilde{(15)}_q,i_g,\widetilde{(25)}_{\bar q})
\Bigg)\,
A^0_{3}(\widetilde{(1i5)}_q,j_g,\widetilde{(2i5)}_{\bar q})\,
{J}_{3}^{(3)}(\widetilde{p_{1i5}},p_j,\widetilde{p_{2i5}})
  \nonumber \\
\ph{(m,n)}&&+ \frac{1}{6}  \,
\Bigg( \tilde{A}_4^0(1_q,3_g,4_g,2_{\bar q}) 
- A_3^0(1_q,3_g,2_{\bar q})\,
 A_3^0(\widetilde{(13)}_q,4_g,\widetilde{(23)}_{\bar q})
\nonumber \\
\ph{(o)}&& \hspace{4mm}- A_3^0(1_{q},4_g,2_{\bar q})\,
 A_3^0(\widetilde{(14)}_q,3_g,\widetilde{(24)}_{\bar q})
\Bigg)\,
A^0_{3}(\widetilde{(134)}_q,5_g,\widetilde{(234)}_{\bar q})\,
{J}_{3}^{(3)}(\widetilde{p_{134}},p_5,\widetilde{p_{234}})
  \nonumber \\
\ph{(p)}&&- \left[{C}^0_4(1_q,3_q,4_{\bar q},2_{\bar q}) 
+ {C}^0_4(2_{\bar q},4_{\bar q},3_q,1_q)\right] \,
A^0_{3}(\widetilde{(134)}_q,5_g,\widetilde{(234)}_{\bar q})
\,\JET_{3}^{(3)}(\widetilde{p_{134}},p_5,\widetilde{p_{234}})
\nonumber \\
\ph{(q)}&& + \frac{1}{2} \sum_{(i,j) \in (3,4)} \Bigg(
d_3^0 (1_q,i_g,j_g) \, A_3^0(\widetilde{(1i)}_q,5_g,2_{\bar q}) 
\, A_3^0 (\widetilde{((1i)5)}_q,\widetilde{(ji)}_g,\widetilde{(25)}_{\bar q})
\, J_3^{(3)} (\widetilde{p_{(1i)5}},\widetilde{p_{ji}},\widetilde{p_{25}})
\nonumber \\
\ph{(r)}&&+ d_3^0 (2_{\bar q},j_g,i_g) \, A_3^0(1_{q},5_g,\widetilde{(2j)}_{\bar q}) 
\, A_3^0 (\widetilde{(15)}_{q},\widetilde{(ij)}_g,\widetilde{((2j)5)}_{\bar q})
\, J_3^{(3)} (\widetilde{p_{15}},\widetilde{p_{ij}},\widetilde{p_{(2j)5}})
\nonumber \\
\ph{(s)}&&- A_3^0(1_q,i_g,2_{\bar q}) \, 
A_3^0(\widetilde{(1i)}_q,5_g,\widetilde{(2i)}_{\bar q}) \,
A_3^0(\widetilde{((1i)5)}_q,j_g,\widetilde{((2i)5)}_{\bar q})
\, J_3^{(3)} (\widetilde{p_{(1i)5}},\widetilde{p_{(2i)5}},p_j)
\Bigg)
\bigg\} .
\end{eqnarray}

\subsection{Four-parton contribution}
The four-parton contribution comes from the subleading colour one-loop 
correction to $\gamma^* \to q\bar q gg$, where the gluonic emissions are
colour-ordered, from the leading colour one-loop 
correction to $\gamma^* \to q\bar q gg$, where the gluonic emissions are
not colour-ordered and from the one-loop correction to the identical-flavour
process $\gamma^* \to q\bar q q \bar q$. It reads:
\begin{eqnarray}
\lefteqn{{\rm d}\sigma_{NNLO,N^0}^{V,1}=
N_{{4}}\, {N^0}\, \left(\frac{\alpha_s}{2\pi}\right)\,
{\rm d}\Phi_{4}(p_{1},\ldots,p_{4};q)\, }\nonumber \\ && \times
\bigg\{ - \frac{1}{2} \bigg(\sum_{(i,j)\in (3,4)}  
A_{4}^{1,b} (1_q,i_g,j_g,2_{\bar q}) - 
\tilde{A}_{4}^{1,a} (1_q,3_g,4_g,2_{\bar q}) +
\tilde{A}_{4}^{1,d} (1_q,3_g,4_g,2_{\bar q})\bigg) \nonumber \\ &&
- C_4^{1,d} (1_q,3_q,4_{\bar q},2_{\bar q})
- C_4^{1,d} (2_{\bar q},4_{\bar q},3_q,1_q)
\bigg\}\,
\JET_{3}^{(4)}(p_{1},p_2,p_3,p_{4}),
\label{eq:sigvB}
\end{eqnarray}

The one-loop single unresolved subtraction term for this colour factor is
\begin{eqnarray}
\lefteqn{{\rm d}\sigma_{NNLO,N^0}^{VS,1}
= 
 {N_{{4}}}\,{N^0}\, \left(\frac{\alpha_s}{2\pi}\right)\,
 {\rm d}\Phi_{4}(p_{1},p_2,p_3,p_{4};q)\,}
\nonumber \\
\ph{(a)}&& \Bigg\{
  \frac{1}{2} \sum_{(i,j)\in (3,4)}\bigg[  \left(
\frac{1}{2}{\cal D}_3^0(s_{1i}) 
+\frac{1}{2}{\cal D}_3^0(s_{2j}) \right)\,    
\tilde{A}_{4}^{0} (1_q,i_g,j_g,2_{\bar q})\,\nonumber \\ 
\ph{(b)}&& \hspace{4mm}
+ {\cal A}_3^0(s_{12}) {A}_{4}^{0} (1_q,i_g,j_g,2_{\bar q})
\bigg]
\JET_{3}^{(4)}(p_{1},p_2,p_3,p_{4})
\nonumber \\
\ph{(c)}&&
- \frac{1}{2} \,  {\cal A}_3^0(s_{12}) 
\tilde{A}_{4}^{0} (1_q,3_g,4_g,2_{\bar q})\JET_{3}^{(4)}(p_{1},p_2,p_3,p_{4})
 \nonumber \\ 
\ph{(d)}&&
+ \left({\cal A}_3^0(s_{13})  + {\cal A}_3^0(s_{24}) \right) 
\left(C_4^0(1_q,3_q,4_{\bar q},2_{\bar q}) + 
C_4^0(2_{\bar q},4_{\bar q},3_q,1_q)
\right)\JET_{3}^{(4)}(p_{1},p_2,p_3,p_{4}) \nonumber \\ 
\ph{(e)}&&
+\frac{1}{2}\sum_{(i,j)\in(3,4)} \bigg\{ -\bigg[
d_3^0(1_q,i_{g},j_{g})\left[
\tilde{A}_3^1(\widetilde{(1i)}_q,\widetilde{(ji)}_g,2_{\bar q})
+{\cal A}_2^1(s_{1234})
A_3^0(\widetilde{(1i)}_q,\widetilde{(ji)}_g,2_{\bar q})\right] 
\nonumber \\ 
\ph{(f)}&&
\hspace{4mm}
\JET_{3}^{(3)}(\widetilde{p_{1i}},\widetilde{p_{ji}},p_2)  
 + (1\leftrightarrow 2) \bigg]
-\tilde{A}_3^1(1_q,i_g,2_{\bar q}) \,
 A_3^0( \widetilde{(1i)}_q,j_g,\widetilde{(2i)}_{\bar q}) \,
\JET_{3}^{(3)}(\widetilde{p_{1i}},p_j,\widetilde{p_{2i}}) \nonumber\\
\ph{(g)}&&
-A_3^0(1_q,i_{g},2_{\bar q})\left[
{A}_3^1(\widetilde{(1i)}_q,j_g,\widetilde{(2i)}_{\bar q})
+{\cal A}_2^1(s_{1234})
A_3^0(\widetilde{(1i)}_q,j_g,\widetilde{(2i)}_{\bar q})\right] 
\JET_{3}^{(3)}(\widetilde{p_{1i}},p_j,\widetilde{p_{2i}}) \nonumber \\ 
\ph{(h)}&&
-{A}_3^1(1_q,i_g,2_{\bar q}) \,
 A_3^0( \widetilde{(1i)}_q,j_g,\widetilde{(2i)}_{\bar q}) \,
\JET_{3}^{(3)}(\widetilde{p_{1i}},p_j,\widetilde{p_{2i}}) \nonumber\\
\ph{(i)}&&- \bigg[
d_3^0(1_q,i_{g},j_{g})\,{\cal A}_3^0(s_{\widetilde{(1i)}2})\, 
{A}_3^0(\widetilde{(1i)}_q,\widetilde{(ji)}_g,2_{\bar q})
\JET_{3}^{(3)}(\widetilde{p_{1i}},\widetilde{p_{ji}},p_2)  
+ (1\leftrightarrow 2) \bigg]\nonumber \\ 
\ph{(j,k)}&& 
- \left[{\cal A}_3^0 (s_{12}) - {\cal A}_3^0 (s_{12i})
\right]\,{A}_3^0(1_q,i_g,2_{\bar q}) \,
 A_3^0( \widetilde{(1i)}_q,j_g,\widetilde{(2i)}_{\bar q}) \,
\JET_{3}^{(3)}(\widetilde{p_{1i}},p_j,\widetilde{p_{2i}}) \nonumber\\
\ph{(l)}&&
-\frac{1}{2} \left[{\cal D}_3^0 (s_{\widetilde{(1i)}j}) +
{\cal D}_3^0 (s_{\widetilde{(2i)}j}) 
\right]\, A_3^0(1_q,i_{g},2_{\bar q})\,
{A}_3^0(\widetilde{(1i)}_q,j_g,\widetilde{(2i)}_{\bar q})\,
\JET_{3}^{(3)}(\widetilde{p_{1i}},p_j,\widetilde{p_{2i}}) \nonumber \\ 
\ph{(m,n)}&&
- \left[\frac{1}{2} {\cal D}_3^0 (s_{1i}) +\frac{1}{2}
{\cal D}_3^0 (s_{2i}) - {\cal A}_3^0 (s_{12i}) 
\right] {A}_3^0(1_q,i_g,2_{\bar q}) \,
 A_3^0( \widetilde{(1i)}_q,j_g,\widetilde{(2i)}_{\bar q}) \,
\JET_{3}^{(3)}(\widetilde{p_{1i}},p_j,\widetilde{p_{2i}}) \nonumber\\
\ph{(o,p)}&&
+\left[ 
 \frac{1}{2} {\cal D}_3^0 (s_{\widetilde{(1i)}j}) +\frac{1}{2}
{\cal D}_3^0 (s_{\widetilde{(2i)}j}) 
- {\cal A}_3^0 (s_{12i})
- \frac{1}{2} {\cal D}_3^0 (s_{1j}) -\frac{1}{2}
{\cal D}_3^0 (s_{2j}) 
+ {\cal A}_3^0 (s_{12})
\right]\nonumber \\ && \hspace{4mm}
 A_3^0(1_q,i_{g},2_{\bar q})\,
{A}_3^0(\widetilde{(1i)}_q,j_g,\widetilde{(2i)}_{\bar q})\,
\JET_{3}^{(3)}(\widetilde{p_{1i}},p_j,\widetilde{p_{2i}}) \nonumber \\ 
\ph{(q)}&&- b_{0} \log\frac{q^2}{s_{12i}} A_3^0(1_q,i_{g},2_{\bar q})\,
A_3^0(\widetilde{(1i)}_q,j_g,\widetilde{(2i)}_{\bar q})
\JET_{3}^{(3)}(\widetilde{p_{1i}},p_j,\widetilde{p_{2i}}) \Bigg\}
 \end{eqnarray}
\ph{The cancellations are as follows: (5a,4b), (5bc,4a), (5d,4c), (5ef,4d),
(5hi,4m), (5klno,4j), (5qr,4p), (5s,4p). The terms (5g,5j,5m,5p) and
(4e,4f,4g,4h,4i,4k,4q) integrate to the three-parton contribution, while 
(4l,4n,4o) cancel each other.}

\subsection{Three-parton contribution}

The three parton contribution to the $N^0$ colour factor receives 
contributions from the  
three-parton virtual two-loop correction and the integrated 
five-parton tree-level and 
four-parton one-loop subtraction terms, which read
\begin{eqnarray}
\lefteqn{{\rm d}\sigma_{NNLO,N^0}^{S}+{\rm d}\sigma_{NNLO,N^0}^{VS,1} = N^0
 \,}\nonumber \\
&&  \times \Bigg\{ - \bigg[
 {\cal A}_4^0  (s_{12}) 
+ \frac{1}{2}\, \tilde{{\cal A}}_4^0(s_{12}) + 2\,  {\cal C}_4^0 (s_{12}) 
 + \frac{1}{2} {\cal A}_3^0(s_{12}) \left(
{\cal D}_3^0  (s_{13}) + {\cal D}_3^0  (s_{23})\right)
\nonumber \\ &&\hspace{4mm}
- {\cal A}_3^0(s_{12})\, {\cal A}_3^0(s_{12}) 
+ {\cal A}_3^1  (s_{12}) 
+  \tilde{{\cal A}}_3^1(s_{12})  
\bigg]\,A_3^0({1}_q,{3}_g, 2_{\bar q}) 
\nonumber \\ &&\hspace{4mm}
- \frac{1}{2}
 \left( {\cal D}_3^0(s_{13}) +
{\cal D}_3^0(s_{23}) \right) \, 
\tilde{A}_3^1(1_q,3_g
,2_{\bar q}) 
-  {\cal A}_3^0(s_{13}) \,
{A}_3^1(1_q,3_g,2_{\bar q})
\nonumber \\ 
&& \hspace{4mm} 
- \frac{b_{0}}{\e} \,    
{\cal A}_3^0(s_{12})
\left(({s_{12}})^{-\e} -  ({s_{123}})^{-\e}\right) 
\,A_3^0({1}_q,{3}_g,
2_{\bar q}) 
\Bigg\} \d \sigma_3\;.
\end{eqnarray}

Combining the infrared poles of this expression with the two loop
matrix element, we obtain the cancellation of all 
infrared poles in this colour factor,
\begin{equation}
\Poles\left({\rm d}\sigma_{NNLO,N^0}^{S}\right) +
\Poles\left({\rm d}\sigma_{NNLO,N^0}^{VS,1}\right) +
\Poles\left({\rm d}\sigma_{NNLO,N^0}^{V,2}\right) = 0 \,.
\end{equation}

\section{Construction of the $1/N^2$ colour factor}
\label{sec:termC}
The $1/N^2$ colour factor receives contributions from five-parton 
tree-level $\gamma^* \to q \bar q g g g$ and $\gamma^* \to q \bar q q \bar qg$,
four-parton one-loop 
$\gamma^* \to q \bar q g g$ and $\gamma^* \to q \bar q q \bar q$ 
as well as tree-level two-loop $\gamma^* \to q \bar q g$.
The gluon emissions are all photon-like, not containing any gluon 
self-coupling. The four-quark processes contribute through the 
identical-quark-only terms.

The construction of the subtraction terms for this colour factor was discussed
in detail in~\cite{our3j:cf2,ourant}. 
  
\subsection{Five-parton contribution}
\label{sec:dsub}

Two different five-parton final states contribute at $1/N^2$ to three-jet final
states at NNLO:  $\gamma^*\to q\bar q ggg$ and 
 $\gamma^*\to q\bar q q\bar qg$ with identical quarks.  

The NNLO radiation term appropriate for the three jet final state is given by
\begin{eqnarray}
{\rm d}\sigma_{NNLO,1/N^2}^R&=& 
\frac{N_{{5}}}{N^2} \,  {\rm d}\Phi_{5}(p_{1},\ldots,p_{5};q) 
\nonumber
\\ 
&\times&\left[ \frac{1}{3!}\,
 \bar A_{5}^0 (1_q,3_g,4_g,5_g,2_{\bar q}) 
+ 2\, \tilde C_{5}^0 (1_q,2_{\bar q},3_q,4_{\bar q},5_g) 
\right]
\JET_{3}^{(5)}(p_{1},\ldots,p_{5})\,,
\label{eq:sigrC}\end{eqnarray}
where the symmetry factor in front of 
$\bar{A}_5^0$ is due to the inherent 
indistinguishability of gluons. 
The factor 2 in front of $\tilde C_{5}^0$ arises 
from the fact that two different momentum arrangements contribute to 
the squared matrix element (\ref{eq:4q1g}). If the quarks and antiquarks are 
not distinguished by the jet functions, these contribute equally.

The real radiation subtraction term for this colour factor is:
\begin{eqnarray}
\lefteqn{{\rm d}\sigma_{NNLO,1/N^2}^{S}
=  \frac{N_{5}}{N^2}\, {\rm d}\Phi_{5}(p_{1},\ldots,p_{5};q)}\, 
\nonumber \\ &&
\Bigg\{ \frac{1}{3!}
\sum_{i,j,k \in P_C(3,4,5)}\;\bigg[
A^0_3(1_q,i_g,2_{\bar q})\,
\tilde{A}^0_{4}(\widetilde{(1i)}_q,j_g,k_g,\widetilde{(2i)}_{\bar q})\,
\,\JET_{3}^{(4)}(\widetilde{p_{1i}},p_j,p_k,\widetilde{p_{2i}})\;
\nonumber \\
&& 
 + \biggl(
\tilde{A}^0_4(1_q,i_g,j_g,2_{\bar q})
-A^0_3(1_q,i_g,2_{\bar q})\;
A^0_3(\widetilde{(1i)}_q,j_g,\widetilde{(2i)}_{\bar q})
\nonumber \\
&&
\hspace{5mm}-A^0_3(1_q,j_g,2_{\bar q})\; 
A^0_3(\widetilde{(1j)}_q,i_g,\widetilde{(2j)}_{\bar q})\;
\biggr)
A^0_{3}(\widetilde{(1ij)}_q,k_g,\widetilde{(2ij)}_{\bar q})\,
\JET_{3}^{(3)}(\widetilde{p_{1ij}},p_k,\widetilde{p_{2ij}}) \bigg]\nonumber \\
&&+ 2 \bigg[
 A^0_3(1_q,5_g,2_{\bar q})\, C_4^0(\widetilde{(15)}_q,
3_q,4_{\bar q},\widetilde{(25)}_{\bar q})\, 
\JET_{3}^{(4)}(\widetilde{p_{15}},p_3,p_4,\widetilde{p_{25}}) \nonumber \\&&
\hspace{5mm}+ A^0_3(1_q,5_g,4_{\bar q})\, C_4^0(\widetilde{(15)}_q,3_q,
\widetilde{(45)}_{\bar q},2_{\bar q})  \,
\JET_{3}^{(4)}(\widetilde{p_{15}},p_2,p_3,\widetilde{p_{45}}) \nonumber \\&&
\hspace{5mm}+  A^0_3(3_q,5_g,2_{\bar q})\, C_4^0(1_q,
\widetilde{(35)}_q,4_{\bar q},\widetilde{(25)}_{\bar q})\, 
\JET_{3}^{(4)}(\widetilde{p_{35}},p_1,p_4,\widetilde{p_{25}}) \nonumber \\&&
\hspace{5mm}+ A^0_3(3_q,5_g,4_{\bar q})\, C_4^0(1_q,\widetilde{(35)}_q,
\widetilde{(45)}_{\bar q},2_{\bar q})  \,
\JET_{3}^{(4)}(\widetilde{p_{35}},p_1,p_2,\widetilde{p_{45}}) \nonumber \\&&
\hspace{5mm}-  A^0_3(1_q,5_g,3_q)\, C_4^0(\widetilde{(15)}_{q},
\widetilde{(35)}_q,4_{\bar q},2_{\bar q}) \,
\JET_{3}^{(4)}(\widetilde{p_{15}},p_2,p_4,\widetilde{p_{35}}) \nonumber \\&&
\hspace{5mm}- A^0_3(2_{\bar q},5_g,4_{\bar q})\, C_4^0(1_q,3_q,
\widetilde{(45)}_{\bar q},\widetilde{(25)}_{\bar q}) \,
\JET_{3}^{(4)}(\widetilde{p_{25}},p_1,p_3,\widetilde{p_{45}}) \nonumber \\&&
\hspace{5mm}+ {C}^0_4(1_q,3_q,4_{\bar q},2_{\bar q}) \,
A^0_{3}(\widetilde{(134)}_q,5_g,\widetilde{(234)}_{\bar q})
\,\JET_{3}^{(3)}(\widetilde{p_{134}},p_5,\widetilde{p_{234}}) \bigg]\Bigg\}\;.
\label{eq:subrC}
\end{eqnarray}
The sum in the first contribution runs only over the three cyclic 
permutations of the gluon momenta to prevent double counting of identical
configurations obtained by interchange of $j$ and $k$.

\subsection{Four-parton contribution}
 \label{sec:vsub}
At one-loop,
there are two contributions to
the colour suppressed contribution proportional to $1/N^2$, one from the four
quark final state and one from the two quark-two gluon final state:
\begin{eqnarray}
{\rm d}\sigma_{NNLO,1/N^2}^{V,1}&=&
\frac{N_{{4}}}{N^2}\, \left(\frac{\alpha_s}{2\pi}\right)\,
{\rm d}\Phi_{4}(p_{1},\ldots,p_{4};q)\, \nonumber \\ &&
\left\{ \frac{1}{2!}\,
\tilde A_{4}^{1,b} (1_q,3_g,4_g,2_{\bar q})
+ 2\, C_{4}^{1,e} (1_q,3_q,4_{\bar q},2_{\bar q}) \right\}
\,
\JET_{3}^{(4)}(p_{1},\ldots,p_{4}),\hspace{4mm}
\label{eq:sigvC}
\end{eqnarray}
where the origin of the symmetry factors is as in the 
real radiation five-parton contributions of the previous section.

The corresponding subtraction term is:
\begin{eqnarray}
{\rm d}\sigma_{NNLO,1/N^2}^{VS,1}
&= &
 \frac{N_{{4}}}{N^2}\, \left(\frac{\alpha_s}{2\pi}\right)\,
 {\rm d}\Phi_{4}(p_{1},\ldots,p_{4};q)\nonumber \\
&\times& \Bigg\{
\, \frac{1}{2!} \sum_{i,j\in P(3,4)} \bigg[ - {\cal A}_3^0(s_{12})
\tilde A_{4}^0 (1_q,i_g,j_g,2_{\bar q})
\JET_{3}^{(4)}(p_{1},p_i,p_j,p_{2})\nonumber \\ &&
+
\bigg(
A_3^0(1_q,i_g,2_{\bar q})\left[
\tilde A_3^1(\widetilde{(1i)}_q,j_g,\widetilde{(2i)}_{\bar q})
+{\cal A}_2^1(s_{1234})A_3^0(\widetilde{(1i)}_q,j_g,\widetilde{(2i)}_{\bar
q})\right]
\nonumber \\
&&\hspace{4mm}+\tilde A_3^1(1_q,i_g,2_{\bar q})A_3^0(\widetilde{(1i)}_q,j_g,\widetilde{(2i)}_{\bar q})
\bigg)
\JET_{3}^{(3)}(\widetilde{p_{1i}},p_j,\widetilde{p_{2i}}) \nonumber \\ &&
+ {\cal A}_3^0(s_{12})
A_3^0(1_q,i_g,2_{\bar q})A_3^0(\widetilde{(1i)}_q,j_g,\widetilde{(2i)}_{\bar q})
\JET_{3}^{(3)}(\widetilde{p_{1i}},p_j,\widetilde{p_{2i}}) \bigg]
\nonumber \\ &&
- 2 \Big[
 {\cal A}_3^0(s_{12}) 
+ {\cal A}_3^0(s_{14}) + {\cal A}_3^0(s_{23})+ {\cal A}_3^0(s_{34})
- {\cal A}_3^0(s_{13}) 
- {\cal A}_3^0(s_{24}) \Big]\nonumber \\ &&\hspace{4mm} \times
 C_{4}^0 (1_q,3_q,4_{\bar q},2_{\bar q})
\JET_{3}^{(4)}(p_{1},p_3,p_4,p_{2})\Bigg\}\;.
\end{eqnarray}

\subsection{Three-parton contribution}
\label{sec:cancel}
The three-parton contribution consists of the two-loop three-parton matrix
element together with the integrated forms of 
the five-parton and four-parton
subtraction terms,
\begin{eqnarray}
{\rm d}\sigma_{NNLO,1/N^2}^{S}+{\rm d}\sigma_{NNLO,1/N^2}^{VS,1} &=& 
\frac{1}{N^2}
 \Bigg\{  \bigg[
\frac{1}{2} \tilde{{\cal A}}_4^0  (s_{12}) 
 + 2\,  {\cal C}_4^0 (s_{12}) 
+  \tilde{{\cal A}}_3^1(s_{12})  
\bigg]\,A_3^0({1}_q,{3}_g, 2_{\bar q}) 
\nonumber \\ &&\hspace{4mm}
+  {\cal A}_3^0(s_{13}) \,
\tilde{A}_3^1(1_q,3_g,2_{\bar q})
\Bigg\} \d \sigma_3\;.
\end{eqnarray}

Combining the infrared poles of this expression with the two loop
matrix element, we obtain the cancellation of all 
infrared poles in this colour factor,
\begin{equation}
\Poles\left({\rm d}\sigma_{NNLO,1/N^2}^{S}\right) +
\Poles\left({\rm d}\sigma_{NNLO,1/N^2}^{VS,1}\right) +
\Poles\left({\rm d}\sigma_{NNLO,1/N^2}^{V,2}\right) = 0 \,.
\end{equation}

\section{Construction of the $N_F\,N$ colour factor}
\label{sec:termD}
The colour factor  $N_F\,N$ receives contribution from the colour-ordered 
five-parton tree-level process $\gamma^* \to q\bar q q'\bar q' g$,
the four-parton one-loop processes  $\gamma^* \to q\bar q q'\bar q'$
and  $\gamma^* \to q\bar q g g$ at leading colour and from the 
two-loop three-parton process $\gamma^* \to q\bar qg$.

\subsection{Five-parton contribution}

The NNLO radiation term appropriate for the three jet final state is given by
\begin{eqnarray}
{\rm d}\sigma_{NNLO,N_FN}^R&=& 
N_{{5}}\,N_FN \,  {\rm d}\Phi_{5}(p_{1},\ldots,p_{5};q) 
\nonumber
\\ 
&\times&\left[ 
B_5^{0,a}(1_q,5_g,4_{\bar q'};3_{q'},2_{\bar q})
+B_5^{0,b}(1_q,4_{\bar q'};3_{q'},5_g,2_{\bar q})
\right]
\JET_{3}^{(5)}(p_{1},\ldots,p_{5}),
\label{eq:sigrD}\end{eqnarray}

The two terms represent the two colour orderings of the leading 
colour amplitude for this process. 
Since the leading colour $q q'\bar q' g$-antenna
subtraction terms allows $q$ to represent either a quark or an antiquark,
both colour orderings are mixed together. Therefore, 
it is not possible to construct a subtraction term 
for an individual contribution, but only for their sum. 

The subtraction term for this contribution is 
\begin{eqnarray}
\lefteqn{{\rm d}\sigma_{NNLO,N_FN}^S= 
N_{{5}}\,N_FN \,  {\rm d}\Phi_{5}(p_{1},\ldots,p_{5};q) }
\nonumber
\\ 
\ph{(a)}&\times&\Big\{
A^0_3(1_q,5_g,4_{\bar q'})\,
{B}^0_{4}(\widetilde{(15)}_q,3_{q'},\widetilde{(45)}_{\bar q'},2_{\bar q})\,
{J}_{3}^{(4)}(\widetilde{p_{15}},p_2,p_3,\widetilde{p_{45}}) 
\nonumber \\
\ph{(b)}&& +
A^0_3(3_{q'},5_g,2_{\bar q})\,
{B}^0_{4}(1_q,\widetilde{(35)}_{q'},4_{\bar q'},\widetilde{(25)}_{\bar q})\,
{J}_{3}^{(4)}(p_1,\widetilde{p_{25}},\widetilde{p_{35}},p_4) 
\nonumber \\
\ph{(c)}&&+  
G^0_3(5_g,3_{q'},4_{\bar q'})
{A}^0_{4}(1_q,\widetilde{(34)}_{g},\widetilde{(54)}_{g},2_{\bar q})\,
{J}_{3}^{(4)}(p_1,p_2,\widetilde{p_{34}},
\widetilde{p_{54}}) 
 \nonumber \\
\ph{(d)}&&+ 
G^0_3(5_g,3_{q'},4_{\bar q'})
{A}^0_{4}(1_q,\widetilde{(54)}_{g},\widetilde{(34)}_{g},2_{\bar q})\,
{J}_{3}^{(4)}(p_1,p_2,\widetilde{p_{34}},
\widetilde{p_{54}}) 
 \nonumber \\
\ph{(e1)}&&+
\bigg( E_{4,a}^0(1_q,3_{q'},4_{\bar q'},5_g) 
\nonumber \\
\ph{(g1)}&&\hspace{4mm}- 
G_3^0(5_g,3_{q'},4_{\bar q'})\,
d_3^0(1_q,\widetilde{(34)}_g,\widetilde{(54)}_{g})
\bigg)\,
A^0_{3}(\widetilde{(134)}_q,\widetilde{(543)}_g,2_{\bar q})
{J}_{3}^{(3)}(\widetilde{p_{134}},p_2,\widetilde{p_{543}})
  \nonumber \\
\ph{(e2,f)}&&+
\bigg( E_{4,b}^0(1_q,3_{q'},4_{\bar q'},5_g) 
- A_3^0(1_q,5_g,4_{\bar q'})\, 
E_3^0(\widetilde{(15)}_q,3_{q'},\widetilde{(45)}_{\bar q'})
\nonumber \\
\ph{(g2)}&&\hspace{4mm}- 
G_3^0(5_g,3_{q'},4_{\bar q'})\,
d_3^0(1_q,\widetilde{(54)}_{g},\widetilde{(34)}_g)
\bigg)\,
A^0_{3}(\widetilde{(154)}_q,\widetilde{(345)}_g,2_{\bar q})
{J}_{3}^{(3)}(\widetilde{p_{154}},p_2,\widetilde{p_{345}})
  \nonumber \\
\ph{(h1)}&&+
\bigg( E_{4,a}^0(2_{\bar q},4_{\bar q'},3_{q'},5_g) 
\nonumber \\
\ph{(j1)}&& \hspace{4mm}
- 
G_3^0(5_g,3_{q'},4_{\bar q'})\,
d_3^0(2_{\bar q},\widetilde{(43)}_g,\widetilde{(53)}_{g})
\bigg)\,
A^0_{3}(1_q,\widetilde{(534)}_g,\widetilde{(243)}_{\bar q})
{J}_{3}^{(3)}(p_1,\widetilde{p_{243}},\widetilde{p_{534}}) \nonumber\\
\ph{(h1,i1)}&&+
\bigg( E_{4,b}^0(2_{\bar q},4_{\bar q'},3_{q'},5_g) 
- A_3^0(3_{q'},5_g,2_{\bar q})\, 
E_3^0(\widetilde{(25)}_{\bar q},4_{\bar q'},\widetilde{(35)}_{q'})
\nonumber \\
\ph{(j2)}&& \hspace{4mm}
- 
G_3^0(5_g,3_{q'},4_{\bar q'})\,
d_3^0(2_{\bar q},\widetilde{(53)}_{g},\widetilde{(43)}_g)
\bigg)\,
A^0_{3}(1_q,\widetilde{(435)}_g,\widetilde{(253)}_{\bar q})
{J}_{3}^{(3)}(p_1,\widetilde{p_{253}},\widetilde{p_{435}})\Big\}\,. \nonumber\\
&&   
\label{eq:nnlosubcanf}
\end{eqnarray}

\subsection{Four-parton contribution}
The four parton contribution to the $N_FN$ colour factor reads:
\begin{eqnarray}
{\rm d}\sigma_{NNLO,N_FN}^{V,1}&=&
N_{{4}}\,{N_FN}\, \left(\frac{\alpha_s}{2\pi}\right)\,
{\rm d}\Phi_{4}(p_{1},\ldots,p_{4};q)\, \nonumber \\ &&
\left\{ \frac{1}{2} \sum_{(i,j)\in (3,4)}
\left( B_{4}^{1,a} (1_q,i_{q'},j_{\bar q'},2_{\bar q}) 
+ A_{4}^{1,c} (1_q,i_g,j_g,2_{\bar q})\,\right)
\right\}
\,
\JET_{3}^{(4)}(p_{1},\ldots,p_{4})
\,.\nonumber \\ 
\label{eq:sigvD}
\end{eqnarray}
The average over the permutations of the momenta $(3)$ and $(4)$ 
has to be made in both contributions to this colour factor. In the one-loop 
correction to the $\gamma^*\to qq'\bar q' \bar q$ final state $B_4^{1,a}$, 
the secondary quark-antiquark pair has to be symmetrised, since the 
quark-gluon antenna functions used in the one-loop subtraction terms 
do not distinguish quarks and antiquarks. 
The summation over the two colour orderings of 
the one-loop correction to the  $\gamma^*\to q gg\bar q$ final state
$A_{4}^{1,c}$ must be 
kept since the one-loop subtraction functions appropriate to this 
term contain both orderings because of their cyclicity.

The corresponding subtraction term is:
\begin{eqnarray}
\lefteqn{{\rm d}\sigma_{NNLO,N_FN}^{VS,1}
= 
 N_{{4}}{N_FN}\, \left(\frac{\alpha_s}{2\pi}\right)\,
 {\rm d}\Phi_{4}(p_{1},\ldots,p_{4};q)}\nonumber \\
\ph{(a)}&\times& \Bigg\{
- \frac{1}{2} \sum_{(i,j)\in (3,4)} 
\left[ {\cal A}_3^0(s_{1j}) + {\cal A}_3^0(s_{2i})
 \right]\,    
B_{4}^{0} (1_q,i_{q'},j_{\bar q'},2_{\bar q})\,
\JET_{3}^{(4)}(p_{1},\ldots,p_{4}) \nonumber \\ 
\ph{(b)}&&
- 
\sum_{(i,j)\in (3,4)} 
{\cal G}_3^0(s_{ij}) 
\, 
 A_{4}^0 (1_q,i_g,j_g,2_{\bar q})\,
\JET_{3}^{(4)}(p_{1},\ldots,p_{4})\nonumber \\
\ph{(c)} &&+\Bigg\{
\frac{1}{2}\,\bigg(
E_3^0(1_q,3_{q'},4_{\bar q'})\left[
A_3^1(\widetilde{(13)}_q,\widetilde{(43)}_g,2_{\bar q})
+{\cal A}_2^1(s_{1234})
A_3^0(\widetilde{(13)}_q,\widetilde{(43)}_g,2_{\bar q})\right]  \nonumber \\
\ph{(d)} &&\hspace{4mm}
+E_3^1(1_q,3_{q'},4_{\bar q'})\,
A_3^0(\widetilde{(13)}_q,\widetilde{(43)}_g,2_{\bar q})  \nonumber \\
\ph{(e)} &&\hspace{4mm}
+ \frac{1}{2} \left({\cal D}_3^0(s_{134}) + 
 {\cal D}_3^0(s_{2\widetilde{(43)}})
\right) \,
E_3^0(1_q,3_{q'},4_{\bar q'})\,
 A_3^0(\widetilde{(13)}_q,\widetilde{(43)}_g,2_{\bar q})
 \nonumber \\
\ph{(f,g)} &&\hspace{4mm}
+ \left( {\cal A}_3^0(s_{13}) + 
{\cal A}_3^0(s_{14}) -   {\cal D}_3^0(s_{134})
\right) \,
E_3^0(1_q,3_{q'},4_{\bar q'})\,
 A_3^0(\widetilde{(13)}_q,\widetilde{(43)}_g,2_{\bar q})
\nonumber\\
\ph{(h)} &&\hspace{4mm}
+ b_{0} \log\frac{q^2}{s_{134}} E_3^0(1_q,3_{q'},4_{\bar q'})\,
A_3^0(\widetilde{(13)}_q,\widetilde{(43)}_g,2_{\bar q})
\bigg)
 \JET_{3}^{(3)}(\widetilde{p_{13}},\widetilde{p_{43}},p_2)
+ (1\leftrightarrow 2)\Bigg\}
\nonumber \\
\ph{(i)} &&+\,\Bigg\{
\frac{1}{2}\,\bigg(
D_3^0(1_q,3_{g},4_{g})
\hat A_3^1(\widetilde{(13)}_q,\widetilde{(43)}_g,2_{\bar q})
+\hat D_3^1(1_q,3_{g},4_{g})\,
A_3^0(\widetilde{(13)}_q,\widetilde{(43)}_g,2_{\bar q})\nonumber\\
\ph{(j)} &&\hspace{4mm}
+ 2 
{\cal G}_3^0(s_{34}) 
\, D_3^0(1_q,3_{g},4_{g})\,
A_3^0(\widetilde{(13)}_q,\widetilde{(43)}_g,2_{\bar q})
\nonumber \\
\ph{(k)} &&\hspace{4mm}
+ b_{0,F} \log\frac{q^2}{s_{134}} D_3^0(1_q,3_{g},4_{g})\,
A_3^0(\widetilde{(13)}_q,\widetilde{(43)}_g,2_{\bar q})
\bigg)
 \JET_{3}^{(3)}(\widetilde{p_{13}},\widetilde{p_{43}},p_2)
+ (1\leftrightarrow 2)\Bigg\}\;. \nonumber \\
\end{eqnarray}
\ph{The cancellations are as follows: (5a,4a), (5b,4a), (5c,4b), (5d,4b),
(5f,4f), (5g,4j), (5i,4f), (5j,4j). 
The terms (5e,5h,4c,4d,4e,4g,4h,4i,4k) integrate to 
the three-parton channel. }

\subsection{Three-parton contribution}

The three parton contribution to the $N_F \, N$ colour factor contains the 
three-parton virtual two-loop correction and the integrated 
five-parton tree-level and 
four-parton one-loop subtraction terms, which read
\begin{eqnarray}
\lefteqn{{\rm d}\sigma_{NNLO,N_FN}^{S}+{\rm d}\sigma_{NNLO,N_FN}^{VS,1} = 
{N_F}\, {N}} \,\nonumber \\
&& \times \Bigg\{ \bigg[
 {\cal E}_4^0  (s_{13}) + 
 {\cal E}_4^0  (s_{23}) 
- \frac{1}{4}
\left({\cal D}_3^0  (s_{13}) \, {\cal E}_3^0  (s_{13})
+ {\cal D}_3^0  (s_{23})\,  {\cal E}_3^0  (s_{23}) \right)
\nonumber \\ && \hspace{4mm} + \frac{1}{4}
\left({\cal D}_3^0  (s_{13}) \, {\cal E}_3^0  (s_{23})
+ {\cal D}_3^0  (s_{23}) \, {\cal E}_3^0  (s_{13}) \right)
+ \frac{1}{2}\, \hat{{\cal D}}_3^1  (s_{13}) +\frac{1}{2} \,
  \hat{{\cal D}}_3^1 (s_{23})
+ \frac{1}{2}\, {\cal E}_3^1 
 (s_{13}) \nonumber \\ && \hspace{4mm} +\frac{1}{2} \,
 {\cal E}_3^1  (s_{23})
\bigg]\,A_3^0({1}_q,{3}_g, 2_{\bar q}) + \frac{1}{2}
 \left( {\cal E}_3^0(s_{13}) +
{\cal E}_3^0(s_{23}) \right) \, 
{A}_3^1(1_q,3_g
,2_{\bar q})\nonumber \\ 
&& \hspace{4mm}
+ \frac{1}{2}
 \left( {\cal D}_3^0(s_{13}) +
{\cal D}_3^0(s_{23}) \right) \,
\hat{A}_3^1(1_q,3_g,2_{\bar q})
\nonumber \\ 
&& \hspace{4mm}
+ \frac{b_{0,F}}{2\e}\, \bigg[  
{\cal D}_3^0(s_{13})
\left(({s_{13}})^{-\e} -  ({s_{123}})^{-\e}\right) 
+{\cal D}_3^0(s_{23})
\left(({s_{23}})^{-\e} -  ({s_{123}})^{-\e}\right) \bigg]
\,A_3^0({1}_q,{3}_g,
2_{\bar q}) 
\nonumber \\ 
&& \hspace{4mm} 
+ \frac{b_{0}}{2\e} \, \bigg[  
{\cal E}_3^0(s_{13})
\left(({s_{13}})^{-\e} -  ({s_{123}})^{-\e}\right) 
+{\cal E}_3^0(s_{23})
\left(({s_{23}})^{-\e} -  ({s_{123}})^{-\e}\right)  \bigg]
\,A_3^0({1}_q,{3}_g,
2_{\bar q}) 
\Bigg\} \d \sigma_3\;.
\nonumber \\ 
\end{eqnarray}

Combining the infrared poles of this expression with the two loop
matrix element, we obtain the cancellation of all 
infrared poles in this colour factor,
\begin{equation}
\Poles\left({\rm d}\sigma_{NNLO,N_FN}^{S}\right) +
\Poles\left({\rm d}\sigma_{NNLO,N_FN}^{VS,1}\right) +
\Poles\left({\rm d}\sigma_{NNLO,N_FN}^{V,2}\right) = 0 \,.
\end{equation}

\section{Construction of the $N_F/N$ colour factor}
The $N_F/N$ colour factor receives contributions from five-parton 
tree-level $\gamma^* \to q \bar q q' \bar q' g$, four-parton one-loop 
$\gamma^* \to q \bar q q'\bar q'$ and $\gamma^* \to q \bar q gg$ at 
subleading colour 
as well as three-parton two-loop $\gamma^* \to q \bar q g$.
The gluon emissions are all photon-like.

This colour factor is part of the QED-type corrections. We described 
the construction of the subtraction terms for this colour factor 
previously in~\cite{our3j:QED}.

\label{sec:termE}
\subsection{Five-parton contribution}
The NNLO radiation term appropriate for the three jet final state is given by
\begin{eqnarray}
{\rm d}\sigma_{NNLO,N_F/N}^R&=& 
N_{{5}}\,\frac{N_F}{N} \,  {\rm d}\Phi_{5}(p_{1},\ldots,p_{5};q) 
\,\Big[ 
B_5^{0,c}(1_q,5_g,2_{\bar q};3_{q'},4_{\bar q'})\nonumber \\
&&\hspace{0.6cm}
+B_5^{0,d}(1_q,2_{\bar q};3_{q'},5_g,4_{\bar q'})
-2\,B_5^{0,e}(1_q,2_{\bar q};3_{q'},4_{\bar q'};5_g)
\Big]
\JET_{3}^{(5)}(p_{1},\ldots,p_{5})\nonumber \\
&=& 
N_{{5}}\,\frac{N_F}{N} \,  {\rm d}\Phi_{5}(p_{1},\ldots,p_{5};q) 
\,\frac{1}{2}\,\sum_{(i,j)\in (3,4)} \Big[ 
B_5^{0,c}(1_q,5_g,2_{\bar q};i_{q'},j_{\bar q'})\nonumber \\
&&\hspace{0.6cm}
+B_5^{0,d}(1_q,2_{\bar q};i_{q'},5_g,j_{\bar q'})
-2\,B_5^{0,e}(1_q,2_{\bar q};i_{q'},j_{\bar q'};5_g)
\Big]
\JET_{3}^{(5)}(p_{1},\ldots,p_{5})\,,\nonumber \\
\label{eq:sigrE}\end{eqnarray}
where the symmetrization over the momenta of the secondary quark-antiquark
pair exploits the fact that the jet algorithm does not distinguish 
quarks and antiquarks. This symmetrisation reduces the number of
non-vanishing unresolved limits considerably, since the 
interference term in $B_5^{0,e}$ is odd under this interchange. As a result, 
the unresolved structure of the symmetrised   $B_5^{0,e}$ equals the 
unresolved structure of $B_5^{0,c}+B_5^{0,d}$.

The subtraction term reads:
\begin{eqnarray}
\lefteqn{{\rm d}\sigma_{NNLO,N_F/N}^S=
N_{{5}}\,\frac{N_F}{N} \,  {\rm d}\Phi_{5}(p_{1},\ldots,p_{5};q) 
\,\frac{1}{2}\,\sum_{(i,j)\in (3,4)} \Bigg\{ } \nonumber \\
\ph{(a)} &&- A^0_3(1_q,5_g,2_{\bar q})\,
{B}^{0}_{4} (\widetilde{(15)}_q,\widetilde{(25)}_{\bar q},
i_{q'},j_{\bar q'})\,
{J}_{3}^{(4)}(\widetilde{p_{15}},\widetilde{p_{25}},p_i,p_{j}) 
\nonumber \\
\ph{(b)} &&- A^0_3(i_{q'},5_g,j_{\bar q'})\,
{B}^{0}_{4} (1_{q},2_{\bar q},\widetilde{(i5)}_{q'},\widetilde{(j5)}_{\bar q'})
\,{J}_{3}^{(4)}(p_1,p_2,\widetilde{p_{i5}},\widetilde{p_{j5}}) 
\nonumber \\
\ph{(c)} &&-  \frac{1}{2} \left\{
E^0_3(1_q,i_{q'},j_{\bar q'})\,
\tilde{A}^0_{4}(\widetilde{(1j)}_q,\widetilde{(ij)}_g,5_g,2_{\bar q})\,
{J}_{3}^{(4)}(\widetilde{p_{1j}},p_2,\widetilde{p_{ij}},p_5) 
+  (1\leftrightarrow 2)\right\}
 \nonumber \\
\ph{(d,e)} &&-  \Bigg(\, B_4^0(1_q,i_{q'},j_{\bar q'},2_{\bar q})
- \frac{1}{2} \left\{E_3^0(1_q,i_{q'},j_{\bar q'})\,
A_3^0( \widetilde{(1i)}_q,\widetilde{(ji)}_g,2_{\bar q})
+  (1\leftrightarrow 2)\right\}
\,\Bigg)
\nonumber \\
&& \hspace{1cm}
\times
{A}^0_{3}(\widetilde{(1ij)}_q,5_g,\widetilde{(2ji)}_{\bar q})
\,
{J}_{3}^{(3)}(\widetilde{p_{1ij}},\widetilde{p_{2ji}},p_5) \nonumber \\
\ph{(f,g)} &&- \frac{1}{2} \Bigg\{
\Bigg( \tilde{E}_4^0(1_q,i_{q'},j_{\bar q'},5_g) 
 - A_3^0(i_{q'},5_g,j_{\bar q'})
 E_3^0(1_{q},\widetilde{(i5)}_{q'},\widetilde{(j5)}_{\bar q'})\Bigg)
\nonumber \\
&& \hspace{1cm}
\times
{A}^0_{3}(\widetilde{(1i5)}_q,\widetilde{(j5i)}_g,2_{\bar q})\, 
{J}_{3}^{(3)}(\widetilde{p_{1i5}},p_2,\widetilde{p_{j5i}})
+ (1\leftrightarrow 2)
\Bigg\}  \nonumber \\
\ph{(h)} &&+ \frac{1}{2} \Bigg\{ E_3^0(1_q,i_{q'},j_{\bar q'}) \,
A_3^0(\widetilde{(1j)_q},5_g,2_{\bar q})\, 
A^0_{3}(\widetilde{((1j)5)}_q,\widetilde{(ij)}_g,
\widetilde{(25)}_{\bar q})\, {J}_{3}^{(3)}
(\widetilde{p_{(1j)5}},\widetilde{p_{25}},\widetilde{p_{ij}}) 
\nonumber \\
&& \hspace{1cm}
+ (1\leftrightarrow 2) \Bigg\} \Bigg\}
\label{eq:nnlosubcfnf}
\end{eqnarray}

\subsection{Four-parton contribution}
The four parton contribution to the $N_F/N$ colour factor reads:
\begin{eqnarray}
{\rm d}\sigma_{NNLO,N_F/N}^{V,1}&=&
N_{{4}}\,\frac{N_F}{N}\, \left(\frac{\alpha_s}{2\pi}\right)\,
{\rm d}\Phi_{4}(p_{1},\ldots,p_{4};q)\, \nonumber \\ &&
\Bigg\{ - \frac{1}{2}\sum_{(i,j)\in (3,4)}
\Big( B_{4}^{1,b} (1_q,i_{q'},j_{\bar q'},2_{\bar q}) 
+ 2 C_{4}^{1,f} (1_q,i_{q},j_{\bar q},2_{\bar q})\nonumber \\
&& \hspace{12mm} +
\tilde A_{4}^{1,c} (1_q,i_g,j_g,2_{\bar q}) \Big)
\Bigg\}
\,
\JET_{3}^{(4)}(p_{1},\ldots,p_{4})
\,.
\label{eq:sigvE}
\end{eqnarray}
Like in the $N_F\, N$ colour factor, the expression is symmetrised 
over the momenta $(3)$ and $(4)$ to remove terms which are 
antisymmetric under charge conjugation, and can not be accounted for 
properly by the quark-gluon antenna functions.

The corresponding subtraction term is:
\begin{eqnarray}
\lefteqn{{\rm d}\sigma_{NNLO,N_F/N}^{VS,1}
= 
 N_{{4}}\,\frac{N_F}{N}\, \left(\frac{\alpha_s}{2\pi}\right)\,
 {\rm d}\Phi_{4}(p_{1},\ldots,p_{4};q)}\nonumber \\
\ph{(a,b)} &\times& \Bigg\{
 \left[ {\cal A}_3^0(s_{12}) + {\cal A}_3^0(s_{34}) \right]\,    
B_{4}^{0} (1_q,3_{q'},4_{\bar q'},2_{\bar q})\,
\JET_{3}^{(4)}(p_{1},\ldots,p_{4}) \nonumber \\ 
\ph{(c)}&&
+ \frac{1}{4} \left[ {\cal E}_3^0(s_{13}) + 
{\cal E}_3^0(s_{14}) + {\cal E}_3^0(s_{23})+{\cal E}_3^0(s_{24}) \right]\,    
\tilde{A}_{4}^{0} (1_q,3_{g},4_{g},2_{\bar q})
\,\JET_{3}^{(4)}(p_{1},\ldots,p_{4})  \nonumber \\ 
\ph{(d)} &&-
\frac{1}{2}\,\bigg\{\bigg(
E_3^0(1_q,3_{q'},4_{\bar q'})\left[
\tilde{A}_3^1(\widetilde{(13)}_q,\widetilde{(43)}_g,2_{\bar q})
+{\cal A}_2^1(s_{1234})
A_3^0(\widetilde{(13)}_q,\widetilde{(43)}_g,2_{\bar q})\right]
\nonumber \\
\ph{(e)} &&\hspace{4mm}+ {\cal A}_3^0(s_{(\widetilde{13})2}) \,
E_3^0(1_q,3_{q'},4_{\bar q'}) 
\,{A}_3^0(\widetilde{(13)}_q,\widetilde{(43)}_g,2_{\bar q}) \nonumber \\
\ph{(f,g)} &&\hspace{4mm}+\left[\tilde{E}_3^1(1_q,3_{q'},4_{\bar q'})
+ {\cal A}_3^0(s_{34}) {E}_3^0(1_q,3_{q'},4_{\bar q'})\right] \,
A_3^0(\widetilde{(13)}_q,\widetilde{(43)}_g,2_{\bar q})\bigg)
\JET_{3}^{(3)}(\widetilde{p_{13}},\widetilde{p_{43}},p_2) \nonumber \\
&&\hspace{4mm}
+ (1\leftrightarrow 2)\bigg\}
\nonumber \\
\ph{(h,i)} &&-
\frac{1}{2}\,\sum_{(i,j)\in (3,4)} \bigg(
A_3^0(1_q,i_{g},2_{\bar q})
\hat A_3^1(\widetilde{(1i)}_q,j_g,\widetilde{(2i)}_{\bar q})
+\bigg[\hat A_3^1(1_q,i_{g},2_{\bar q})
\nonumber \\
\ph{(j)} 
&&\hspace{6mm}+\frac{1}{2}\left({\cal E}_3^0(s_{1i})+{\cal E}_3^0(s_{1j})
+{\cal E}_3^0(s_{2i})+{\cal E}_3^0(s_{2j})\right)
A_3^0(1_q,i_{g},2_{\bar q})
\bigg]\,
A_3^0(\widetilde{(1i)}_q,j_g,\widetilde{(2i)}_{\bar q})\nonumber \\
\ph{(k)} &&
\hspace{4mm}+ b_{0,F} \log \frac{q^2}{s_{12i} }A_3^0(1_q,i_{g},2_{\bar q})
A_3^0(\widetilde{(1i)}_q,j_g,\widetilde{(2i)}_{\bar q})\bigg)
\JET_{3}^{(3)}(\widetilde{p_{1i}},\widetilde{p_{2i}},p_j)
\Bigg\}
\end{eqnarray}
\ph{The cancellations are as follows: (5a,4a), (5b,4b), (5c,4c), (5e,4j),
(5g,4g), (5h,4j). The terms (5d,5f,4d,4e,4f,4h,4i,4k) integrate to 
the three-parton channel. }

\subsection{Three-parton contribution}
The three parton contribution to the $N_F/N$ colour factor consists of the 
three-parton virtual two-loop correction and the integrated 
five-parton tree-level and 
four-parton one-loop subtraction terms, which read
\begin{eqnarray}
\lefteqn{{\rm d}\sigma_{NNLO,N_F/N}^{S}+{\rm d}\sigma_{NNLO,N_F/N}^{VS,1} = 
\frac{N_F}{N}} \,\nonumber \\
&& \times \Bigg\{  -  \bigg[{\cal B}_4^0  (s_{12})
+ \frac{1}{2}\, \tilde{{\cal E}}_4^0  (s_{13}) +
 \frac{1}{2}\,\tilde{{\cal E}}_4^0  (s_{23}) + \frac{1}{2}
{\cal A}_3^0  (s_{12}) \, \left({\cal E}_3^0  (s_{13})
+ {\cal E}_3^0  (s_{23}) \right)
\nonumber \\ &&
+  \hat{{\cal A}}_3^1  (s_{12})
+  \frac{1}{2}\, \tilde{{\cal E}}_3^1  (s_{13}) +  \frac{1}{2}\, 
 \tilde{{\cal E}}_3^1  (s_{23}) 
\bigg] 
\,A_3^0({1}_q,{3}_g, 2_{\bar q}) - \frac{1}{2}
 \left( {\cal E}_3^0(s_{13}) +
{\cal E}_3^0(s_{23}) \right) \, 
\tilde{A}_3^1(1_q,3_g
,2_{\bar q})\nonumber \\ 
&&
- {\cal A}_3^0(s_{12})  \, 
\hat{A}_3^1(1_q,3_g,2_{\bar q})
- \frac{b_{0,F}}{\e} \,  
{\cal A}_3^0(s_{12})
\left(({s_{12}})^{-\e} -  ({s_{123}})^{-\e}\right) \,A_3^0({1}_q,{3}_g,
2_{\bar q}) 
\Bigg\} \d \sigma_3\;.
\end{eqnarray}

Taking the infrared pole part of this expression, we obtain
cancellation of all infrared poles in this channel:
\begin{equation}
\Poles\left({\rm d}\sigma_{NNLO,N_F/N}^{S}\right) +
\Poles\left({\rm d}\sigma_{NNLO,N_F/N}^{VS,1}\right) +
\Poles\left({\rm d}\sigma_{NNLO,N_F/N}^{V,2}\right) = 0 \,.
\end{equation}

\section{Construction of the $N_F^2$ colour factor}
\label{sec:termF}
The 
 $N_F^2$ colour factor receives contributions only from the four-parton 
one-loop process $\gamma^*\to q\bar q q'\bar q'$ and from the 
three-parton two-loop process  $\gamma^*\to q\bar q g$. 

This colour factor is also part of the QED-type corrections, 
described
previously in~\cite{our3j:QED}.

\subsection{Four-parton contribution}
The four-parton one-loop contribution to this colour factor is 
\begin{eqnarray}
{\rm d}\sigma_{NNLO,N_F^2}^{V,1}&=&
N_{{4}}\,N_F^2\, \left(\frac{\alpha_s}{2\pi}\right)\,
{\rm d}\Phi_{4}(p_{1},\ldots,p_{4};q)\,
B_{4}^{1,c} (1_q,3_{q'},4_{\bar q'},2_{\bar q})\; 
\JET_{3}^{(4)}(p_{1},\ldots,p_{4}).\hspace{4mm}
\label{eq:sigvNF2}
\end{eqnarray}
This contribution is free of explicit infrared poles (as can be inferred 
from the absence of a five-parton contribution to this colour structure). 

The subtraction term corresponding to this contribution is 
\begin{eqnarray}
\label{eq:VS1NF2}
{\rm d}\sigma_{NNLO,N_F^2}^{VS,1}
&= &  N_{{4}}\, N_F^2\, \left(\frac{\alpha_s}{2\pi}\right)\,
{\rm d}\Phi_{4}(p_{1},\ldots,p_{4};q)\,
\frac{1}{2}\,\Bigg\{ \nonumber \\
&& 
\bigg[\left(
\hat E_3^1(1_q,3_{q'},4_{\bar q'})
+   b_{0,F} \, \log\frac{q^2}{s_{134}} \,
E_3^0(1_q,3_{q'},4_{\bar q'})\right)
\,
A_3^0(\widetilde{(13)}_q,\widetilde{(43)}_g,
2_{\bar q}) 
\nonumber \\ && \hspace{3mm} + E_3^0(1_q,3_{q'},4_{\bar q'})\,
\hat{A}_3^1(\widetilde{(13)}_q,\widetilde{(43)}_g
,2_{\bar q})
\bigg] \JET_{3}^{(3)}(\widetilde{p_{13}},\widetilde{p_{43}},p_2)
\nonumber \\ &+&
\bigg[ \left(\hat E_3^1(2_{\bar q},3_{q'},4_{\bar q'})
+  b_{0,F}\, \log\frac{q^2}{s_{234}}\,
E_3^0(2_{\bar q},3_{q'},4_{\bar q'})\right)
\,A_3^0(1_q,\widetilde{(43)}_g,
\widetilde{(23)}_{\bar q}) \nonumber \\ &&
\hspace{3mm} + E_3^0(2_{\bar q},3_{q'},4_{\bar q'})\,
\hat{A}_3^1(1_q,\widetilde{(43)}_g,\widetilde{(23)}_{\bar q})
 \bigg]
\JET_{3}^{(3)}(p_1,\widetilde{p_{43}},\widetilde{p_{23}})\Bigg\}\;.
\end{eqnarray}
Although $\hat E_3^1$ and $\hat{A}_3^1$ contain explicit infrared poles, 
these cancel in their sum, as can be seen from (5.16) and (6.32) 
of~\cite{ourant}. ${\rm d}\sigma_{NNLO,N_F^2}^{VS,1}$ is therefore free of 
explicit infrared poles.

\subsection{Three-parton contribution}

The three parton contribution to the $N_F^2$ colour factor consists of the 
three-parton virtual two-loop correction and the integrated 
four-parton one-loop subtraction term, which reads
\begin{eqnarray}
\lefteqn{{\rm d}\sigma_{NNLO,N_F^2}^{VS,1} = 
N_F^2\, 
\frac{1}{2}} \nonumber \\
&\times& 
\bigg[\left(
\hat{\cal E}_3^1(s_{13}) + \hat{\cal E}_3^1(s_{23})\right)
\,A_3^0({1}_q,{3}_g, 2_{\bar q}) + 
 \left( {\cal E}_3^0(s_{13}) +
{\cal E}_3^0(s_{23}) \right) \, 
\hat{A}_3^1(1_q,3_g
,2_{\bar q})\nonumber \\ 
&&
+ \frac{b_{0,F}}{\e} \, \left[ 
{\cal E}_3^0(s_{13})
\left(({s_{13}})^{-\e} -  ({s_{123}})^{-\e}\right) +
{\cal E}_3^0(s_{23})
\left(({s_{23}})^{-\e} -  ({s_{123}})^{-\e}\right) \right] A_3^0({1}_q,{3}_g,
2_{\bar q}) 
\bigg] \d \sigma_3\;.\nonumber \\
\end{eqnarray}

Combining the infrared poles of this expression with the two loop
matrix element, we obtain the cancellation of all 
infrared poles in this colour factor,
\begin{equation}
\Poles\left({\rm d}\sigma_{NNLO,N_F^2}^{VS,1}\right) +
\Poles\left({\rm d}\sigma_{NNLO,N_F^2}^{V,2}\right) = 0 \,.
\end{equation}

\section{Construction of the $N_{F,\gamma}$ colour factor}
\label{sec:termG}
The $N_{F,\gamma}$ colour factor comes from the interference of amplitudes 
in which  the external vector boson couples to different quark lines. It 
receives contributions from five-parton 
tree-level $\gamma^* \to q \bar q q' \bar q' g$, four-parton one-loop 
$\gamma^* \to q \bar q q'\bar q'$ and $\gamma^* \to q \bar q gg$  
as well as three-parton two-loop $\gamma^* \to q \bar q g$ and 
$\gamma^* \to ggg$. This colour factor is absent in three-jet production at 
NLO and four-jet production at LO 
because of Furry's theorem~\cite{ERT}, and its contribution to four-jet 
production at NLO is numerically tiny~\cite{dixonsigner}. The numerical 
magnitude of this term in the two-loop corrections to the three-parton channel 
is equally very small~\cite{3jme,nigeljochum}. A detailed discussion of this 
colour factor, and of the effects leading to its numerical suppression is 
contained in~\cite{dixonsigner}.  
All partonic 
channels contributing to this colour factor are individually finite. 

In this section, we document this colour factor for completeness. Given that 
is numerical impact can be safely expected to be negligible, we refrain 
from its implementation.

\subsection{Five-parton contribution}
The NNLO radiation term appropriate for the three jet final state is given by
\begin{eqnarray}
\lefteqn{{\rm d}\sigma_{NNLO,N_{F,\gamma}}^R= 
N_{{5}}\, N_{F,\gamma} \,  {\rm d}\Phi_{5}(p_{1},\ldots,p_{5};q) }\nonumber \\
&&\times \Big[ -N
\left(
 \hat B_5^{0,a}(1_q,5_g,4_{\bar q'};3_{q'},2_{\bar q})
+\hat B_5^{0,b}(1_q,4_{\bar q'};3_{q'},5_g,2_{\bar q})
-\hat B_5^{0,e}(1_q,4_{\bar q'};3_{q'},2_{\bar q},5_g) \right) \nonumber \\
&& \hspace{6mm}
+ \frac{1}{N}
\left(
 \hat B_5^{0,c}(1_q,5_g,2_{\bar q};3_{q'},4_{\bar q'})
+\hat B_5^{0,d}(1_q,2_{\bar q};3_{q'},5_g,4_{\bar q'})
+\hat B_5^{0,e}(1_q,2_{\bar q};3_{q'},4_{\bar q'};5_g) \right) \Big]
\nonumber \\ &&  
\times J_3^{(5)}(p_1,\ldots,p_5)\;.
\label{eq:sigrG}\end{eqnarray}
Once symmetrised  over the quark and antiquark momenta, this term can be 
integrated safely without the need for an infrared subtraction. It 
is free from infrared singularities associated with gluon $5_g$ unresolved, 
since the corresponding four-parton tree-level term $\hat{B}_4^0$ 
vanishes after symmetrisation over the quark and antiquark momenta. 
Double unresolved singularities can not appear since there is no tree-level 
three-parton process proportional to $N_{F,\gamma}$.

\subsection{Four-parton contribution}
The four parton contribution to the $N_{F,\gamma}$ colour factor reads:
\begin{eqnarray}
{\rm d}\sigma_{NNLO,N_{F,\gamma}}^{V,1}&=&
N_{{4}}\,N_{F,\gamma}\, \left(\frac{\alpha_s}{2\pi}\right)\,
{\rm d}\Phi_{4}(p_{1},\ldots,p_{4};q)\, \nonumber \\ &&
\times \Bigg\{
N  \hat B_4^{1,a}(1_q,3_q,4_{\bar q},2_{\bar q})
- \frac{1}{N} \hat B_4^{1,b}(1_q,3_q,4_{\bar q},2_{\bar q})
\nonumber \\ && \hspace{6mm}
+N\, A_{4}^{1,e} (1_q,i_g,j_g,2_{\bar q})
- \frac{1}{N}
\tilde A_{4}^{1,e} (1_q,3_g,4_g,2_{\bar q})
\Bigg\}
\,
\JET_{3}^{(4)}(p_{1},\ldots,p_{4})
\,.\nonumber \\ 
\label{eq:sigvG}
\end{eqnarray}
Terms which vanish  under symmetrisation of the quark and antiquark momenta,
arising from $\hat{B}_4^{1,c}$ in (\ref{eq:4q1lall})
have been omitted here. After this symmetrisation, all explicit infrared poles 
present in individual terms in the above expression cancel. Moreover, 
(\ref{eq:sigvG}) is finite in all single unresolved limits, such that 
no antenna subtraction is needed. 

\subsection{Three-parton contribution}
The three-parton contribution to the $N_{F,\gamma}$ 
colour factor consists of the 
three-parton virtual two-loop correction to $\gamma^*\to q\bar q g$~\cite{3jme}
and the one-loop squared correction to 
$\gamma^* \to ggg$~\cite{nigeljochum}. Both are individually finite, and 
were shown to be numerically tiny. Since no subtractions were carried out in 
the five-parton and four-parton channels, there are no integrated subtraction 
terms in the three-parton channel.

\section{Numerical implementation}
\label{sec:numer}

Using the matrix elements and 
antenna subtraction terms derived in the 
previous sections, NNLO corrections to 
any infrared-safe three-jet observable in $e^+e^-$ annihilation 
(jet cross section, 
event shape variable) can be computed numerically. We implemented this 
numerical evaluation into a parton-level event generator program,
which we name {\tt EERAD3}. 

This program is based on the 
program {\tt EERAD2}~\cite{cullen}, which computes four-jet
production at NLO. {\tt EERAD2}
 contained already the five-parton and four-parton 
matrix elements relevant here, as well as the NLO-type subtraction terms 
${\rm d} \sigma_{NNLO}^{S,a}$ and ${\rm d} \sigma_{NNLO}^{VS,1,a}$. 

The implementation contains three channels, classified 
by their partonic multiplicity: 
\begin{itemize}
\item
in the five-parton channel, we
integrate
\begin{equation}
{\rm d}\sigma_{NNLO}^{R} - {\rm d}\sigma_{NNLO}^{S}\;.
\end{equation}
\item in the four-parton channel, we integrate
\begin{equation}
{\rm d}\sigma_{NNLO}^{V,1} - {\rm d}\sigma_{NNLO}^{VS,1}\;.
\end{equation}
\item in the three-parton channel, we integrate
\begin{equation}
{\rm d}\sigma_{NNLO}^{V,2} +{\rm d}\sigma_{NNLO}^{S}
+ {\rm d}\sigma_{NNLO}^{VS,1}\;.
\end{equation}
\end{itemize}
The numerical integration over these channels is carried out by Monte Carlo 
methods using the {\tt VEGAS}~\cite{vegas} implementation. The structure of 
the programme is displayed in Figure~\ref{fig:mc}.
\FIGURE[t]{
\caption{Structure of the EERAD3 parton-level Monte Carlo event generator 
programme.}
\label{fig:mc}
\epsfig{file=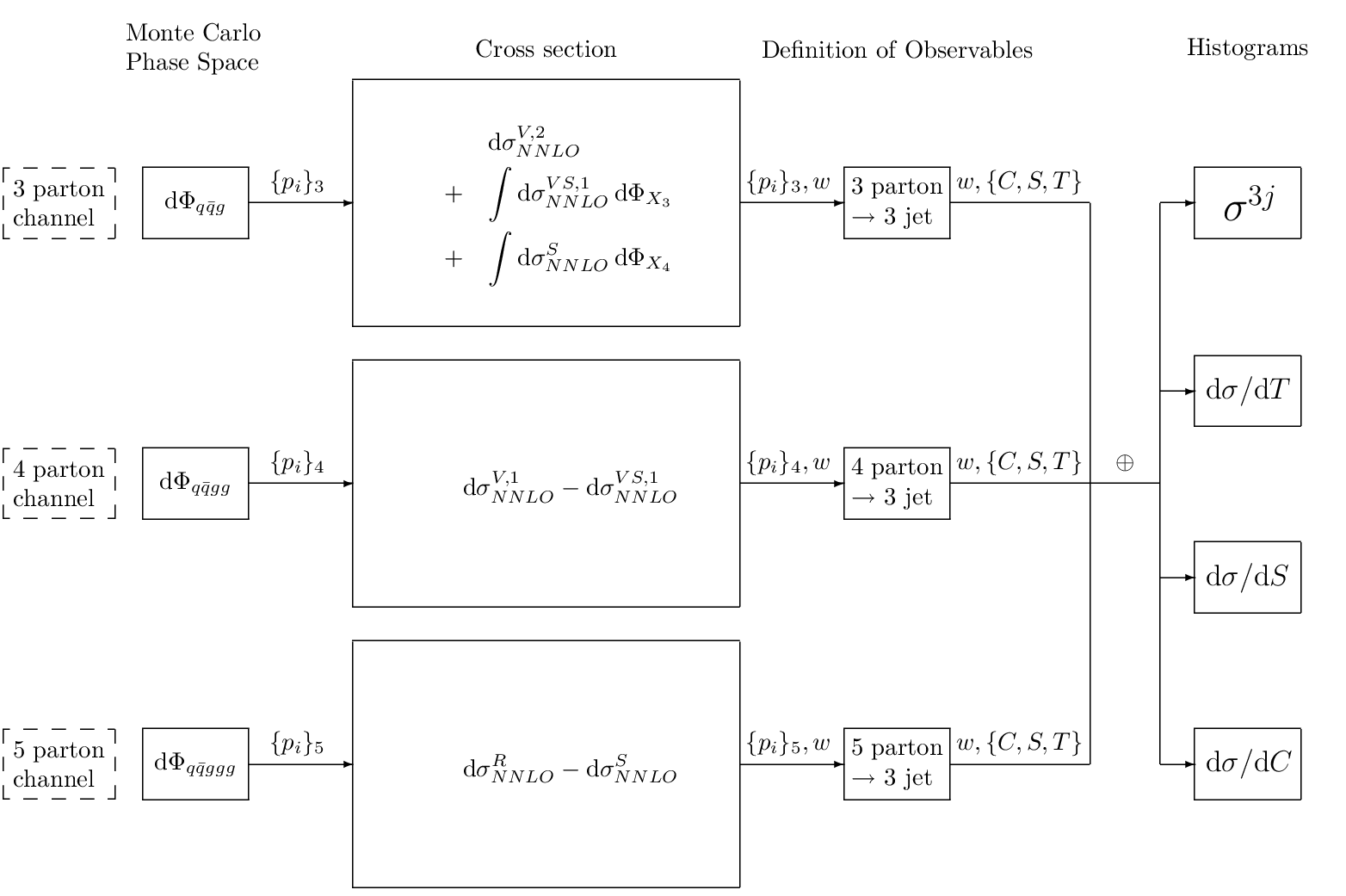,width=14cm}
}

The phase space in the 
four-parton and five-parton channel is decomposed into wedges which are 
constructed such that two of the invariants are smaller than any of 
the other 
invariants. This decomposition allows an optimal generation of phase space 
points in the unresolved limits. The full four-parton phase space is 
obtained by summing (a) 12 wedges with ($s_{ij},s_{ik}$) smallest, plus (b) 3
wedges with ($s_{ij},s_{kl}$) smallest. To obtain the 
 full five-parton phase space, we sum (a) 30 wedges with 
($s_{ij},s_{ik}$) smallest, and (b) 15 wedges with ($s_{ij},s_{kl}$) smallest.
The phase space integration in either channel is carried out 
by integrating only over a single wedge of type  (a) and a single wedge
of type (b), while summing the integrands appropriate to all wedges of 
the given type. In doing this summation, we combine (in the exact unresolved 
limits) phase space points which are related to each other by a rotation of 
the system of unresolved partons, thereby largely
cancelling the angular-dependent terms. In all colour factors 
containing angular-dependent terms, the combination of phase space wedges 
yields a substantial improvement of the numerical stability of the results. 

It was already demonstrated above that the integrands in the four-parton and 
three-parton channel are free of explicit infrared poles. In the 
five-parton and four-parton channel, we tested the proper implementation of 
the subtraction by generating trajectories of phase space points approaching 
a given single or double unresolved limit using the {\tt RAMBO}~\cite{rambo}
phase space generator. Along these trajectories, we observe that the 
antenna subtraction terms converge towards the physical matrix 
elements, and that the cancellations among individual 
contributions to the subtraction terms take place as expected in 
the antenna subtraction method.

Moreover, we checked the correctness of the 
subtraction by introducing a 
lower cut (slicing parameter) $y_0$ 
on all phase space variables, and observing 
that our results are independent of this cut (provided it is 
chosen small enough). This behaviour indicates that the 
subtraction terms ensure that the contribution of potentially singular 
regions of the final state phase space does not contribute to the numerical 
integrals, but is accounted for analytically.

\section{Thrust distribution as an example}
\label{sec:thrust}

To illustrate the implementation and to study the numerical impact of the
individual NNLO contributions, we consider the thrust distribution. 
We already reported the NNLO results on this event shape distribution 
in a previous paper~\cite{ourthrust}, where the phenomenological 
implications are discussed in detail.

The thrust variable for a hadronic final state in $e^+e^-$ annihilation is 
defined as~\cite{farhi} 
\begin{equation}
T=\max_{\vec{n}}
\left(\frac{\sum_i |\vec{p_i}\cdot \vec{n}|}{\sum_i |\vec{p_i}|}\right)\,,
\end{equation}
where $p_i$ denotes the three-momentum of particle $i$, with the sum running 
over all particles. The unit vector $\vec{n}$ is varied to find  the 
thrust direction $\vec{n}_T$ which maximises the expression in parentheses 
on the right hand side.

It can be seen that a two-particle final state has fixed $T=1$,
consequently the thrust distribution receives its first non-trivial 
contribution from three-particle final states, which, at order $\alpha_s$, 
correspond to three-parton final states. Therefore, 
both theoretically and experimentally, the thrust distribution is 
closely related to three-jet production.

A study of the phenomenological implications of the NNLO corrections to 
the thrust distribution was presented in~\cite{ourthrust}, illustrating that 
the NNLO corrections amount to about 15\% of the total result 
over the experimentally relevant range $0.02<1-T<0.3$, and that 
inclusion of these corrections results in a considerable stabilization of the 
renormalisation scale dependence of the theoretical prediction. 
In the present context, we use the thrust distribution only as 
an example to 
illustrate certain features of our calculation.

The three-jet rate and 
event shapes related to it can be expressed in 
perturbative QCD by dimensionless coefficients.
These coefficients depend, for 
non-singlet QCD corrections, only on the jet resolution parameter 
(respectively on the event shape variable). Typically, one denotes these
coefficients by $A,B,C,\ldots$ at LO, NLO, NNLO, etc.

The perturbative expansion of thrust  distribution up to NNLO 
for renormalisation scale $\mu^2 = s$ and 
$\alpha_s\equiv \alpha_s(s)$  is then given by
\begin{eqnarray}
\frac{1}{\sigma^{{\rm had}}}\, \frac{\d\sigma}{\d T} &=& 
\left(\frac{\alpha_s}{2\pi}\right) \frac{\d \bar A}{\d T} +
\left(\frac{\alpha_s}{2\pi}\right)^2 \frac{\d \bar B}{\d T}
+ \left(\frac{\alpha_s}{2\pi}\right)^3 
\frac{\d \bar C}{\d T} \,.
\label{eq:NNLO}
\end{eqnarray}
Here we define the effective coefficients in terms of the 
perturbatively calculated  coefficients $A$, $B$ and $C$, which are 
all normalised to the tree-level cross section 
\begin{equation}
\sigma_0 = \frac{4 \pi \alpha}{3 s} N \, e_q^2\;.
\end{equation}
 for $e^+e^- \to q \bar q$. Using 
  \begin{equation}
  \sigma_{\rm{had}}=\sigma_0\,
\left(1+\frac{3}{2}C_F\,\left(\frac{\alpha_s}{2\pi}\right)
+K_2\,\left(\frac{\alpha_s}{2\pi}\right)^2+{\cal O}(\alpha_s^3)\,
\right) \;,
\end{equation}
with ($C_F=(N^2-1)/(2N)$, $C_A=N$, $T_R=1/2$ for $N=3$ colours and $N_F$ 
light quark flavours)
\begin{equation}
  K_2=\frac{1}{4}\left[- \frac{3}{2}C_F^2
+C_FC_A\,\left(\frac{123}{2}-44\zeta_3\right)+C_FT_RN_F\,(-22+16\zeta_3)
 \right] \;,
\end{equation}
we obtain:
\begin{eqnarray}
\bar{A} &=& A\;,\nonumber \\
\bar{B} &=& B - \frac{3}{2}C_F\,A\;,\nonumber \\
\bar{C} &=& C -  \frac{3}{2}C_F\,B+ \left(\frac{9}{4}C_F^2\,-K_2\right)\,A 
\;.\label{eq:ceff}
\end{eqnarray}
These coefficients depend only on the 
jet resolution parameter or the event shape variable 
under consideration, and are independent of electroweak 
couplings, centre-of-mass energy and renormalisation scale. 

The above coefficients include only QCD corrections with non-singlet 
quark couplings. At ${\cal O}(\alpha_s^2)$, these amount to the full 
corrections, while the ${\cal O}(\alpha_s^3)$ corrections also 
receive a singlet contribution. 
As discussed above, this singlet contribution 
arises from the interference of diagrams where the external gauge boson 
couples to different quark lines. In four-jet observables at 
 ${\cal O}(\alpha_s^3)$, these singlet contributions were found to be 
extremely small~\cite{dixonsigner}.
Also, the singlet contribution from three-gluon final states 
to three-jet observables was found to be  negligible~\cite{nigeljochum}.

\FIGURE[t]{
\epsfig{file=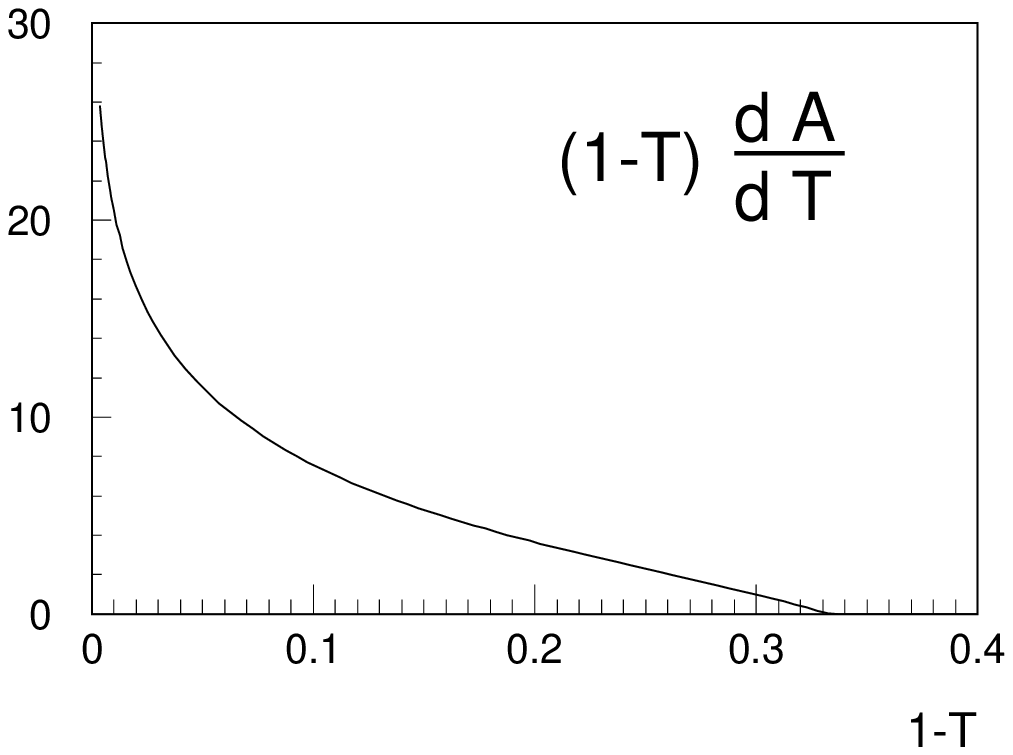,angle=-90,width=7cm} 
\epsfig{file=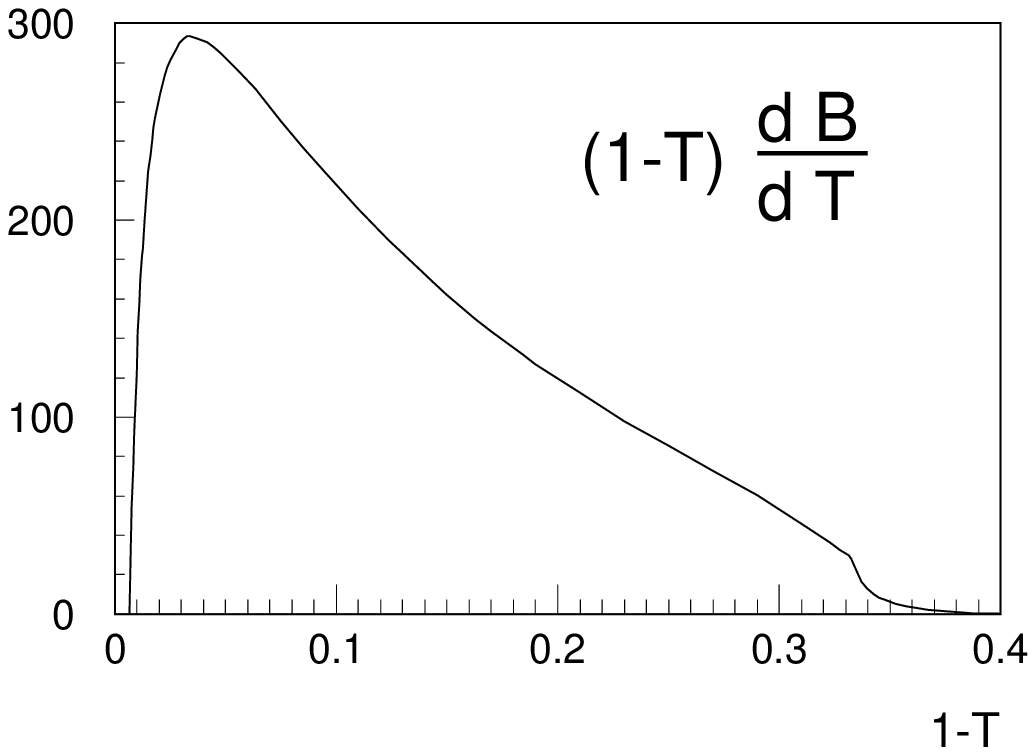,angle=-90,width=7cm} 
\caption{Coefficients of the leading order and next-to-leading order 
contributions to the thrust distributions.}
\label{fig:nlo}}
\FIGURE[t]{
\parbox{12cm}{\begin{center}
\epsfig{file=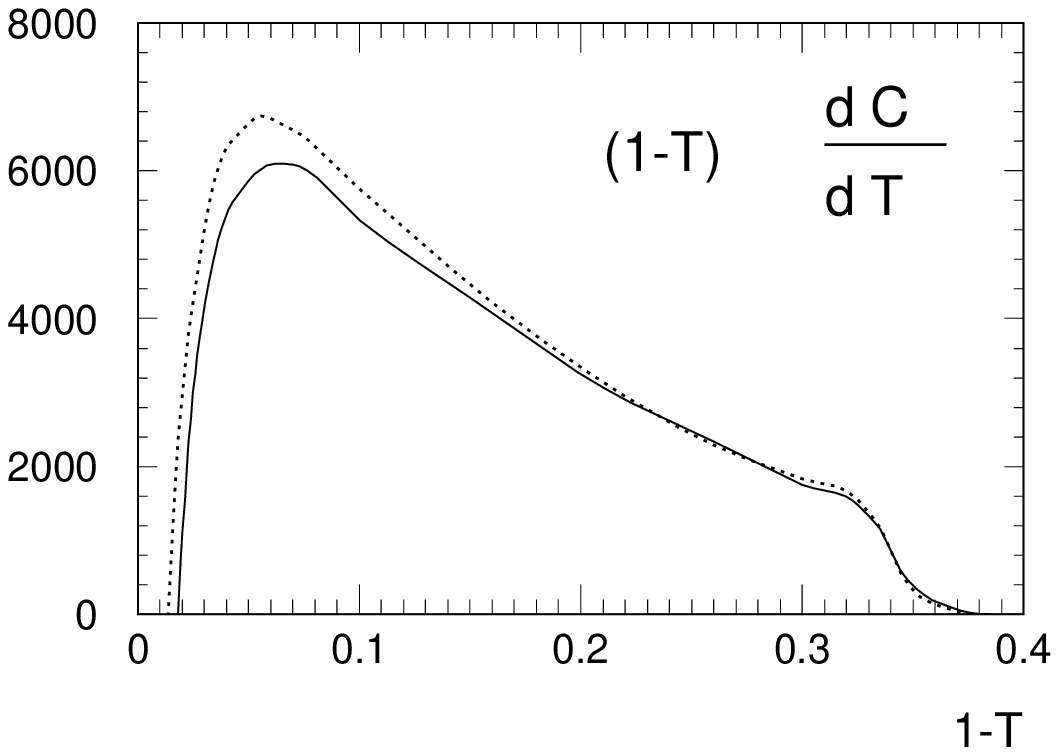,angle=-90,width=7cm}\end{center}} 
\caption{Coefficient of the next-to-next-to-leading order 
contribution to the thrust distribution. Solid: corrected
for large-angle soft terms;
dotted: original result.}
\label{fig:nnlo}}
We determine $A,B,C$ from the perturbative contributions to the 
differential cross section, normalised to the tree-level hadronic cross
section:
\begin{equation}
\frac{\d A}{\d T} = \frac{1}{\sigma_0}\, 
\frac{\d \sigma_{LO}}{\d T} \, ,\qquad
\frac{\d B}{\d T} = \frac{1}{\sigma_0}\, 
\frac{\d \sigma_{NLO}}{\d T} \, ,\qquad
\frac{\d C}{\d T} = \frac{1}{\sigma_0}\, 
\frac{\d \sigma_{NNLO}}{\d T} \, .
\end{equation}
For the determination of the non-singlet coefficients, it is sufficient to 
consider $\sigma_0$ for pure photon exchange, since any electroweak coupling
constant cancels out in the above ratio.
The LO and NLO coefficients $A(T)$ and $B(T)$ were computed in the 
literature long ago~\cite{ERT,ert2,ert3,kn,cs}. 
They are displayed for comparison 
in Figure~\ref{fig:nlo}.

\FIGURE[t]{
\epsfig{file=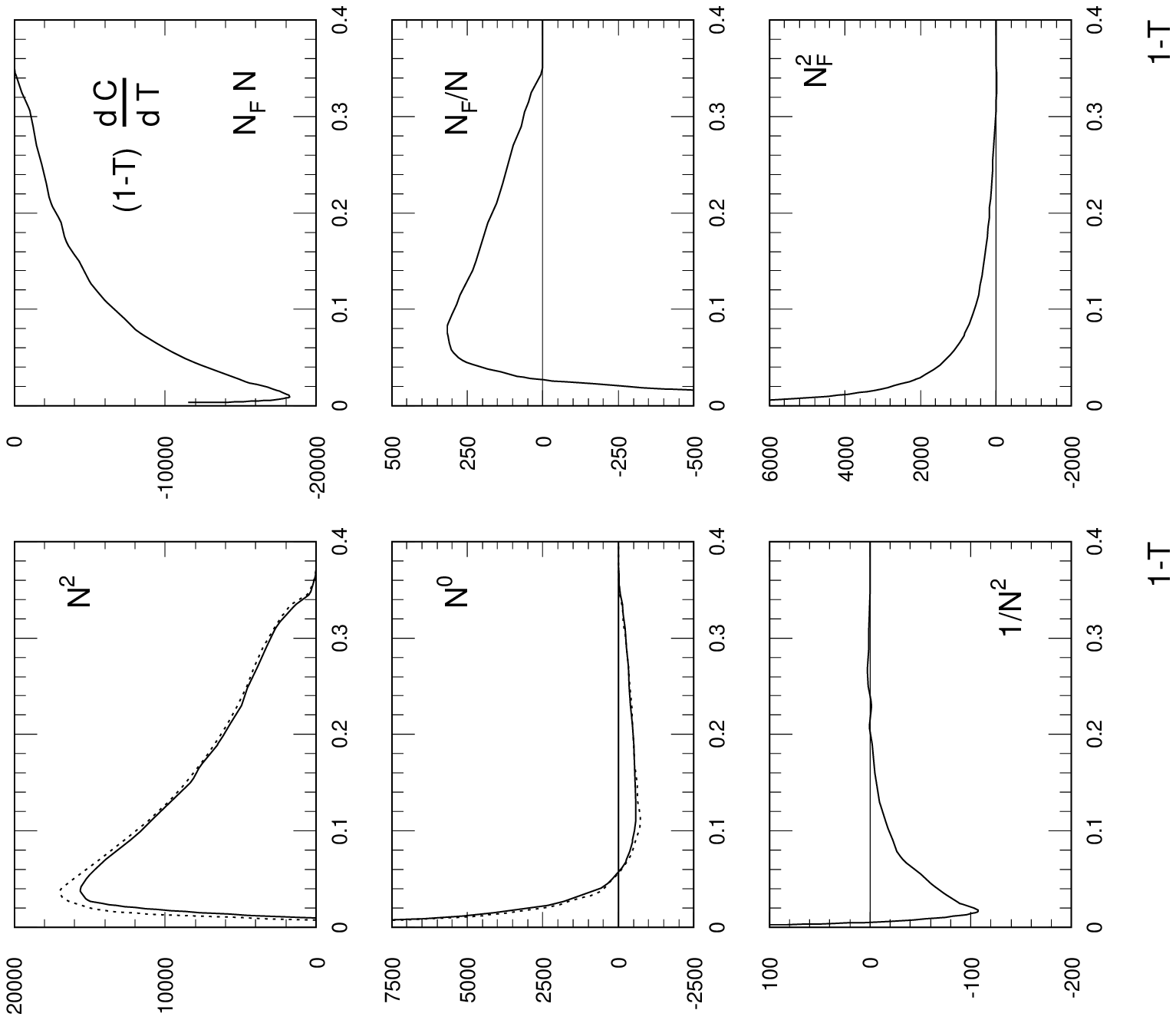,angle=-90,width=15cm} 
\caption{Different colour factor contributions to NNLO coefficient of the 
thrust distribution. In $N^2$ and $N^0$ colour factors: solid: corrected
for large-angle soft terms;
dotted: original result.}
\label{fig:icol}}
The total NNLO 
coefficient $C(T)$ is displayed in Figure~\ref{fig:nnlo}.
The six different colour factor contributions 
to it are shown in Figure~\ref{fig:icol}. It can be seen that the 
numerically dominant contributions come from the $N^2$ and $N_F\, N$ colour 
factors. The contributions of these two colour factors are of opposite sign, 
with $N^2$ being of larger absolute magnitude, thus resulting in a total 
positive result. Contributions at the 10\% level of the total come from 
the $N_F^2$ and $N^0$ colour factors, $N_F/N$ amounts to  about 5\%, while 
the most subleading $1/N^2$ colour factor is below 1\%.  
\FIGURE[t]{
\epsfig{file=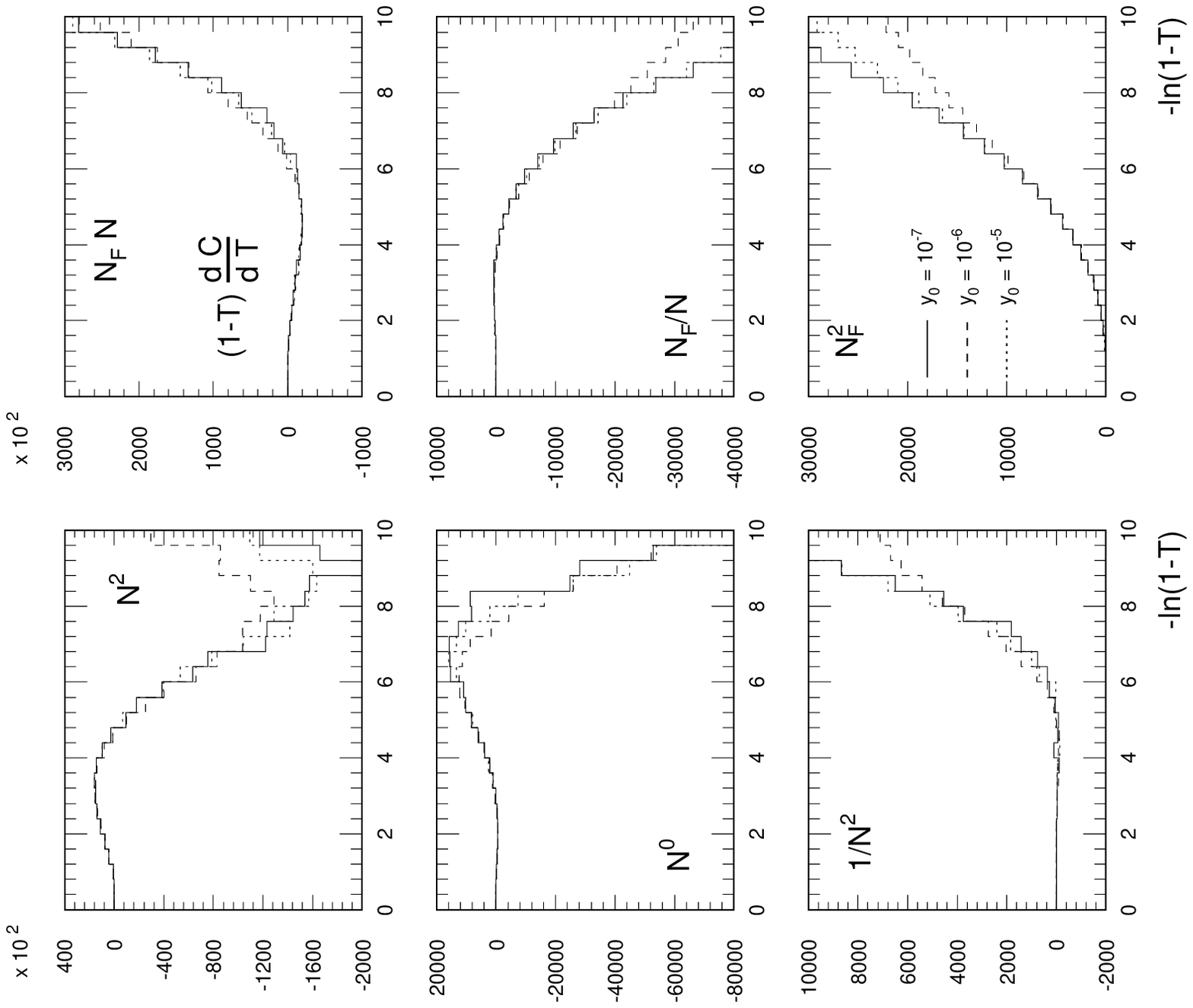,angle=-90,width=15cm} 
\caption{Dependence on phase space cut $y_0$ in different colour factors.
We see that the results are independent of $y_0$ for $-\ln(1-T) < 4$, but, as
explained in the text, differ at larger values of $-\ln(1-T)$}
\label{fig:y0}}

To illustrate the independence of our results on $y_0$, we display the 
different colour factor contributions to $C(T)$ as function of 
$-\ln(1-T)$ in Figure~\ref{fig:y0} for different values of the phase space 
cut $y_0= 10^{-5},10^{-6},10^{-7}$. 
By rescaling all phase space invariants to the total 
centre-of-mass energy squared, $y_0$ becomes dimensionless. 
Since the value of $(1-T)$ determines the typical scale of the smallest resolved 
invariant, one must require $y_0$ to be several orders of magnitude smaller 
than $(1-T)$ for the cancellation between matrix element and subtraction term 
to be accurate. 
Figure~\ref{fig:y0}  shows very clearly that over the phenomenologically relevant 
range, i.e. $0.02<1-T$ or equivalently, $-\ln(1-T) < 4$,
our results do not depend on $y_0$. As $(1-T)$ approaches $y_0$
(starting at about $(1-T)\approx {\cal O}(1000) y_0$), 
the calculation becomes unreliable as expected. This behaviour can be 
understood to arise from the fact that the subtraction terms converge 
to the full matrix element only once all unresolved invariants are much 
smaller than any of the resolved invariants. 

The numerical convergence of our calculation deteriorates for lower values 
of $y_0$ for two reasons.
\begin{itemize}
\item the 
absolute magnitude of matrix elements and subtraction terms increases for 
decreasing $y_0$ both in the 
 five-parton and four-parton channel. Consequently, numerical cancellations 
between matrix elements and subtraction terms happen over larger orders of 
magnitude, thereby enhancing numerical rounding errors. 
\item the four-parton 
one-loop matrix elements start themselves to become numerically unstable 
because of the presence of inverse Gram determinants, which can become 
singular inside the integration region. 
\end{itemize}

Therefore, for all phenomenological applications to the thrust 
distribution~\cite{ourthrust}, we choose
$y_0=10^{-5}$. For applications to other event shapes, one expects a similar
behaviour, and one must first 
determine the value of $y_0$ required for reliable predictions in the
phenomenologically relevant range for that observable.

\noindent {\it The following has been added compared to the original 
version of the paper:} 
The terms of the form $\d\sigma^A_{NNLO}$ in the
five-parton and four-parton contributions to the $N^2$ and $N^0$ colour
factors
were only implemented in this revised version.
 They lead to changes in the numerical values of the NNLO
coefficients which are most pronounced in the approach to the two-jet
region. In a recent work, 
Becher and Schwartz~\cite{becher} have computed the logarithmically
enhanced terms which dominate the thrust distribution in the two-jet region
using soft-collinear effective theory.   They identified a disagreement with
our original numerical results
for the thrust distribution
in the two-jet region for these two colour factors. Our new
results are
displayed in Figures 9,~10 and 11, and are
now in full agreement with the results obtained in~\cite{becher}.
Our numbers also agree with the results obtained in the implementation 
of~\cite{weinzierl}.

In the genuine three-jet
region, which is relevant for precision phenomenology, the changes have a
minor numerical impact. The corrections to the NNLO $N^2$ and $N^0$ colour
factors also affect all other event shape distributions~\cite{event} in a
similar manner; minor numerical effects in the three-jet region, but more
significant effects in the two-jet region.

\section{Conclusions and Outlook}
\label{sec:conc}
In this paper, we provide a detailed description of the calculation of 
NNLO QCD corrections to three-jet production and related event shapes 
in electron-positron annihilation. At this order, three-parton, 
four-parton and five-parton subprocesses contribute. The three-parton 
and four-parton subprocesses contain explicit infrared singularities from 
loop corrections. Four-parton and five-parton subprocesses contain
singularities which only become explicit after integrating the contributions 
over the phase space relevant to the three-jet final states.
Those singularities arise when one or two partons become 
unresolved (collinear or soft).
For an infrared-safe observable, adding together all
infrared singularities, one observes  
a complete cancellation, resulting in an infrared-finite result. 

To implement the four-parton and 
five-parton processes in a numerical programme, one has to devise 
a procedure for extracting the implicit infrared singularities from them.
 We extract these 
singularities using subtraction terms which numerically subtract all 
infrared singularities from the five-parton and four-parton channels.
The subtraction terms are then integrated analytically and combined 
with the three-parton channel,
where they cancel all explicit infrared poles. The subtraction terms 
are derived using the antenna subtraction method, which is based on 
antenna functions encapsulating all unresolved partonic radiation emitted 
from a pair of hard radiator partons. 

Three-jet production at NNLO receives contributions from seven different 
colour factors. Among these, only the six colour factors of non-singlet 
configurations require subtraction, while the singlet colour 
factor is separately finite in all three partonic channels. We describe 
the construction of the antenna subtraction terms for the six non-singlet 
colour factors in detail, and demonstrate the 
cancellation of infrared poles. 

All partonic channels have been implemented in a parton-level event generator 
programme {\tt EERAD3}, which can be used to compute any infrared-safe 
observable related to three-jet final states in $e^+e^-$ annihilation. 
We devised various tests of the implementation, demonstrating in particular 
the correct numerical 
cancellation between matrix elements and subtraction terms and the 
independence on phase space restrictions deep inside the subtraction regions. 

We observe that the largest part of the NNLO correction is contained 
in the two leading colour factors $N^2$ and $N_F\, N$. The remaining 
four colour factors yield corrections at or below the ten per cent level.

First phenomenological results on the thrust distribution at NNLO were 
obtained already in an earlier paper~\cite{ourthrust}. In the thrust 
distribution, the NNLO corrections amount to about 15\% of the total result.
They are lower in magnitude than the NLO corrections, indicating the 
perturbative stability of this observable. Inclusion of the NNLO corrections 
considerably reduces the dependence of the result on the renormalisation 
scale.

At LEP, a wide variety of QCD event shapes was measured to high 
precision~\cite{lep}. The accurate extraction of the strong coupling 
constant $\alpha_s$ from these data sets was up to now limited by the 
theoretical uncertainty inherent to the available NLO calculations. We 
expect that our new NNLO results will improve  this situation considerably. 
Numerical studies of other event shape variables and of the three-jet 
rate are ongoing, and will be reported elsewhere. 

On approaching the two-jet limit ($(1-T)\to 0$ in the thrust distribution), 
one observes that the perturbative fixed-order expansion starts to break down 
due to the emergence of large logarithmic corrections at all orders in 
perturbation theory. By matching NLO calculations and NLL resummation 
of these logarithmic corrections~\cite{ctwt}, a reliable description 
of distributions over the full kinematic range could be accomplished. 
Such a matching is also possible at NNLO, where it requires the derivation 
of a number of new matching constants for each event shape. 

The subtraction terms derived here for $e^+e^- \to 3$~jets at NNLO 
can be transcribed to the crossed reactions $ep \to (2+1)$~jets and 
$pp\to V+1$~jet at NNLO
without much modification. In these cases, which 
involve partons in the initial state~\cite{hadant}, the same antenna functions
are used with different antenna phase spaces. To accomplish the 
above-mentioned NNLO calculations therefore still requires the analytical 
integration of the relevant antenna functions over the phase spaces
relevant to their initial-state kinematics, which appears feasible
with present technology.  
 NNLO calculations of other 
exclusive observables at hadron colliders, such as $pp \to 2$~jets, 
could also be carried out using the antenna subtraction method by 
constructing their subtraction terms along the lines described here.

\section*{Acknowledgements}
We wish to thank Zoltan Kunszt for many comments and 
for his continuous encouragement throughout the whole project. Much of this 
work has its foundations in earlier projects, carried out in a very
pleasant and fruitful collaboration with 
Ettore Remiddi, whom we would like to thank for these early contributions, and 
for many discussions.

This research was supported in part by the Swiss National Science Foundation
(SNF) under contracts PMPD2-106101  and 200020-109162, 
 by the UK Science and Technology Facilities Council and 
by the European Commission under contract MRTN-2006-035505 (Heptools).

First steps in this project were taken while the authors 
attended the KITP Programme 
``QCD and Collider Physics'', which was supported by the National Science
Foundation under Grant No.\ PHY99-07949.

\noindent {\bf Note added:} 
The results  presented in this paper were checked by two subsequent studies: 
the calculation of all logarithmically enhanced contributions to the 
thrust distribution by Becher and Schwartz~\cite{becher}, and an 
independent implementation of our subtraction formulae by 
Weinzierl~\cite{weinzierl}. 
These works uncovered numerical discrepancies in the two-jet limit 
of the observables 
in two of the six colour factor contributions: $N^2$ and $N^0$. 
In~\cite{weinzierl}, it was shown that the origin of these discrepancies is
in an oversubtraction of large-angle soft radiation. 
The present  version of the paper includes the content 
of the erratum to the journal version. Modifications
 were made in formula (3.23), sections 3.4 and 
14, and are clearly indicated in the text and in the figures. 
We would like to thank Thomas Becher and Stefan Weinzierl for 
useful discussions and for numerical
comparisons with the results of~\cite{becher} and~\cite{weinzierl}.

\bibliographystyle{JHEP}

\end{document}